\documentclass[journal=jpcck,manuscript=article]{achemso}

\usepackage[usenames]{color}

\usepackage{amsmath,amssymb,graphicx,textcase}
\usepackage{dcolumn}
\usepackage{bm}

\usepackage{graphicx}
\usepackage{dcolumn}
\usepackage{bm}

\usepackage{framed}

\newcommand{\nG}{n_\textrm{G}}
\newcommand{\dotchem}{CsPbBr$_3$ } 

\title{Intermittency of \dotchem perovskite quantum dots analyzed by an unbiased statistical analysis}

\author{Isabelle M. Palstra}
\affiliation{Institute of Physics, University of Amsterdam, Science Park 904, 1098 XH Amsterdam, The Netherlands}
\alsoaffiliation{Center for Nanophotonics, AMOLF,  Science Park 104, 1098 XG, Amsterdam
The Netherlands}
\author{Ilse Maillette de Buy Wenniger}
\affiliation{Center for Nanophotonics, AMOLF,  Science Park 104, 1098 XG, Amsterdam
The Netherlands}\author{Biplab K. Patra}
\affiliation{Center for Nanophotonics, AMOLF,  Science Park 104, 1098 XG, Amsterdam
The Netherlands}\author{Erik C. Garnett}
\affiliation{Center for Nanophotonics, AMOLF,  Science Park 104, 1098 XG, Amsterdam
The Netherlands}\author{A. Femius Koenderink}
\affiliation{Center for Nanophotonics, AMOLF,  Science Park 104, 1098 XG, Amsterdam
The Netherlands}
\email{f.koenderink@amolf.nl}

\date{\today}

\begin{document}

\begin{abstract}
We analyze intermittency in intensity and fluorescence lifetime of \dotchem perovskite quantum dots by applying unbiased Bayesian inference analysis methods.    We apply changepoint analysis (CPA) and a Bayesian state clustering algorithm to determine the timing of switching events and the number of states between which switching occurs in a statistically unbiased manner, which we have benchmarked particularly to apply to highly multistate emitters. We conclude that  perovskite quantum dots display a plethora of gray states in which brightness broadly speaking correlates inversely with decay rate, confirming the multiple recombination centers model. We leverage the CPA partitioning analysis to examine aging and memory effects. We find that dots tend to return to the bright state before jumping to a dim state, and that when choosing a dim state they tend to explore the entire set of states available.
\end{abstract} 

\maketitle 

\section{Introduction}
\label{sec:Intro}
Cesium lead halide perovskite nanocrystals, introduced in a seminal paper by Protesescu et al.\cite{Protesescu2015} have emerged as highly attractive quantum dots, with advantageous properties in comparison to traditional colloidal  II-VI semiconductor quantum dots. These include very large photon absorption cross sections\cite{Chatterjee2018},  a wide degree of tunability by both size and halide (Br, I, Cl) composition\cite{Protesescu2015}, and reportedly a very high luminescence quantum yield without the need of protecting the nanodot core with epitaxial shells, as is required for   CdSe quantum dots\cite{Swarnkar2015,Huang2017Angewandte}. Furthermore, inorganic halide perovskite materials generally show an exceptionally high tolerance to defects\cite{Kang2017}. Owing to these properties  perovskite nanocrystals are intensively pursued as solar cell materials\cite{Ling2019},  as emitters for LEDs,  display technologies and lasers\cite{Yakunin2015,LeVan2018,Zhang2019},  and could be interesting as single photon sources.  For the purpose of single photon sources, emitters need to satisfy a variety of requirements beyond brightness, tuneability and high quantum efficiency,  which includes single-photon purity,  tight constraints on inhomogeneous spectral broadening, and stability in spectrum, decay rate and intensity\cite{Lounis2005}. 

Perovskite nanocrystals unfortunately follow the almost universally valid rule that solid-state single emitters at room temperature show intermittency\cite{Swarnkar2015,Park2015,LiHuang2018,Gibson2018,Seth2016,Yuan2018,HouKovalenko2020}. In the field of II-VI quantum dots,  intermittency has been studied for over two decades, with the aim of identifying the nature of the usually two or three  distinct bright, dark, and gray states,  and the mechanism by which switching occurs, by analysis of the apparently discrete switching events between dark and bright states\cite{Cichos2007,Efros2016}, and concomitant jumps in spectrum and lifetime. For instance, for II-VI quantum dots a popular model (reviewed in Ref.\cite{Cordones2013}) \textcolor{black}{is the charging/discharging model whereby quantum dots turn from bright to dark upon acquiring a single charge. Many efforts have been made to explain the  typically power-law distributed residence times for on and off states, for instance through hypothesized mechanisms by which charges are exchanged with the environment\cite{CuiBawendi,Cichos2007,Frantsuzov2008,ToddKrauss2010}.  In this respect another powerful model is the socalled Multiple Recombination Center (MRC) model proposed by Frantsuzov et al. in 2009 \cite{Frantsuzov2009,Frantsuzov2010}, which argues that the wide distribution of on/off times underlying binary blinking is due to typically of order 10 available recombination centers. This model furthermore is applicable to a wide array of systems such as quantum dots, rods and wires, as it can explain also  qualitatively different intermittency behavior, such as systems that do not show two but multiple intensity levels, in function of assumed underlying recombination center physics\cite{Frantsuzov2013}}
 For  perovskite nanocrystals several groups studied  intermittency\cite{Swarnkar2015,Park2015,LiHuang2018,Gibson2018,Seth2016,Yuan2018,HouKovalenko2020} and found quite different physics. A set of works observe that perovskite quantum dots do not show bimodal behavior, like II-VI quantum dots do, but instead  a continuous distribution of states between which they switch\cite{Park2015,Seth2016,LiHuang2018}. \textcolor{black}{These observations are difficult to rationalize in a charging-discharging model, but can be described within the MRC model of Frantsuzov et al.\cite{Frantsuzov2009,Frantsuzov2013}, as pointed out for \dotchem dots by Li et al. \cite{LiHuang2018}.    Within this model  activation of individual recombination centers can provide a wide distribution over intensity and rate. An important observation  consistent with the MRC model is a linear dependence between emitted intensity and fluorescence lifetime.  Further evidence for multiple recombination center physics in the context of perovskite PL has been reported in the context of emitting perovskite microparticles that show no quantum confinement but nonetheless blink,  for instance in a recent report by Merdasa et al. that evidences   extremely efficient dynamic quenching sites  that can  appear and disappear \cite{Merdasa2017}.  In contrast to Refs. \cite{Park2015,Seth2016,LiHuang2018}, another group has analyzed intermittency on basis  of changepoint analysis and cluster analysis, which are Bayesian inference tools for the unbiased estimate of the number of states, reporting  that just of order 2-4 states are involved instead of a continuum \cite{Gibson2018}. }Finally, a recent study points at memory effects in  intermittency, visible in that work as  correlations between subsequent dwell times in the brightest state\cite{HouKovalenko2020}. \textcolor{black}{These reported memory effects for perovskite dots are similar to those observed over 15 years ago for II-VI dots by Stefani and coworkers \cite{Stefani2005}, which were explained by the MRC model \cite{Frantsuzov2010}.} 
 
Intermittency analysis is a field known to be fraught by statistical bias in analysis methods\cite{CuiBawendi}, primarily due to binning of data prior to analysis. This is a recognized problem already for interpreting data from bimodal dots. These artefacts may be even more severe for multilevel dots. In this work we report a study of cesium-lead-bromide nanocrystal intermittency, analyzing the photon statistics of a large number of dots using unbiased Bayesian statistics analysis tools, tracing brightness and fluorescence lifetimes simultaneously, and screening for memory effects. These Bayesian statistics methods were first developed by Watkins and Yang\cite{Watkins2005,Ensign2009,Ensign2010} and  have since been applied in a small set of papers to two/few-level  II-VI dots, and in one recent work to \dotchem dots~\cite{Gibson2018}. Our implementation is  through a freely available Python-based analysis toolbox \cite{Methodspaper}, which we have specifically benchmarked by Monte Carlo methods for application to highly multi-state, instead of bi-modal, systems.   In this work, a first main purpose  is to obtain statistically unbiased estimates, or at least lower bounds,  for the number of dark/gray states of perovskite quantum dots from a large number of single dot measurements. \textcolor{black}{Our conclusions solidly support \cite{Park2015,Seth2016,LiHuang2018}, but not Ref.\cite{Gibson2018}, since we find blinking between  a single well-defined bright state and a continuum --- or at least over 10 --- gray/darker states. These are findings that fall within the class of phenomena explainable by the MRC model \cite{Frantsuzov2009,Frantsuzov2013,LiHuang2018}.}  Next, our purpose is to screen for memory effects in residence times, intensity levels and decay rate sequences in data that has been separated in segments by unbiased changepoint analysis, thereby extending Ref.\cite{HouKovalenko2020}, which did not leverage the benefit of CPA analysis. 
We find no evidence for memory in residence times, but do find that a substantial fraction of dots tend to switch back and forth repeatedly between the quite uniquely defined bright state and the band of gray  states,  instead of jumping through all states in an uncorrelated random fashion.

\begin{figure*}
  \includegraphics[width=\linewidth]{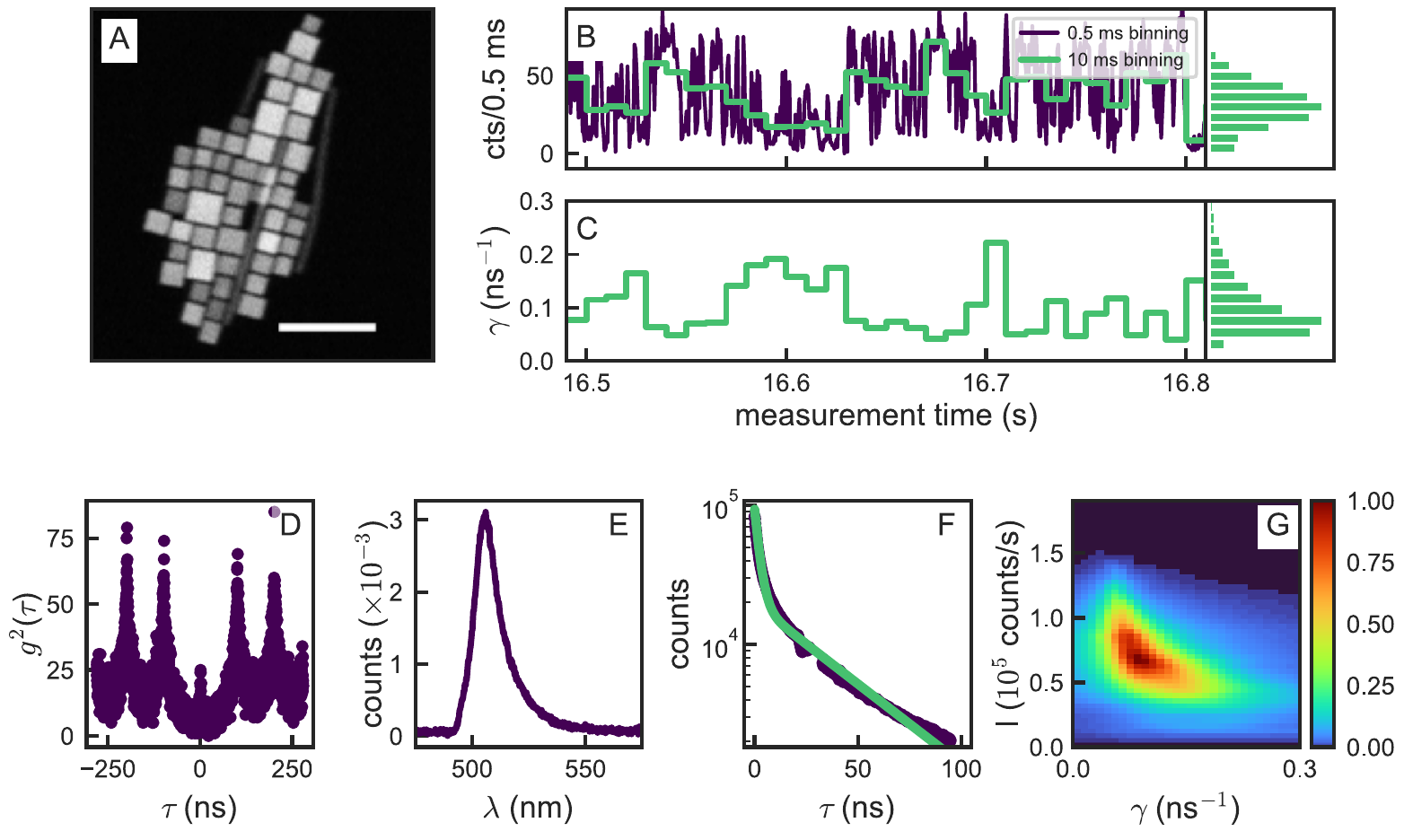}
  \caption{Properties of a \dotchem quantum dot.
  (A) a SEM image showing a cluster of \dotchem quantum dots. The scale bar is 100 nm.
  A time trace of the (B) intensity and (C) fluorescence decay rate  of a typical quantum dot when split into bins of 10 ms (green). We find a single peak in both the intensities and lifetimes around 60 counts/ms and 0.05 ns$^{-1}$, respectively. For visualization purposes, we also show the photon events binned into 0.5 ms bins (purple).
  (D) The $g^2(\tau)$ of this qdot. The dots used in this analysis were selected for having $g^2(0) < 0.5\cdot g^2(100 \mathrm{ns})$.
  (E) The spectrum of this qdot. We find a peak in the emission at 505 nm.
  (F) The decay trace of all the photon events combined. We have excluded an electronic artefact between 20 and 30 ns. We have a reasonable fit to a bi-exponential decay with rates of $\gamma_1, \, \gamma_2=0.43,\,0.03$ ns$^{-1}$, respectively.
  (G) The FDID diagram of this dot. We see a main peak at $I,\gamma=$ $0.7\times 10^5$ cts/s and 0.09 ns$^{-1}$
  }
  \label{fig:binned_pp}
\end{figure*}

\section{Experimental methods}
\label{sec:binning}
To introduce our measurement protocol and the photophysics of the \dotchem dots at hand we first present in Figure~\ref{fig:binned_pp}  the typical behavior of a \dotchem quantum dot,  as analyzed with the standard approach of plotting time binned data. \textcolor{black}{We prepare quantum dots according to a modified literature report~\cite{Protesescu2015,Patra2020}.}
\textbf{Preparation of cesium oleate.}
We load 0.814 g of Cs$_2$CO$_3$ into a 100 mL 3-neck flask along with 40 ml of octadecene (ODE) and 2.5 ml of oleic acid (OA) and dry this for 1 hour at 120 $^\circ$C. This is then heated under an N$_2$ atmosphere to 150 $^\circ$C until all Cs$_2$CO$_3$ has reacted with OA. To prepare for the next step, we preheat the resulting cesium oleate to 100 $\circ$C before injection. This is necessary as it precipitates out from ODE at room temperature.

\textbf{Synthesis of CsPbBr$_3$ nanocubes.}
We load 0.188 mmol of PbBr$_2$ in 5 ml of ODE, 0.5 ml of oleylamine and 0.5 ml of OA into a three-neck round bottom flask and dry this under vacuum at 120 $^{\circ}$C for an hour, after which the reaction atmosphere is made inert by flushing the flask with N$_2$. After complete solubilization of PbBr$_2$, the temperature is raised to 200 $^{\circ}$C and 0.4 ml of the preheated cesium oleate is injected into the three-neck flask. After the injection, the color of the solution turns from colorless to greenish yellow indicating the formation of perovskite cubes. Then we lower the temperature to 160 $^{\circ}$C and anneal the solution at that temperature for 10 min to get uniform size dispersion of the cubes. After that we cool down the solution using ice water bath for further use.

\textbf{Isolation and purification of CsPbBr3 cubes.}
After the synthesis, we centrifuge our solution twice to collect the cubes.
First, we take 1 ml from the stock solution just after the synthesis and centrifuge at 8000 rpm for 20 min to collect all CsPbBr3 particles from the solution. We discard the supernatant, gently wash the inner wall of the tube using tissue paper and add 2 ml of toluene to disperse the CsPbBr3 solid. 
The second step of centrifugation is run at 2000 rpm for 5 min to get rid of all the particles that are too large. In the supernatant, we have 2 ml of toluene containing \dotchem nanocubes having a size distribution around 10-15 nm.  As the scanning electron micrograph in {Figure~\ref{fig:binned_pp}(A)} shows  our quantum dots are essentially cubic in shape.  

Before the measurement, about 400 $\mu$L of the solution is spin coated at 1000 rpm on glass coverslips that had been cleaned in a base piranha solution. In order to protect the quantum dots from moisture in the air, the quantum dots were covered by a layer of PMMA (8\% solid weight in anisole), by spincoating for 60 s at 4000 rpm. Quantum dots stored in solution were found to be unchanged in their properties over $\gtrsim 6$ months. For the optical experiments we prepared microscope slides with samples from solutions no more than 1 month old, and then performed microscopy on a given substrate within a time span of 7 days. We found no differences between data taken directly, and data taken after 7 days. 

\textbf{Single emitter microscope.}\label{subsec:optic_setup}
For optical characterisation and measurements, we use an inverted optical microscope to confocally pump the dots at 450 nm (LDH-P-C-450B pulsed laser, PicoQuant) at 10 MHz repetition rate of $<70$~ps pulses, with 90 nW inserted into the microscope. An oil objective (Nikon Plan APO VC, NA=1.4) focuses the pump laser onto the sample and collects the fluorescence. The excitation provides similar pulse energy density as in\cite{Gibson2018}at the lowest energy density $\langle N\rangle \ll 1$ quoted in that work.  With the estimated efficiency of our set up, the excitation probability per optical pulse is estimated at $<$0.1 from the count rate. The fluorescence from the sample is directed either to a camera (PCO.edge 4.2, PCO AG), a spectrometer (PI Acton SP2300) or two fiber-coupled avalanche photodiodes (APDs) (SPCM-AQRH-14, Excelitas) in a Hanbury-Brown \& Twiss configuration. The APDs are coupled to a photon correlator (Becker \& Hickl DPC-230) that measures the absolute photon arrival times.

\textbf{Measurement protocol.} Using the camera and wide-field pump illumination, we select an emitter that appears to be diffraction-limited. After driving it to the laser spot, we do a time-correlated single-photon counting (TCSPC) measurement to collect photon arrival times.  To calculate the photon correlations we use a home-built TCSPC toolkit that utilizes the algorithm developed by Wahl et al.\cite{Wahl2003} to calculate $g^2(\tau)$ and the lifetimes for the different emitters and for the individual CPA segments.  From  $g^2(\tau)$ we select the emitters with a strong anti-bunching signal (normalized $g^2(\tau=0)<0.5$) \textcolor{black}{to ensure single quantum emitter behavior}. Of the 75 dots measured, 40 passed this test. \textcolor{black}{We note that within those 40 dots we found no systematic correlations between any of the variables (brightness, decay rates, apparent number of levels, residence time power law exponent) and the normalized $g^2(\tau=0)$ value. }Our TCSPC measurements are taken over 120 seconds of acquisition time. We note that in our decay traces taken using a Becker-Hickl DPC 230 photon-counting and correlator card in reverse start-stop configuration, a small time interval centered at around 30 ns is subject to an electronic artefact which we attribute to a ringing in the DPC-230 TDS timing chip. Therefore we exclude this time interval for decay rate fitting.

\textbf{Initial characterization.} Figure~\ref{fig:binned_pp} presents initial characterization of an exemplary single dot on basis of standard timebinned analysis, where the data is sliced in 10 ms long segments, to each of which intensity and decay rate is fitted.  Throughout this work we consider photon counting data, in which absolute time-stamps are collected with 0.165 ns resolution for all collected photons and concomitant excitation laser pulses, on two avalanche photodiodes (APDs) in  a Hanbury-Brown and Twiss configuration. This allows to construct \emph{a posteriori} from one single data set  the intensity, fluorescence decay rate, and the $g^{(2)}(\tau)$ photon-photon correlation.  In our optical measurements we post select all single nanoparticles on basis of photon antibunching ($g^{(2)}(\tau=0)<0.5$).  For the example at hand, the selected emitter shows clear intermittency in intensity and decay rate (panels Fig. \ref{fig:binned_pp}(B,C) discussed further below), while   {Fig.~\ref{fig:binned_pp}(D)} shows a marked anti-bunching at zero time delay in the $g^{(2)}(\tau)$ that is constructed from the full photon record. The quantum dot in  {Fig.~\ref{fig:binned_pp}(E)} shows a time-averaged emission spectrum that peaks at around 505~nm and has a spectral FWHM of 20 nm, which is consistent with reports by Protesescu et al.\cite{Protesescu2015}, and together with the antibunching photon statistics points at quantum confinement.  The time-integrated  fluorescence decay trace ({Fig.~\ref{fig:binned_pp}(F)}) is markedly non-single exponential. Fitted to a double exponential decay we find decay rates of $\gamma_1, \gamma_2=0.43,\,0.03$ ns$^{-1}$. We must note, however, that a double exponential is often not sufficient to fit these emitters, and typical decay rates for our dots range from $0.05$ to $0.9$~ns$^{-1}$. At these decay rates, the fastest decay rate component of the quantum dots generally span at least 10 timing card bin widths.

  {Figure~\ref{fig:binned_pp}(B, C)} shows just a fraction of the intensity and decay rate time trace, plotted according to  the common practice of partitioning the single photon data stream in bins. The fluorescence decay rate for each bin is obtained by fitting data within each 10 ms bin to a single exponential decay law employing a maximum likelyhood estimator method that is appropriate for Poissonian statistics~\cite{Bajzer1991}.
As expected from prior reports on single pervoskite nanocrystal blinking \cite{Swarnkar2015,Park2015,LiHuang2018,Gibson2018,Seth2016,Yuan2018,HouKovalenko2020},  the intensity and decay rate time trace show clear evidence for intermittency. The intensity varies from essentially zero to 150 counts per ms.  {Figure~\ref{fig:binned_pp}(B, right panel)} shows a histogram of intensities, binned over the entire time trace (for all dots in this work, 120 s, or till bleaching occurred). The histogram shows a broad distribution of intensities with most frequent intensities around 60 cts/ms. This is in contrast with the  typical bi- or trimodal physics of II-VI quantum dots, which usually show distinct bright and dark states\cite{Kuno2000,Cichos2007,Frantsuzov2008,ToddKrauss2010,Cordones2013,Efros2016}. 
However, the width of the peak well exceeds the Poisson variance expected at these count rates, suggesting that there are many intensity levels. The decay rate histogram also displays intermittent behavior, in step with the intensity blinking. The most frequent decay rate is around 0.07~ns$^{-1}$.  {Fig.~\ref{fig:binned_pp}(G)} displays a \textit{Fluorescence Decay Rate Intensity Diagram} [FDID], a 2D histogram displaying the frequency of occurrence of intensity-decay rate combinations.
This type of visualization was first introduced by~\cite{Bae2016, Galland2011, Rabouw2013} to identify \textit{correlations} between intensity and fluorescence decay rate (FLIDs in those works, using lifetime instead of decay rate).  For II-VI quantum dots, FDID diagrams typically separate out  bright and slowly decaying states from dark, quickly decaying states\cite{Galland2011,Rabouw2013}.   Instead, for the perovskite quantum dot at hand, the FDID diagram presents a broad distribution with a long tail towards dim  states with a fast decay.

The picture that emerges from  {Fig.~\ref{fig:binned_pp}} is consistent with recent observations of several groups \cite{Park2015,Seth2016,LiHuang2018}, showing a a continuous distribution of dark, gray states.  This should be contrasted to typical II-VI quantum dot behavior in which blinking usually involves just two or three apparent intensity levels, and also the  recent report by~\cite{Gibson2018}  on very similar \dotchem dots, but taken under very low repetition rate excitation conditions (fs pulses at very low repetition rate, as opposed to picoseond pulses at $\geq 10$ MHz --- at similar $\langle N \rangle < 0.1$).

\section{Computational methods}
\label{sec:method}
Since extreme caution is warranted when scrutinizing photon counting statistics to determine quantitative intermittency metrics due to artefacts of binning\cite{Watkins2005, Bae2016, Crouch2010, Ensign2009,Ensign2010}, we proceed to analyze the data of a large number of dots with  state-of-the-art bias-free statistical analysis  to determine a lower bound to the number of involved states, and the switching dynamics and memory effects therein. We apply tools of Bayesian statistics, specifically, changepoint analysis (CPA) to partition the data in segments separated by switching events, and level-clustering to determine (a lower bound to) the  number of states, as a rigorous and bias-free approach to investigate the intermittency of quantum dots. These tools were first proposed  by Watkins and Yang\cite{Watkins2005}, and later also used and extended  in the context of quantum dot intermittency by \cite{Zhang2006,Rubinsztein2009,Ensign2009,Ensign2010,Cordones2011,Schmidt2012,Schmidt2014,Bae2016,Gibson2018,Rabouw2019}.  
We refer to Ref.~\cite{Methodspaper} for our freely available implementation and a detailed description of benchmarking of this tool set.  Here we summarize just the salient outcomes  relevant for this work, obtained by extensive Monte Carlo based benchmarking to determine the performance of CPA and clustering for highly multilevel emitters.

CPA performs segmentation of the time record of single photon counting events into intervals within which the count rate is most likely a constant value, delineated by switching events or `changepoints' at which the count rate changes,  in as far as can be judged given the shot noise in the data. Since CPA works on a full time series with many jumps by finding a single jump at a time, and successively subdividing the time stream until segments with no further jumps are found, the ultimate performance is ultimately set by how well CPA can pinpoint in the last stage of the subdivision single jumps in short fragments of the photon stream.  For significant intensity contrasts  CPA detects changepoint in very short fragments (e.g.,  to accurately resolve a jump with a  5-fold count rate contrast, a record of just 200 photons suffice), with single-photon event accuracy. Smaller jumps are missed unless fragments are longer (e.g., factors 1.5 contrast jumps require fragments of ca. $10^3$ photons for near sure ($>90$\%) detection. At typical \textit{practical} count rates of $10^5$ cts/s this means that switching events further apart than 10 ms are accurately identified as long as jump contrasts exceed a factor 1.5 (ca. 100 ms for contrasts as small as 1.2).  Switching events that are even closer in time are missed by CPA. This is  intrinsic to the photon budget, i.e., the ultimate information content in the discrete event time stream fundamentally does not allow pinpointing even more closely spaced switching evens.

After dividing the time trace into segments spaced by changepoints, one is left with sequences containing the residence times $T_q$ for each segment,  photon counts $N_q$ and instantaneous segment intensities ($I_q=N_q/T_q$), as well as decay rates $\gamma_q$, obtained by maximum likelihood fitting of the decay trace from each segment to a single exponential decay.  The question how many actual intensity levels most likely underlie the measured noisy sequence $\mathcal{I}_{m_r}$ can be determined using Watkins \& Yang's clustering algorithm\cite{Watkins2005}. While Watkins and Yang considered Poisson distributed noise, as in this work, we recommend also the work of Li and Yang \cite{LiYang2019} as a very clear explanation of the method, though applied to Gaussian distributed noise. The idea is that expectation-maximization is used to group the most similar segments together into $\nG$ intensity levels, where $\nG=1,2,3,\ldots$. After this, the most likely number of levels, $\nG=m_r$, required to describe the data, given that photon counts are Poisson distributed, can be determined by a so-called Bayesian Information Criterion (BIC)~\cite{Watkins2005,LiYang2019}. We have extensively verified by Monte Carlo simulations the performance of CPA and level clustering for dots with \textit{many} assumed discrete intensity levels in a separate work\cite{Methodspaper}. 
In brief, at  small photon budgets in a total time series, only few levels can be detected, but  conversely at the total photon budgets in this work, exceeding $5\cdot 10^6$ events, clustering has a $>95\%$ success rate in pinpointing the exact number of levels in dots with at least 10 assumed intensity levels. Moreover, for photon budgets that are too small to pinpoint all levels exactly (e.g., at $10^4$ counts in a total measurement record, only up to 4 levels can be accurately discerned),  clustering always returns a \textit{lower bound} for the actual number of intensity states.

\section{Results and discussion} 
\subsection{Changepoint analysis and FDID diagrams}
\label{sec:CPArealdata}
\begin{figure}
  \includegraphics[width=\linewidth]{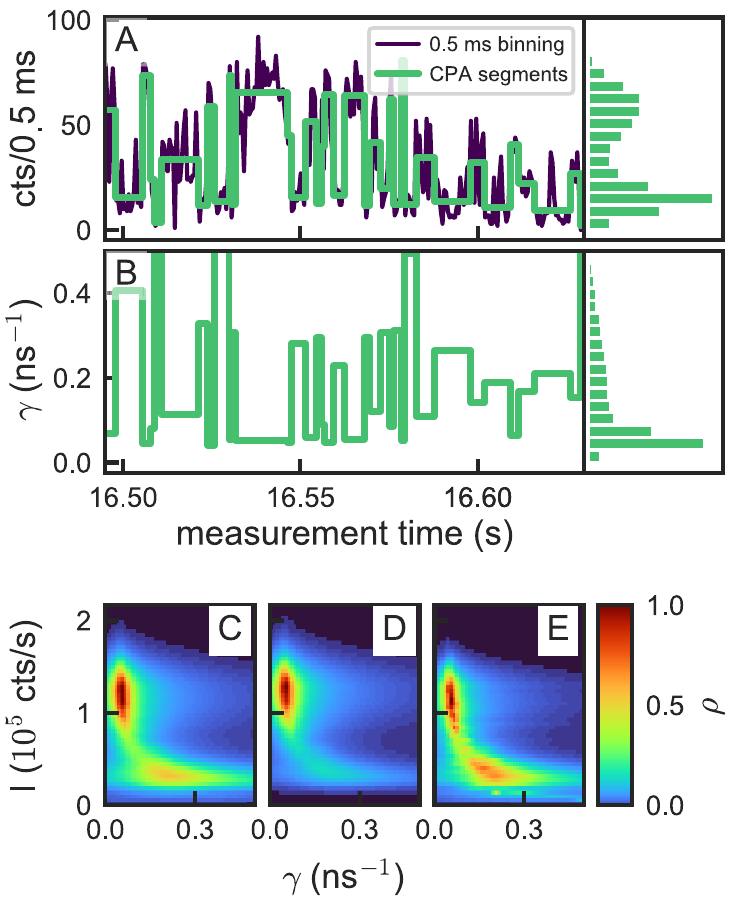}
  \caption{(A) An example of the intensity time trace of a measured quantum dot (purple, binned in 0.5 ms bins for visualisation purposes), and the intensity segments found by CPA (green). In the lower panel the lifetime for the found CPA segments is shown. On the right are histograms of the occurrences of the intensities for both treatments with segments weighted by their duration.
  (B-D) Three FDID plots weighting each CPA segment (B) equally, (C) by their number of counts, (D) by their duration. The choice of weights puts emphasis on different parts of the intensity-decay rate diagram, as they report on differently defined probability density functions.
}
  \label{fig:timetrace_FDIDs}
\end{figure}

We have applied the unbiased CPA analysis and Bayesian inference tools to data from 40 single  \dotchem quantum dots. We first discuss an exemplary single dot as example, and then discuss statistics over many single dots. The example dot is identical to the one considered in Figure~\ref{fig:binned_pp} and refer to the supporting information for results on all dots.
In  {Figure~\ref{fig:timetrace_FDIDs}A} we see that CPA is able to  accurately follow the intensity trace of a typical \dotchem quantum dot. We show only a section of the total measurement for clarity, and strictly for plotting purposes only,  binned the photon arrival times in 0.5 ms intervals. Note that this binning is only for visualization, and does not enter the CPA algorithm. {Figure~\ref{fig:timetrace_FDIDs}B}  displays the fitted decay rates for the same selected time interval, obtained by fitting each of the identified segments.   
The right-hand panels of  {Figures~\ref{fig:timetrace_FDIDs}A} and B show histograms of intensity and lifetime as accumulated over the full time trace. It should be noted that these histograms are intrinsically different from those in Figure~\ref{fig:binned_pp} for two reasons. First, binned data has entries from bins containing jumps, leading to a smearing of the histogram. Second,  since histogramming of segment values $I_q$ is agnostic to segment \textit{duration}, events are differently weighted.  Thus the histogram of intensities now shows a bimodal distribution.  The histogram of the decay rates still exhibits only a single  peak at ca. 0.05  ns$^{-1}$.

Next we construct correlation diagrams of fluorescent decay rate versus intensity (FDIDs) from CPA data. Customarily FDIDs are 2D histograms of intensity and decay rate as extracted from equally long time bins in binned data. As the length of segments found by CPA can vary over many orders of magnitude, an important questions is with what weight a given segment should contribute to a CPA-derived FDID. A first approach is to give all segments an equal contribution to the FDID, which emphasizes the probability for a dot to jump to a given intensity-decay rate combination.  Alternatively, one could weight the contribution of each segment to the histogram  by the amount of counts it contributes.  This histogram hence emphasizes those entries that contribute the most to the time-integrated observed photon flux. Lastly, if one uses the segment durations as weights for contribution of segments to the FDID  one obtains an FDID closest in interpretation to the conventional FDID diagram, which presents the probability density for being in a certain state at a given time. 
  {Figures~\ref{fig:timetrace_FDIDs}D, E and F}  provide all three visualizations. The data shows variations in intensity levels over approximately a factor 10, with concomitant decay rates also varying over an order of magnitude. Overall, all diagrams suggest an inverse dependence  qualitatively consistent with the notion that the dots experience a fixed radiative rate, yet a dynamic variation in the number of available non-radiative decay channels, that make the dot both darker and faster emitting.  This inverse dependence was also observed for perovskite dots by \cite{LiHuang2018}, and can be explained by the MRC model \cite{Frantsuzov2009,Frantsuzov2013}. The unweighted and photon count weighted FDIDs show a peak at similar intensity and decay rate at $\gamma,\,I$ = 0.06 ns$^{-1}$, 12$\times 10^4$ s$^{-1}$, indicative of the most frequently occuring intensity/rate combination that is simultaneously the apparent bright state. The different FDID weightings emphasize different aspect of the data.  For instance, weighting by counts highlights mainly the emissive states and underrepresents the long tail of darker state, with respect to the other weighting approaches. This qualitative difference can result in a quantitative difference for extracted parameters, such as the apparently most frequently occurring combination of intensity and decay rate.

\begin{figure}
  \includegraphics[width=\columnwidth]{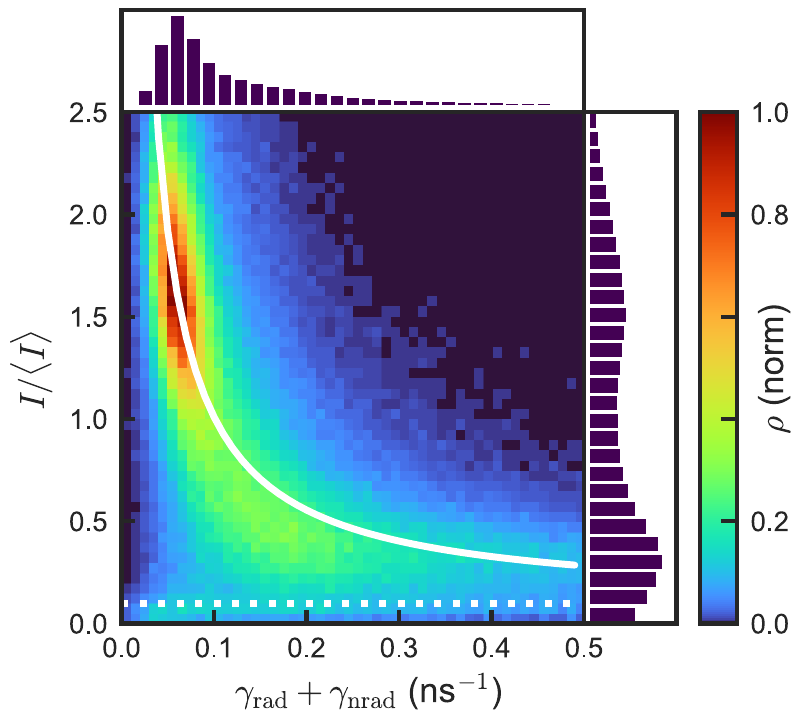}
  \caption{FDID of all 40 single dots, obtained by summing single dot FDIDs for which the segment intensities were normalized to the mean intensity. A simple histogramming was used (no specific weighting of entries by duration or  counts). Overplotted is a parametric curve  of the form $(\gamma_r+\gamma_{nr}, B+ I_0\gamma_r/(\gamma_r+\gamma_{nr}))$ with as input a fixed value $\gamma_r$, and a background $B=0.06I_0$, with $I_0$ adjusted to match the peak in the FDID, and $\gamma_{nr}$ scanned.
  }
 \label{fig:superfdid}
\end{figure}

FDIDs for essentially all dots (see supporting information) are much like the example shown in Figure~\ref{fig:timetrace_FDIDs}, showing a slow decaying bright state with a long tail towards both lower intensity and faster decay.  In fact, we can collapse the  FDIDs of all 40 dots onto each other by summing histograms (no weighting by, e.g., segment duration) of \textit{normalized} intensity $I/\langle I\rangle$ versus  $\gamma$, which further underlines this generic behavior, see Figure~\ref{fig:superfdid}). An appealing explanation for the observed dynamics is if the perovskite  dots are characterized by always emitting from one unique bright state that is efficient and has a slow rate of decay  $\gamma_r$ [labelled as radiative decay rate],  while suffering fluctuations in both brightness and rate through  jumps in a nonradiative rate $\gamma_{nr}$, as in the MRC model \cite{Frantsuzov2009,Frantsuzov2013,LiHuang2018}. In this picture, one would expect the FDID feature to be parametrizable as $I\propto B+ I_0\gamma_r/(\gamma_r+\gamma_{nr})$.   The feature in the collapsed FDID plot can indeed be reasonably parametrized as such a hyperbola. This parametrization is consistent with Ref.\cite{LiHuang2018} in which a linear relation between intensity and fluorescence lifetime was reported.  The required radiative decay rate for the bright state is  $\gamma_r\sim 0.075$~ns$^{-1}$, while the parametrization requires a residual background $B=0.06I_0$.  This residual background is not attributable to set up background or substrate fluorescence, suggesting a weak, slow luminescence component from the dots themselves. Moreover, we note that the FDID feature clearly has a somewhat stronger curvature then the hyperbolic parametrization (steepness of feature at $\gamma < 0.1$ ns$^{-1}$, and $I/
\langle I \rangle > 1.0$).

\subsection{Clustering analysis}

\begin{figure}
  \includegraphics[width=\linewidth]{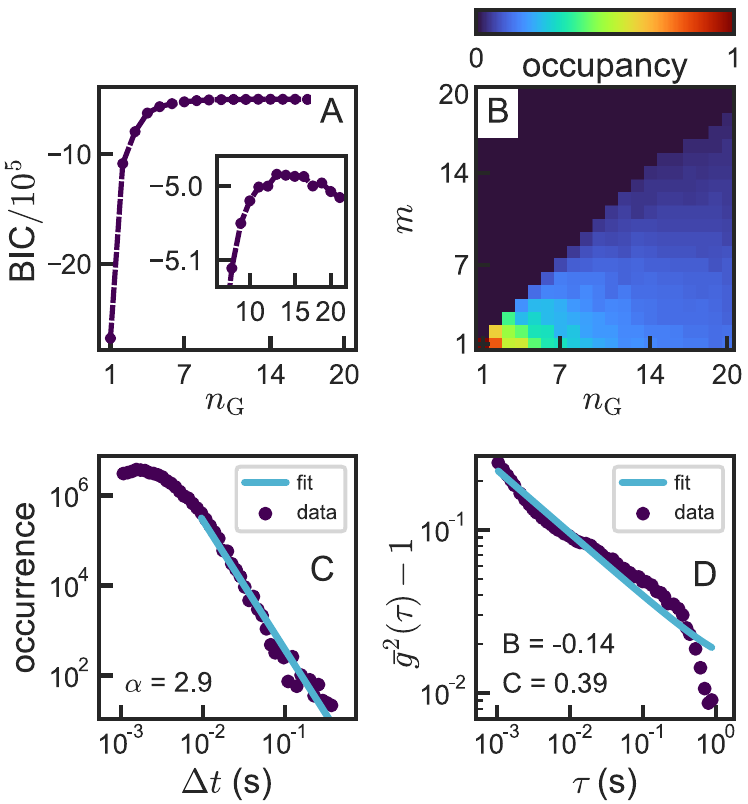}
  \caption{
  (a) BIC criterion for level clustering analysis of a single \dotchem quantum dot. We see that the BIC of this dot peaks at $\nG=13$.
  (b) The occupancy diagram of the same quantum dot. The number of occupied states keeps growing with the number of available states, saturating around $\nG=15$.
  (c) Histogram showing the durations of CPA segments of the \dotchem quantum dots as scatterplot. For this dot we fit (line) a powerlaw tail with an exponent of $\alpha=2.9$.
  (d) Long-term autocorrelation trace of a single \dotchem quantum dot.  Quoted coefficients  $B,C$ refer to parameters in $At^{-C}exp(-Bt)$ fitted (line) to the data (scatter plot).
  }
  \label{fig:switch_longcorr_BIC_weight}
\end{figure}

The FDID diagrams at hand qualitatively support the continuous distribution of states also observed by Refs.\cite{Park2015,Seth2016,LiHuang2018}. As quantification of the number of states involved we perform  clustering analysis \cite{Watkins2005, Ensign2009, Methodspaper} to estimate  the most likely number of intensity states describing the data on basis of Bayesian inference. A plot of the Bayesian Information Criterion as function of the number of levels $\nG$ allowed for describing the data of the specific example dot at hand is shown in Figure~ {\ref{fig:switch_longcorr_BIC_weight}A.} Strikingly, the BIC does not exhibit any maximum in the range $\nG=1\ldots 5$, but at $\nG=13$. Recalling that the BIC criterion in clustering analysis for multistate dots at finite budget generally report a lower bound, this finding indicates that the data for this dot requires at least as many levels to be accurately described, if a discrete level model is at all appropriate.  

Similar conclusions can be drawn from Figure~ {\ref{fig:switch_longcorr_BIC_weight}B}.  We have found in Monte Carlo simulations that if one allows the level clustering algorithm  to find the best description of intensity traces in $\nG$ levels for dots that in fact have just $m< \nG$ levels, then the returned description of the data utilizes just $m$ levels, with the remaining levels having zero occupancy in the best description of the data returned by the algorithm.  Figure~ {\ref{fig:switch_longcorr_BIC_weight}B} shows the occupancy assigned by the clustering algorithm for our measured quantum dot as function of the number of states offered to the algorithm for describing the segmented intensity trace. Each additional state offered to the clustering algorithm is in fact used by the algorithm, whereas Monte Carlo simulations have shown that at the photon budgets involved ($5.5\times10^6$ photons) the clustering algorithm generally does not assign occupancy to more than $m$ levels to simulated $m$-levels dots \cite{Methodspaper}. The occupancy diagram hence confirms the conclusion from the BIC criterion that the dot at hand requires many levels, or even a continuous set of levels, to be described. 

For all 40 dots we extracted wavelength, brightness, and performed the same  CPA and clustering analysis as for the example dot. Moreover, we examined segment duration  statistics for power law exponents. The supporting information contains a detailed graphical overview of the CPA results for each of the 40 dots, while summarized results  are shown in  {Figure~\ref{fig:alldots}}.    {Figure~\ref{fig:alldots}A}  shows that the dots have a low dispersion in peak emission wavelength, with emission between 500 and 510 nm.  All considered dots offered between 2 and 8$\times 10^6$ photon events ({Figure~\ref{fig:alldots}B}) for analysis (120 seconds collection time, or until photobleaching). The mean intensity per measured dot  (histogram  {Figure~\ref{fig:alldots}C}) is typically in the range from  $15\times10^3$ and $80\times10^3$ cts/s, with one single dot as bright as  $110\times10^3$ cts/s.  According to the Monte Carlo analysis in \cite{Methodspaper} the total collected photon count for all dots therefore provides a sufficient photon budget to differentiate with high certainty at least up to 10 states.  We can  thus with confidence  exclude that intermittency in these perovskite quantum dots involves switching between just two or three states as in usual quantum dots. Instead  any physical picture that invokes a set of $m$ discrete levels requires a description in upwards of $m=10$ levels.  In how far further distinctions between $>10$ discrete levels, or instead a continuous band can be made on basis of data is fundamentally limited by the finite photon budget that can be extracted from a single emitter. This  quantification matches the observation  in Ref. \cite{HouKovalenko2020, LiHuang2018, Seth2016, Park2015} (based on examining time-binned FDID diagrams).\textcolor{black}{The main other work that applied CPA tools to perovskite dot by Gibson et al. \cite{Gibson2018}, however,  arrived at an estimate $m_r=2.6$,  which is at variance with our findings as well as with  Ref. \cite{HouKovalenko2020, LiHuang2018, Seth2016, Park2015}.}  

\textcolor{black}{This difference may be attributable  to the different  excitation conditions that are unique to Gibson et al.\cite{Gibson2018} relative to all other works.   Gibson et al.\cite{Gibson2018} report that the lower excitation duty cycle resulting from both lower repetition rate (sub-MHz)  and shorter pulse  (order 0.1 ps versus 10 ps) excitation, promotes photostability.  We note that from a purely experimental point of view, this benefit is not immediately clear to us,  at least not when expressing photostability in number of excitation cycles as we observe dots for 2 minutes at 10 Mhz repetition rate, versus 10 minutes at 0.3 MHz in Ref. \cite{Gibson2018}, at similar count rates per excitation pulse.  Among possible explanations, we can exclude effects purely due to thermal load: according to established thermal analysis of nanoparticles under pulsed excitation \cite{Baffou2013} a nanocrystal and its environment cool down within nanoseconds after excitation, meaning that although our work and Refs. \cite{HouKovalenko2020, LiHuang2018, Seth2016, Park2015} use  higher repetition rates (up to 20 MHz), there is no ground to believe that heating effects build up more strongly than in Ref. \cite{Gibson2018}.   Regarding electronic processes, several works recently claimed that intermittency in perovskite dots arises not from one, but from several competing mechanisms including non-radiative bandgap carrier recombination,  trion-mediated recombination, and hot carrier blinking \cite{Samanta2018,Samanta2019}.  There is a wide range of involved time constants,   some of which are hypothesized to be slower than  the typical MHz  laser repetition rates.  For instance Ref. \cite{Samanta2019} argues that  there is evidence for shallow trap strates with long lifetimes ($>250$ ns), and some reports claim microsecond timescale delayed emission for lead halide perovskite quantum dots \cite{Chirvony2017,Yinthai2017,Vonk2020} which is hypothesized to originate from  carrier trapping/detrapping between the band edge state and energetically shallow structural disorder states.   We note that this means that laser repetition rate is ideally a variable in experiments. However it is not trivial to extend CPA studies to deep sub-MHz  repetition rates as the concomitant fall in overall count rate means that the tail of dark states will become comparable in strength to the fixed background of the single photon detectors (which contribute of order 250 cts/s in our work, summing over both Excelitas detectors).}

\subsection{Residence times}
In Figure  {\ref{fig:switch_longcorr_BIC_weight}C} we show a histogram of the segment lengths found by CPA. In other works, on-states and off-states are often  separated explicitly by thresholding following which on-times and off-times are separately analyzed, for instance to ascertain the almost universally observed power-law dependencies and their exponents. In the case of our \dotchem quantum dots a level assignment in on and off states is not obvious. Therefore we simply combine \textit{all} segment lengths irrespective of intensity level in a single histogram. These switching times are  power-law distributed, at least from minimum time durations of $10$ ms onwards.  The short-time roll off is consistent with the limitations of the information content of the discrete photon event data stream: for segments shorter than 50 photons or so, even if physically there would be  a jump, the photon number would not suffice to resolve it. Thus the roll-off does not exclude that power-law behavior  also occurs for shorter times,  but instead signifies that the testability of such a hypothesis is fundamentally limited. Fitting the power law $t^{-\alpha}$ for time $>10$~ms indicates a power-law exponent of $\alpha=2.9\pm0.1$. 

\begin{figure*}
  \includegraphics[width=\linewidth]{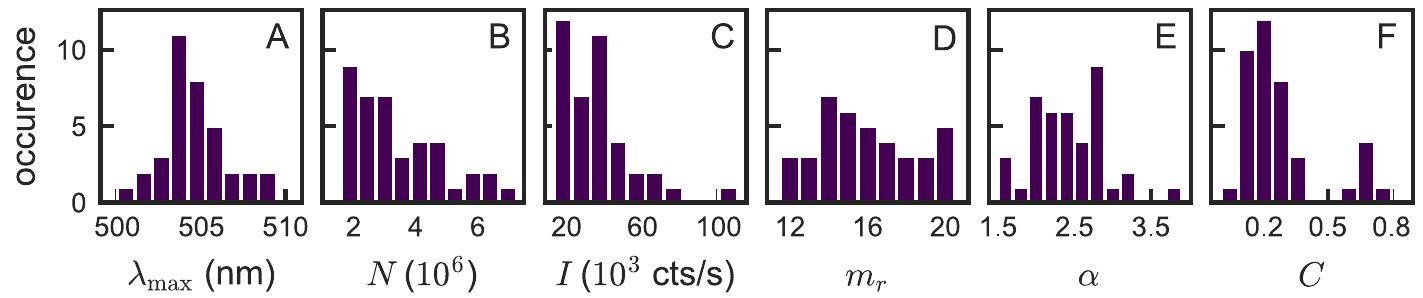}
  \caption{Summary of the behavior of the 40 measured single quantum dots. We show the distribution of found (A) peak wavelengths, (B) total photon count, (C) intensities, (D) most likely number of states, (E) powerlaw exponents of the switching time $\alpha$, and (F) the power-law exponent of the autocorrelation $C$.
  }
  \label{fig:alldots}
\end{figure*}

The peculiar segment-duration power law statistics with exponent $\alpha\sim 2.5$ of our example dot also extends to the full ensemble.   {Figure~\ref{fig:alldots}E} shows the distribution of power-law exponents  that were fitted to the tail of the switching time histograms. We find a broad distribution of power-law exponents ranging from 1.5 to 3.0, with the bulk of the dots showing exponents in the range of $\alpha=2.0$ to 3.7.  These values are significantly higher than the values found for many semiconductor quantum dots, which generally are close to 1.5\cite{Frantsuzov2008}.   Also, these values are significantly higher than the exponents reported for on-times of \dotchem dots extracted from intensity-thresholded time-binned data. We note that  one can (somewhat arbitrarily)  threshold CPA-segmented data in an attempt to isolate 'on-times' for the bright state from the `residence times' associated with the long dark/gray tail of states. Doing so with thresholds $I/\langle I \rangle > \ 1.3$ (on-state) and $I/\langle I \rangle  < 0.7$  (tail of gray/dark states) estimated from  Fig.~\ref{fig:superfdid}, resulted in residence-time histograms for on- and off-times with similarly high slopes as we obtain for the full set. We thus find no support in our data for power laws generally being close to 1.5 or even below, as reported in other recent reports\cite{HouKovalenko2020,Gibson2018}. We note that apart from the methodological difference of not working with binned thresholded data but with CPA analysis,  also the selection of dots reported on may matter. In this work we report on all dots identified as single photon emitters by their $g^{(2)}(0)$. Instead in Ref.~\cite{HouKovalenko2020}  dots are reported to have been selected as those for which inspection of binned time traces suggested the most apparent contrast between bright and dim states, qualitatively appearing closest to bimodal behavior.  According to our analysis of FDIDs and in light of the MRC model, this post-selection may not single out the most representative dots.

An alternative approach to quantifying blinking statistics and power law exponents that requires neither thresholding binned data nor CPA is to simply determine intensity autocorrelation functions $g^{(2)}$ for time scales from ms to seconds, as proposed by Houel et al.\cite{Houel2015}.  According to Houel et al.\cite{Houel2015} the normalized autocorrelation minus 1 may be fit with the equation $At^{-C}\exp(-Bt)$.   { Figure~ {\ref{fig:switch_longcorr_BIC_weight}(D)}} show such an analysis for the exemplary dot at hand, for which we find a reasonable fit with $C=0.39$.  As  {Figure~\ref{fig:alldots}F} shows, across our collection of dots we generally fit exponents $C$ in the range 0.10 to 0.75 to intensity autocorrelation traces.  We note that the relation $\alpha=2-C$ put forward by  Houel et al.\cite{Houel2015} is only expected to hold for two-state quantum dot, and  $C$ does not relate directly to $\alpha$ for quantum dots in which more than two states are at play.

\subsection{Memory effects, aging and correlations in CPA sequences} 
\begin{figure*}
  \includegraphics[width=\linewidth]{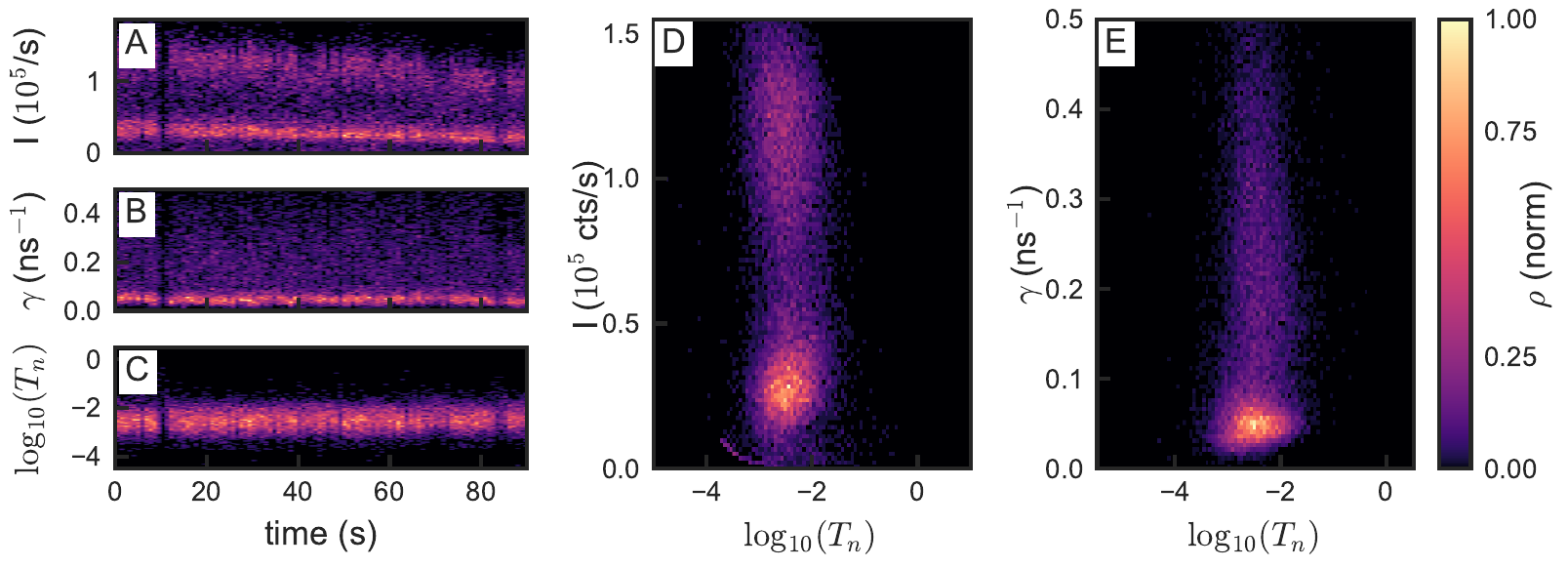}
  \caption{Analysis of (absence of) aging during photocycling of a single perovskite quantum dots,  histogramming intensity (A), decay rate (B) and segment duration (C) in slices of 0.9 sec for a total measurement of 90 sec. (D, E) correlation histograms of intensity versus segment duration,  and decay rate versus segment duration, evidencing that these are uncorrelated quantities.}
  \label{fig:aging_lengthcorrelations}
\end{figure*}
\begin{figure*}
  \includegraphics[width=\linewidth]{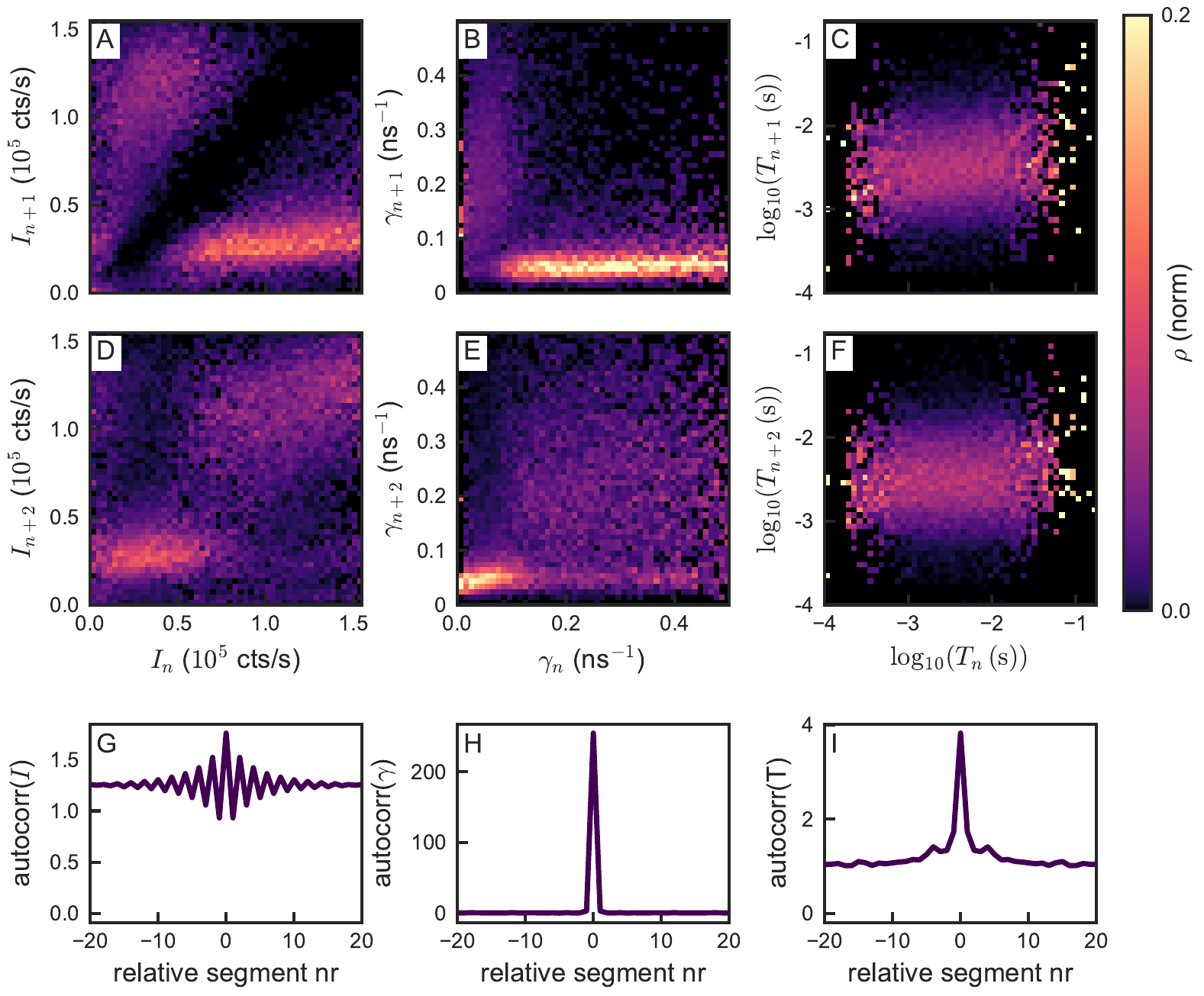}
  \caption{
  (A-F) Conditional probability of observing a value for intensity (A, D), rate (B, E) or segment duration (C, F), given the value of the same obersvable one or two steps earlier, respectively. (G,H,I) Normalized autocorrelation (difference from 1)
 of the sequence of intensity, decay rate and segment durations. This data is for the same single dot as considered in Fig.~\ref{fig:aging_lengthcorrelations}.}
 \label{fig:conditionalprbs_memory}
\end{figure*}

Finally we examine the dots for aging and memory effects,  leveraging the fact that CPA gives an unbiased data segmentation  into segments $n=1\ldots N$ that are classified by segment duration   $T_1,T_2,\ldots$, intensity in counts/sec $I_1,I_2,\ldots$, and decay rate $\gamma_1,\gamma_2,\ldots$  that is established without any distorting temporal binning. \textcolor{black}{Memory effects were first studied by Stefani et al. \cite{Stefani2005} for II-VI quantum dots, and later for perovskite dots in \cite{HouKovalenko2020}, in both systems evidencing memory effects in on/off times.}
We present results again for the same  example dot as in Fig.~\ref{fig:binned_pp} in Figures~\ref{fig:aging_lengthcorrelations} and \ref{fig:conditionalprbs_memory}. With regard to aging, one can ask if over the full measurement time in which a dot undergoes of order $10^8$ excitation cycles,  the distribution of segment duration, intensity and decay rate show any sign of change. To this end, we subdivide the total measurement period (e.g.,  {Fig.\ref{fig:aging_lengthcorrelations}(A-C)}, total measurement time 60 s for this dot) in 100 slices that are equal length in terms of wall-clock time, and examine the evolution of histograms of $I_n$, $\gamma_n$ and $T_n$ for these short measurement intervals as function of their occurrence in the measurement time. As the residence times are very widely distributed, we plot histograms of $\text{log}_{10} T_q$, with $q$ the index of the segments.  There is no evidence that any of these observables change their statistical distribution over the time of the measurement.  While   {Fig.\ref{fig:aging_lengthcorrelations}(A-C)} shows an example for just one dot, this conclusion holds for all dots in our measurement sets, with the caveat that for some dots drifts in microscope focus caused a small gradual downward drift in intensity. We observed no photobrightening of dots during the experiment.

Clustering allows us to ask questions that are not accessible with simple binning of data, as we can examine the datasets for correlations between parameters and between  subsequent segments.  In terms of cross-correlating different observables,  beyond FDIDs that correlate intensity and decay rate, one can also examine correlations between intensity and segment duration, and between  decay rate and segment duration. Histogramming the clustered data to screen for such correlations ({Fig.\ref{fig:aging_lengthcorrelations}(D,E))} show that both the distribution of intensities and of decay rates are uncorrelated, or only very weakly correlated, with the segment duration.  In other words, we find no evidence that within the distribution of states between which the dot switches, some states have different residence time distributions than others. 

Memory effects\cite{Stefani2005,HouKovalenko2020} should appear as  correlations in the values for any given observable in subsequent segments, i.e. in conditional probabilities that quantify what the probability $P_{\Delta n} (A|B)$ is that a chosen observable to obtain a value $A$ is given that it had a value $B$ in  the previous segment ($\Delta n=1$), or generally counting $\Delta n$ events further back into the history  of  previous segments.  {Figure~\ref{fig:conditionalprbs_memory}} shows such conditional probabilities for $\Delta n=1$ (panels A-C) and $\Delta n=2$ (panels D-F) , for intensity (panels A,D), decay rate (panels B,E) and segment duration  (panels C,F). These diagrams are obtained by applying a simple 2D  histogramming approach, listing the value of $B$ as $x-$axis, the value of $A$ as $y-$axis, and normalizing the sum of each of the columns to obtain a conditional probability. 
We note that this approach means that at the extremes  of the histograms (far left, and far right), there are few events to normalize to, leading to large uncertainty.  When screening for memory in intensities, it's important to consider that the CPA algorithm itself  \textit{selects} for intensity jumps. Due to this the intensity after one jump ($\Delta n=1$) is \textit{a priori} very unlikely to achieve a similar value, which leads to a near-zero conditional probability at the diagonal of  {Figure~\ref{fig:conditionalprbs_memory}(a)}. Nonetheless, the distinct features in the diagram at $\Delta n=2$  ({Fig.~\ref{fig:conditionalprbs_memory}(d)}) do suggest that the dot generally alternates repeatedly back and forth between a bright and a dark state.  More telling than diagrams for intensity are those for decay rate. They show that if, in a given step the decay rate is low (slow, bright feature in FDID at $< 0.1$~ns$^{-1}$), then in the subsequent step the decay rate is usually fast, yet widely distributed from 0.1 to 1 ns$^{-1}$,  and vice versa from any of the fast decaying states, the dot is likely to jump to the quite narrowly defined slow rate of the bright state. If one considers the conditional rate at $\Delta n=2$, the conclusion is that if the dot is in the bright state with its slow decay rate at a given step, then likely after two jumps it comes back to this bright, slowly decaying state. If, however, the rate was fast anywhere in the interval from 0.1 to 1 ns$^{-1}$, then after an excursion to the slow rate at $\Delta n=1$, the dot likely in the second step again takes on a fast rate in the interval from 0.1 to 1 ns$^{-1}$ but without a particularly clear preference for any value in that wide interval.  Finally we note that there is no indication in our data that subsequent residence times $(P(T_{m+\Delta n}|T_m)$ show any memory ({Fig.~\ref{fig:conditionalprbs_memory}(c,f)}, showing result for $\text{log}_{10} T_m$). \textcolor{black}{Thus our data do not confirm the observation of Hou et al. \cite{HouKovalenko2020} that there are memory effects in subsequent on-off times. Those memory effects mirror the mirror effects observed by Stefani et al. in 2005 for II-VI quantum dots \cite{Stefani2005}, and indeed the   MRC model \cite{Frantsuzov2010} predicts memory effects in subsequent  on/off durations.  We note that the analysis in these previous works is contingent on thresholding to define on-off states and times, a process counter to the findings of CPA analysis that there are not simply two intensity levels. Moreover we note that in this work we indiscriminately report on all dots we identified as single photon emitters by their $g^{(2)}(0)$,  instead of post selecting those that qualitatively appear closest to bimodal behavior as in Ref. \cite{HouKovalenko2020}.  The fact that the very definition of on-off time is unclear for these quantum dots rather defies analysis of memory in these quantities in the terms used by \cite{Stefani2005,Frantsuzov2010,HouKovalenko2020}. Since it appears that the dots at hand switch between a reasonably unique bright state and the entire tail of dark gray states suggests to define on-times, as selected from CPA by thresholding at circa $I/\langle I \rangle  > 1.3$. With this approach we   found no memory effects for the sequence of on-times.}

One could speculate that the information gleaned from such conditional probability diagrams could be advantageously  condensed in autocorrelations of the traces $I_q$, $\gamma_q$ and $T_q$. We plot normalized autocorrelation traces $G(\Delta q)-1$ where for any sequence $H \in I_q, \gamma_q, T_q$, one defines 
$$G(\Delta q)=\langle H(q) H(q+\Delta q)\rangle/\langle H(q)\rangle^2$$ 
(where $\langle . \rangle$ denotes the mean over $q$ are all segment indices and $\Delta q$ is $1,2,\ldots$), so that at long times $G_m-1$ vanishes. 

In Fig.~\ref{fig:conditionalprbs_memory}(G-I) we plot $G(\Delta q)-1$ for intensity, rate and segment duration. Such segment-autocorrelations are distinct from, e.g., the usual intensity autocorrelation traces that one might examine to determine blinking power laws, since here one autocorrelates subsequent intensity segments \textit{without any regard for their time duration}.  For a conventional two-level dot, the autocorrelation trace $I_q$ would oscillate with large contrast up to very large $q$. Instead, we find that the dot at hand shows an oscillation with a distinct contrast in the intensity segment autocorrelation contrast for up to 5---10 cycles.  In the normalized autocorrelation for decay rates the memory is far less evident. We attribute this not to a lack of memory,  but note that if a dot switches between a state of well defined slow rate, and an array of states with highly distributed fast rates, then upon averaging the wide distribution of fast rates washes out any autocorrelation signature. Finally, the  residence times, which we already found to be uncorrelated between subsequent jumps, show no autocorrelation signature for $z\neq 0$.  
A similar behavior to that shown in Figure~\ref{fig:conditionalprbs_memory}(G) was observed for circa 30\% of the dots studied, with other dots showing no clear intensity autocorrelation.  

\section{Conclusion}  
To conclude, we have reported on intermittency properties of a large number of \dotchem quantum dots on basis of a Bayesian inference data analysis. This approach works with raw, unbinned, photon counting data streams and thereby avoids artefacts commonly associated with the analysis of time binned data.  We find that dots have in addition to their bright emissive state a tail of gray states that qualitatively appears continuous in FDID diagrams, and that according to clustering analysis requires at least 10 to 20 levels to describe, if a discrete-level description would be appropriate. Thereby our work  provides a confirmation of claims in earlier  works\cite{Park2015,Seth2016,LiHuang2018} under similar excitation conditions, with the distinction that we do not use time binned data but rigorously exploit all the information in the data stream to the level that its intrinsic noise allows. We note that the same type of dots have displayed a different behavior, indicative of 2 to 3 levels, in Ref.~\cite{Gibson2018}. Since  that work uses almost identical Bayesian inference methods, we conclude that this distinction is really due to the different physical realization. \textcolor{black}{Alongside possible differences in sample preparation, we note that Ref.~\cite{Gibson2018} also stands out from all other reports due to its quite different excitation conditions, particularly using shorter pulses and significantly lower pulse repetition rates. While Ref.~\cite{Gibson2018} states that this choice improves photostability, when expressed in number of excitation cycles, our experiment is not actually at a disadvantage in terms of photostability since we follow dots for 2 minutes at 10 MHz repetition rate, versus for 10 minutes at 0.3 MHz. Also our estimates exclude the idea that higher pulse repetition rates but at similar per pulse excitation densities, would   cause a more significant heating of the dot that would explain thermally activated modifications since nanoparticles under pulsed excitation loose their energy to the environment in nanoseconds\cite{Baffou2013}.  A possible explanation might lie in the fact that perovskite quantum dots have been reported to have slow-time constant electronic processes, such as delayed exciton emission (microsecond time scales) \cite{Chirvony2017,Yinthai2017,Vonk2020}, and shallow trap states with lifetimes exceeding 250 ns \cite{Samanta2018,Samanta2019}. These observations imply that there are photophysical processes that may be involved in blinking and flickering, and that may not fully relax at higher laser repetition rates.  Finally, a caveat on experimental limitations in the effort to determine dim intensity levels is that, even if the physics is unchanged, lower repetition rate experiments are less likely to identify many gray /dark states once the dark state count rates approach detector background levels. In our set up, the combined dark count rate of both detectors is of order 200---250 counts per second, meaning that the darker levels would be comparable in count rate if we would reduce the excitation rate 30-fold.}

\textcolor{black}{Overall, our results support \cite{Park2015,Seth2016,LiHuang2018}, and as in the first report proposing the validity of the MRC model \cite{Frantsuzov2009,Frantsuzov2013} for perovskite dots \cite{LiHuang2018}, we  find that the tail of gray states display an inverse correlation between intensity and rate, suggesting that the dots have a unique bright state with a given decay rate, to which random activation of recombination centers add nonradiative decay channels.}  However, we note that this observation merits further refinement of models: while plotting intensity versus lifetime may point at strict proportionality,  plotting rate instead of lifetime accentuates deviations, noteably a  deviation in curvature of our data relative to inverse proportional  dependence. Finally, we have analyzed correlations in the measured CPA-segmented sequences of intensity-levels, decay rates and segment lengths.  We  find no evidence for aging, i.e., gradual shifts in e.g., decay rate or blinking dynamics during photocycling of dots through $10^6$ to $10^7$ detected photons (i.e., well over $10^8$ cycles). Also, our data indicate that residence times are not correlated to the state that a dot is in.  The residence times can be fiducially extracted for a limited time dynamic range from ca. 5-15 ms,  limited intrisically by count rate,   to ca. 10 s, limited by the length of the photon record.  We note that in residence time histograms determined by CPA, according to Monte Carlo simulations at long times the analysis fiducially reports on power laws without introducing artefacts, such as apparent  long-time roll offs.  The exponents that we find are in the range from 1.5 to 3.0, which appears high compared to the near-universal value of 1.5 observed for II-VI single photon sources.   In the domain of \dotchem dots, reports have appeared of even lower exponents (down to 1.2)\cite{Gibson2018,HouKovalenko2020} with  exponential roll offs at times $\sim 0.1$~s that cause a steepening of the residence time histogram at longer times.  We note that although exponential roll off certainly steepens slopes in the residence time histogram, our histograms do not point at exponential, but at high-exponent power law behavior. 
 
Regarding memory, we found a distinct memory effect in intensity and rate in the sense that dots  appear to switch between a quite unique bright state with slow decay rate that is evident as the bright pocket in FDIDs,  and the entire tail of dim states in the FDID. Moreover,  the dots do not appear to return preferentially to this  dim state, but explore the entire tail anew at each transition from the bright state.   This is an important refinement on the MRC model which in itself  leaves open if dots return at all to the bright state before choosing another dim state, and which does not specify if dots make repeated visits to the same dim state or not.  In terms of analysis, this memory is only partially visible in  autocorrelation traces of sequences of CPA intensity and rates, as the dim states are so widely distributed. The averaging involved in evaluating autocorrelations washes out some memory effects that do appear more clearly in conditional probability histograms reporting on subsequent jumps. Finally we found no evidence in our data set for the  apparent memory in residence times (segment lengths $T_q$) reported by Hou et al.\cite{HouKovalenko2020} for on-times.   

\textcolor{black}{In our view, this rich data set will stimulate further theory development in the domain of inorganic quantum dot intermittency.  Compared to the case of II-VI quantum dots, a host of different effects could be at play in perovskite quantum dots. For instance, vacancy concentrations in perovskites are orders of magnitude higher than those in II-VI materials, and vacancies are highly mobile, which may affect photoluminescence\cite{Patra2020}. Also, halide perovskites are know to undergo reversible surface (photo)chemical reactions. Given the role of surface defects in blinking (as understood for II-VI dots), this may be highly important for perovskites. Blinking studies in different environmental ga ses could elucidate this\cite{Fang2016}. Also, one could speculate that the strong polaronic effects in perovskites affect blinking, through involvement with the screening of trapped charges \cite{Miyata2017}. Finally, in terms of electronic structure perovskite materials are different from II-VI dots not just in weak versus strong confinement, but also in having strongly anharmonic potentials, near-equal hole and electron effective masses, and a band structure that causes defect levels to be at shallow trap levels, instead of deep trap levels (see Ref.\cite{Samanta2018,Samanta2019,Chirvony2017,Yinthai2017,Vonk2020} for possible relations to intermittency). }

\begin{suppinfo}
PDF containing CPA-based intermittency analysis report for  40 single dots - appended to this preprint. For select dots, data is posted with the Python analysis code of Ref.\cite{Methodspaper,github} at github.
\end{suppinfo}

\begin{acknowledgement}
This work is part of the research program of the Netherlands Organization for Scientific Research (NWO). We would like to express our gratitude to  Tom Gregorkiewicz who passed away in 2019, and whose encouragement and guidance in the early stages of the project were invaluable.
\end{acknowledgement}
 
\bibliography{PalstraIntermittency}

\clearpage

\section{Supporting information}

The following  40 pages,  and lists summary sheets of analysis on 40 single quantum dots. Figures 1, 2, 4, 6,7 in the main text are taken from dot 6.  Python code to generate these results from raw data,  and raw data for select dots is available through Ref 26 of the main text, and the associated Github project at \url{https://github.com/AMOLFResonantNanophotonics/CPA/}.

\listoffigures

 \clearpage
 
\begin{figure*}\begin{framed} 
  \includegraphics[angle=90,width=1.0\linewidth]{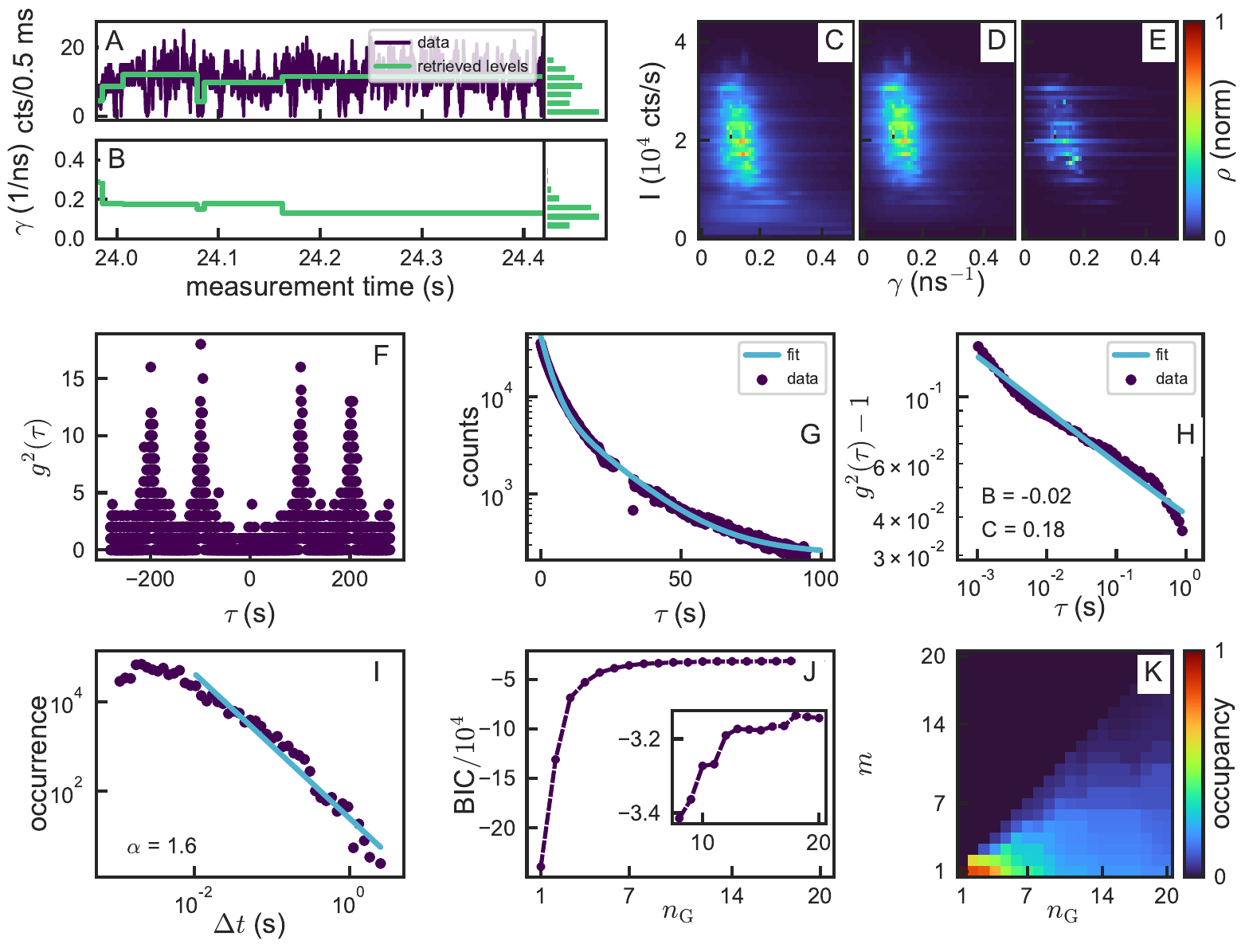}
\end{framed}\caption{Summary of observations on dot 1 of 40}
\end{figure*}\clearpage 

\begin{figure*}\begin{framed}
  \includegraphics[angle=90,width=1.0\linewidth]{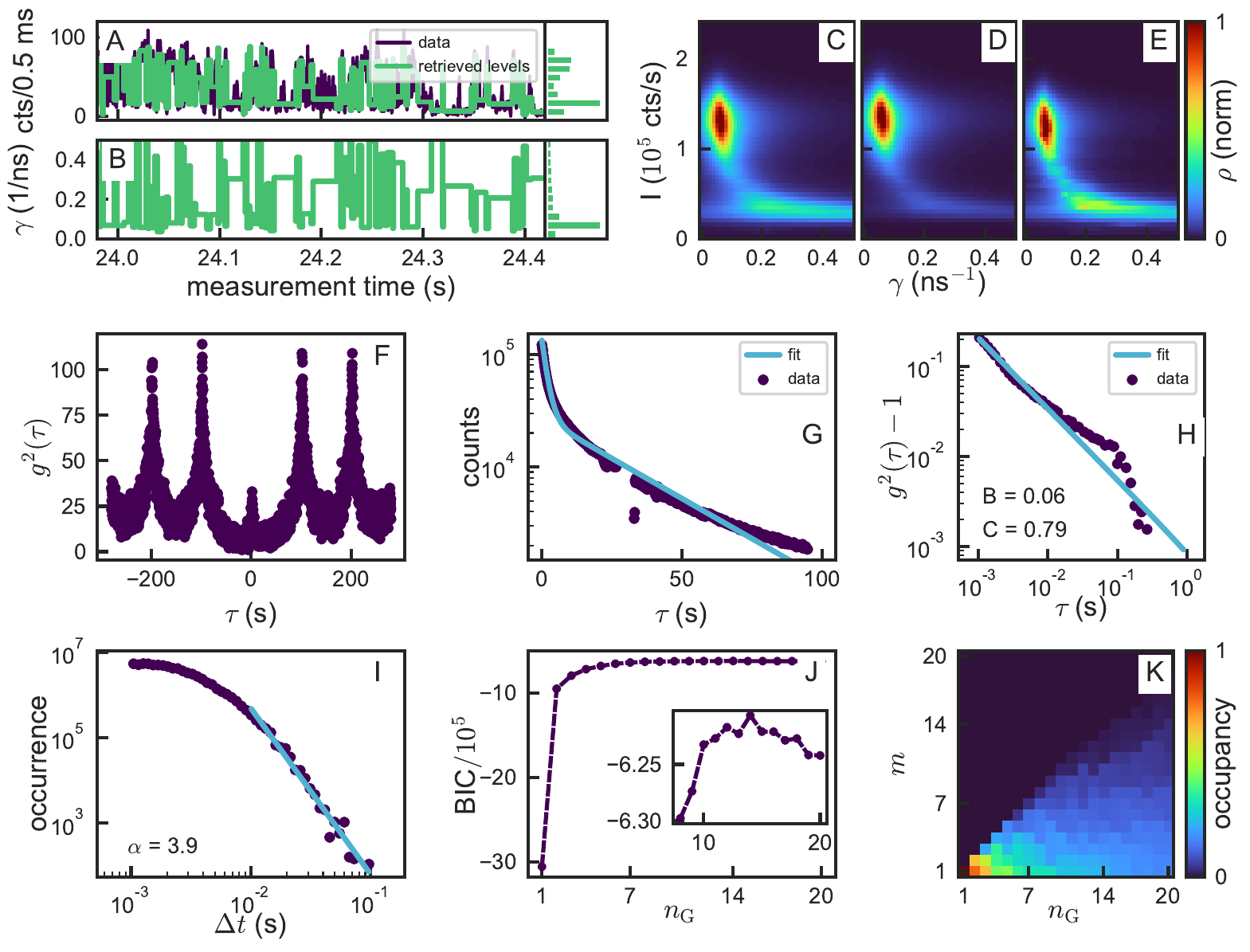}
 \end{framed}\caption{Summary of observations on dot 2 of 40}
\end{figure*}\clearpage 
 
\begin{figure*}\begin{framed}
  \includegraphics[angle=90,width=1.0\linewidth]{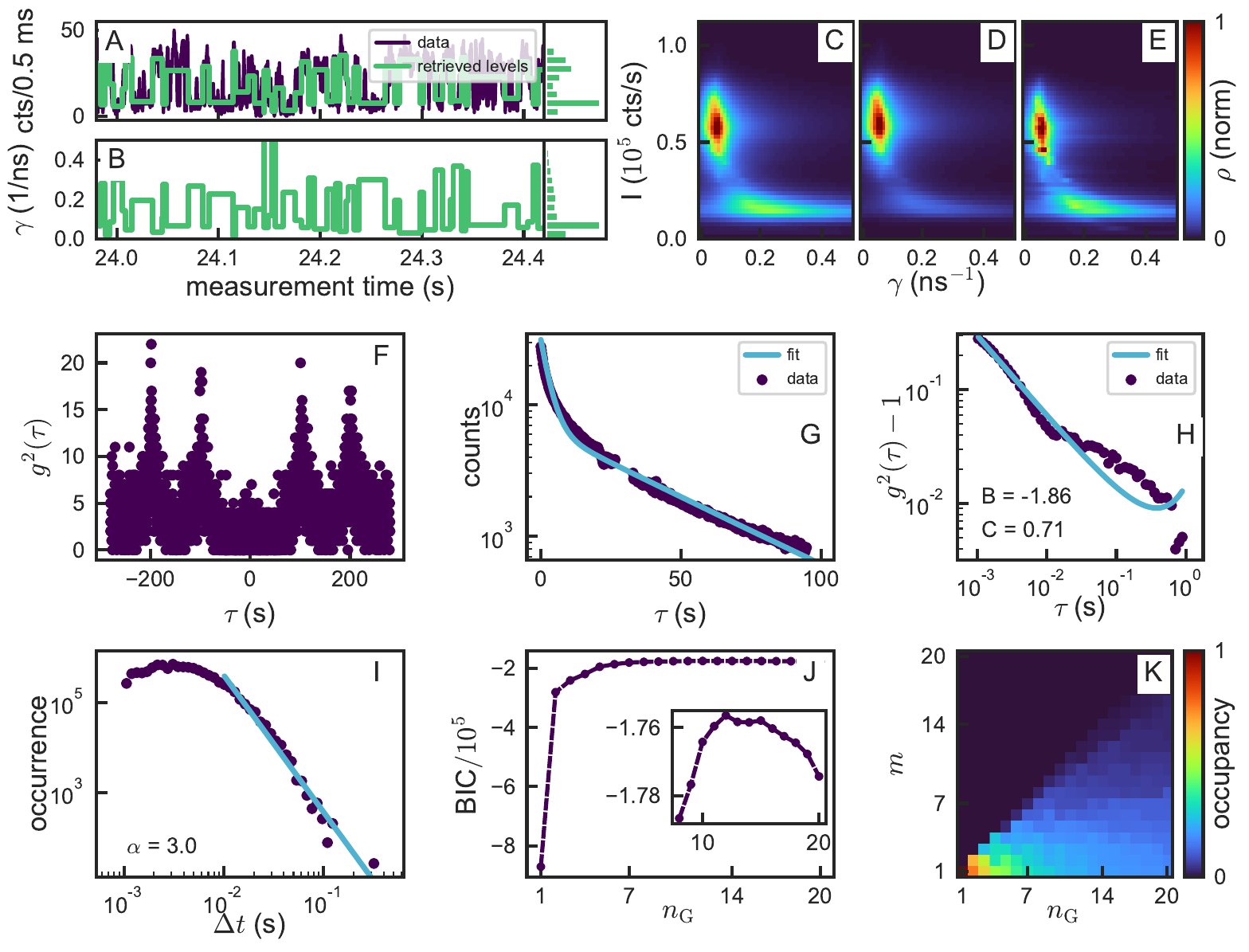}
 \end{framed}\caption{Summary of observations on dot 3 of 40}
\end{figure*}\clearpage 

\begin{figure*}\begin{framed}
  \includegraphics[angle=90,width=1.0\linewidth]{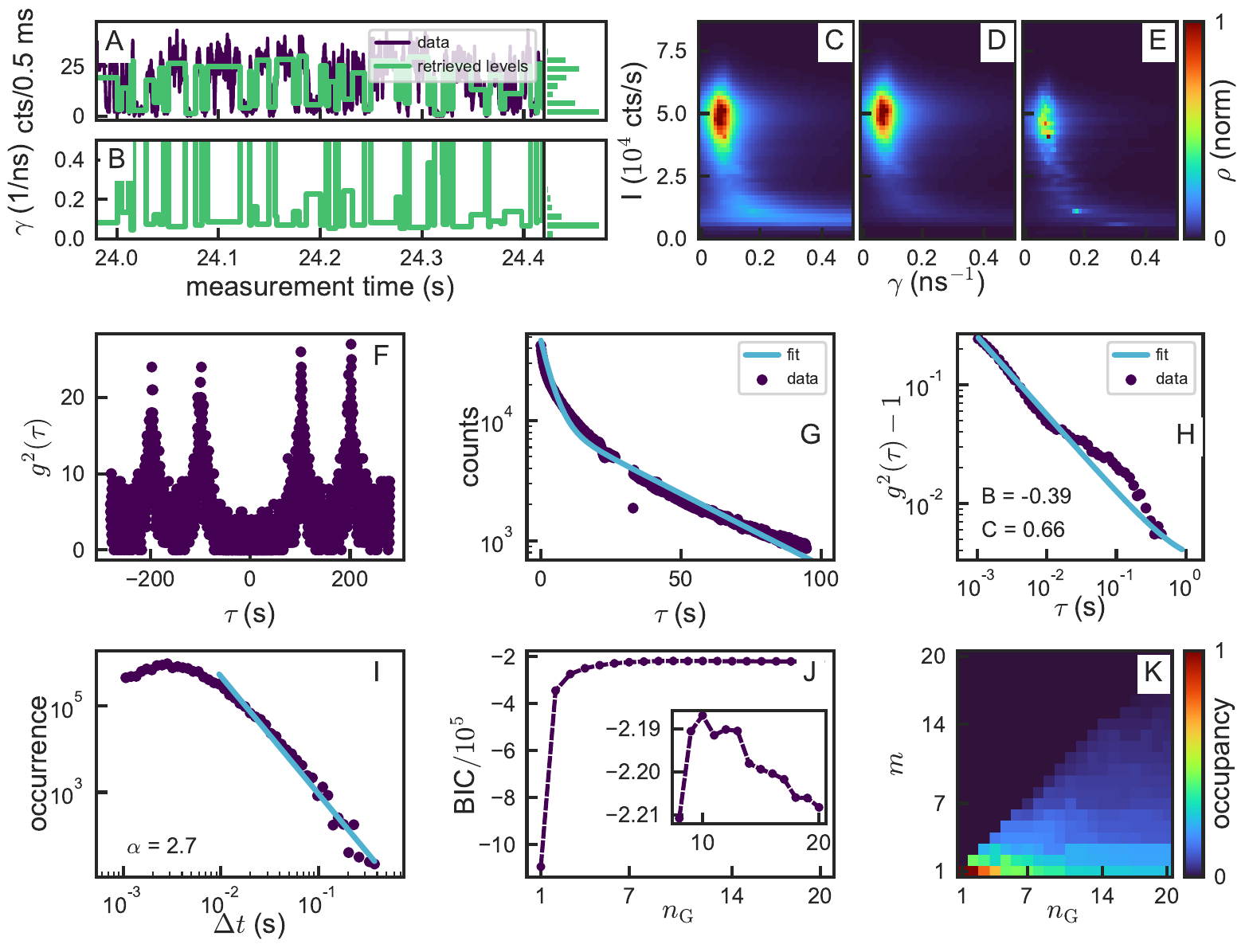}
 \end{framed}\caption{Summary of observations on dot 4 of 40}
\end{figure*}\clearpage 

\begin{figure*}\begin{framed}
  \includegraphics[angle=90,width=1.0\linewidth]{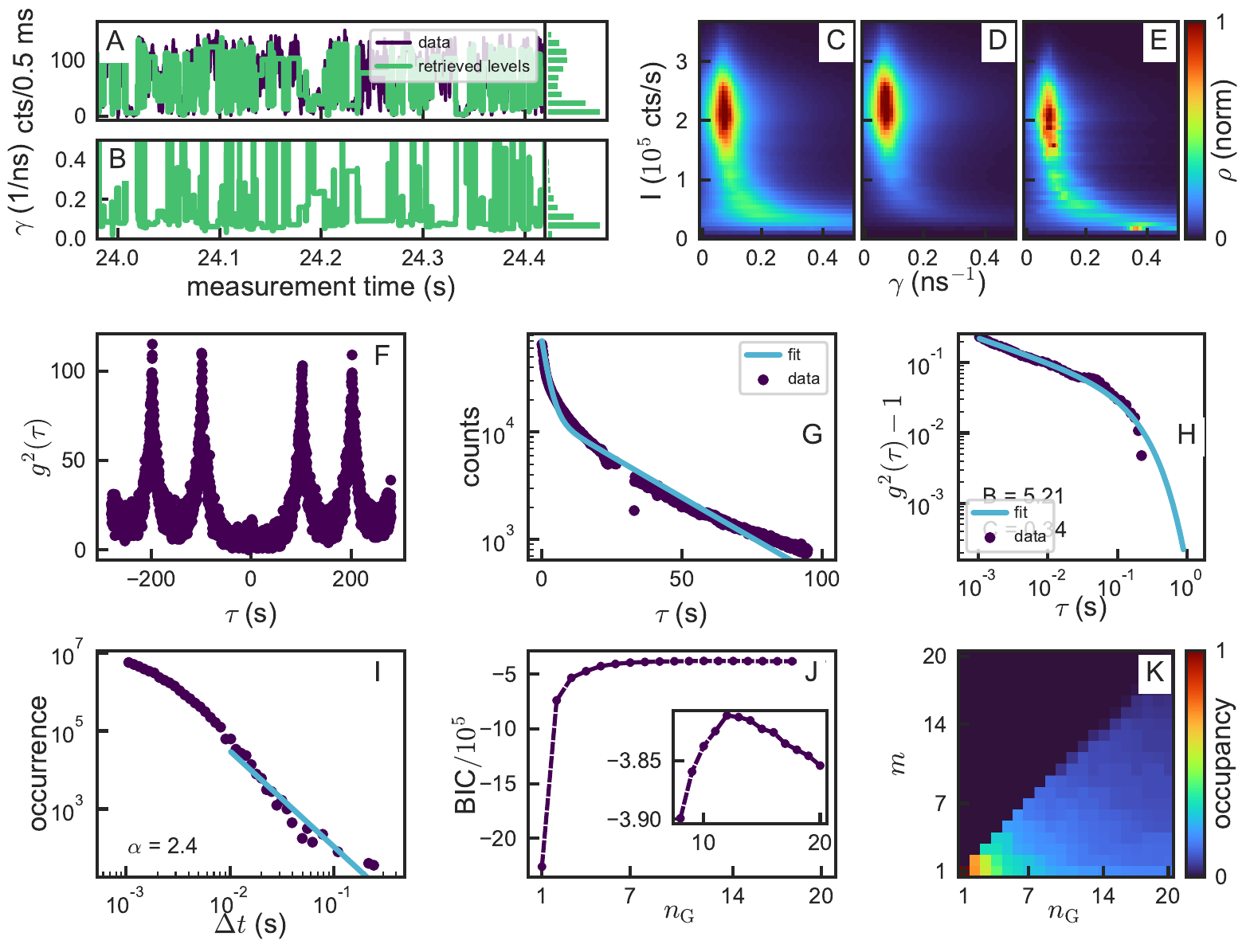}
 \end{framed}\caption{Summary of observations on dot 5 of 40}
\end{figure*}\clearpage 

\begin{figure*}\begin{framed}
  \includegraphics[angle=90,width=1.0\linewidth]{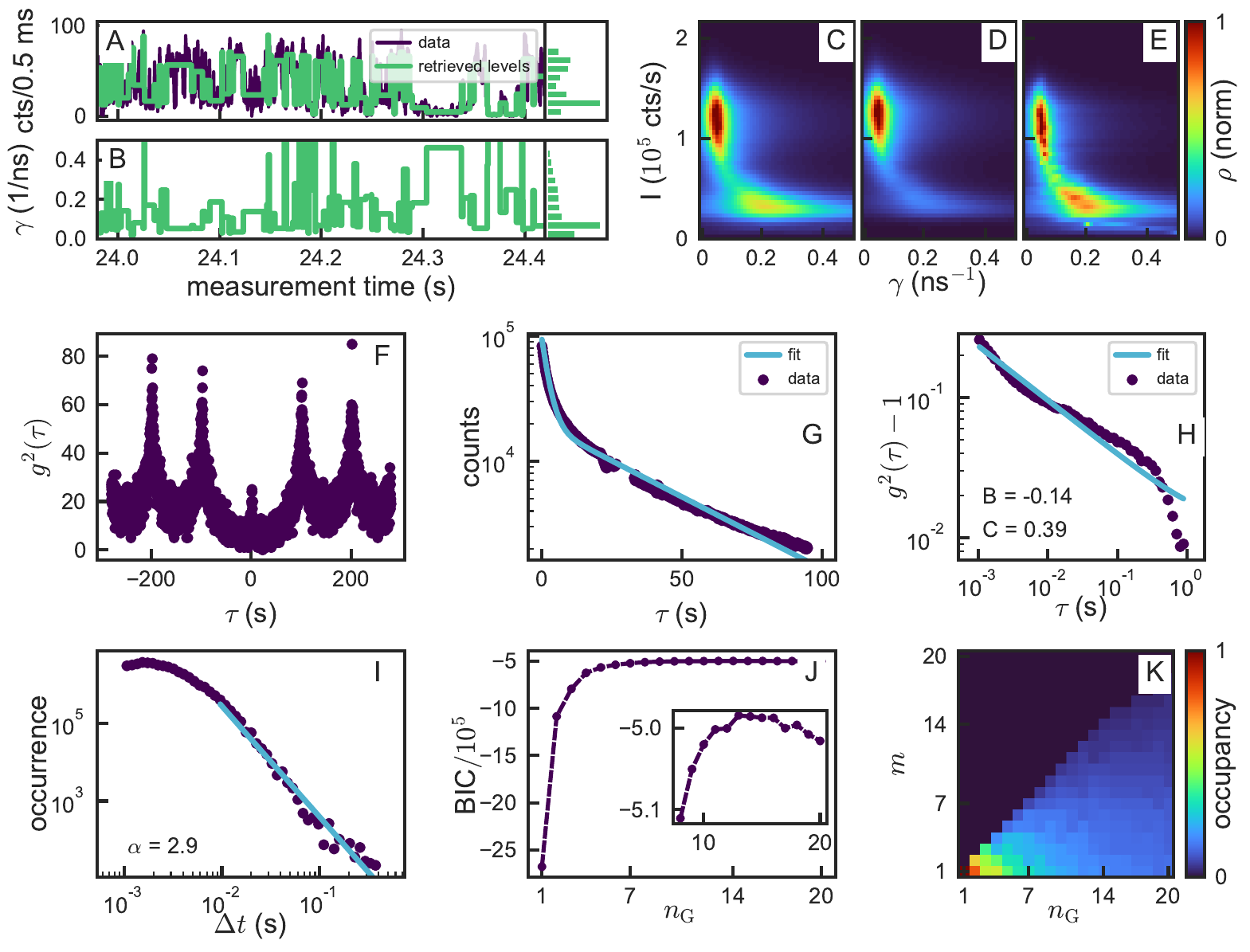}
 \end{framed}\caption{Summary of observations on dot 6 of 40}
\end{figure*}\clearpage 

\begin{figure*}\begin{framed}
  \includegraphics[angle=90,width=1.0\linewidth]{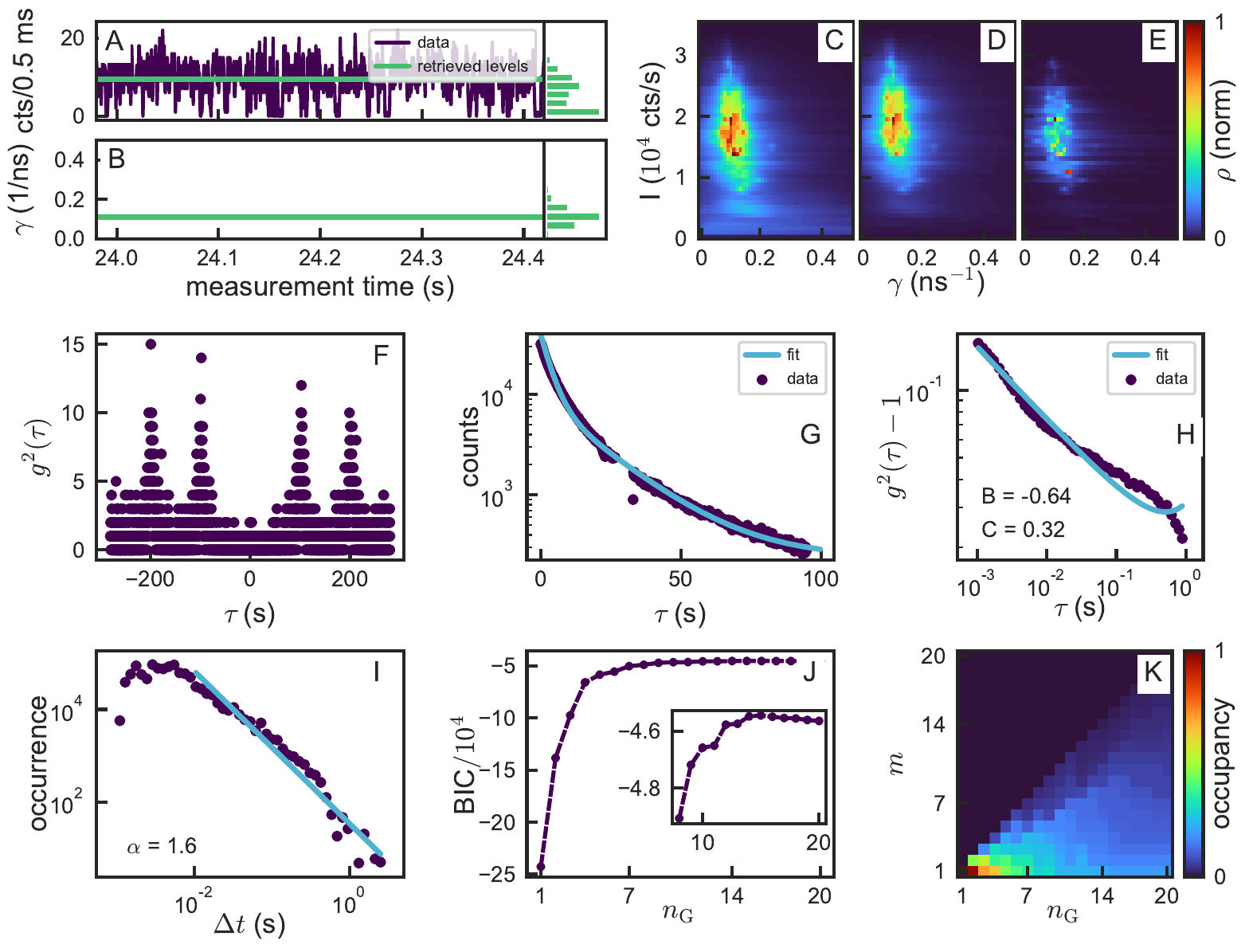}
 \end{framed}\caption{Summary of observations on dot 7 of 40}
\end{figure*}\clearpage 

\begin{figure*}\begin{framed}
  \includegraphics[angle=90,width=1.0\linewidth]{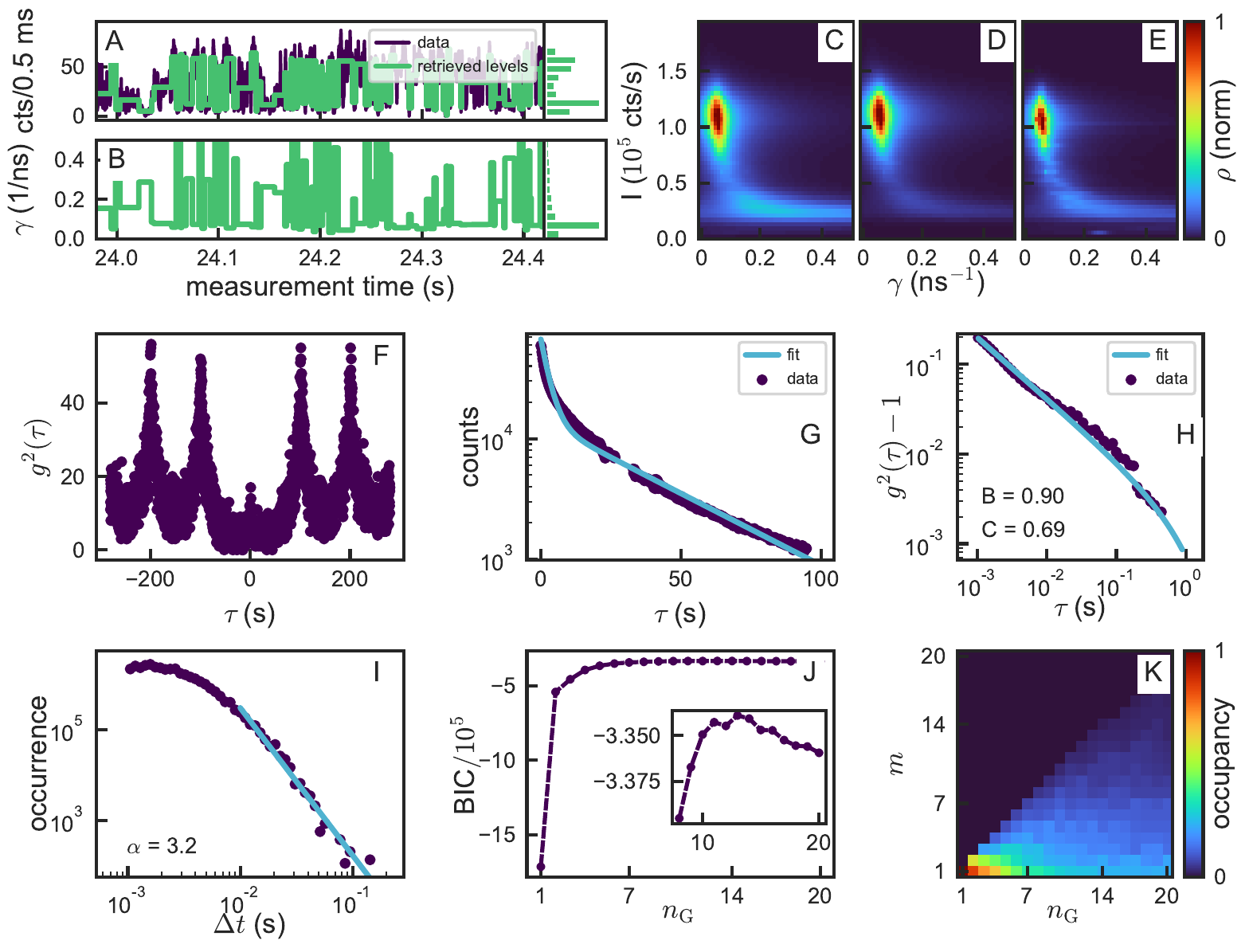}
 \end{framed}\caption{Summary of observations on dot 8 of 40}
\end{figure*}\clearpage

\begin{figure*}\begin{framed}
  \includegraphics[angle=90,width=1.0\linewidth]{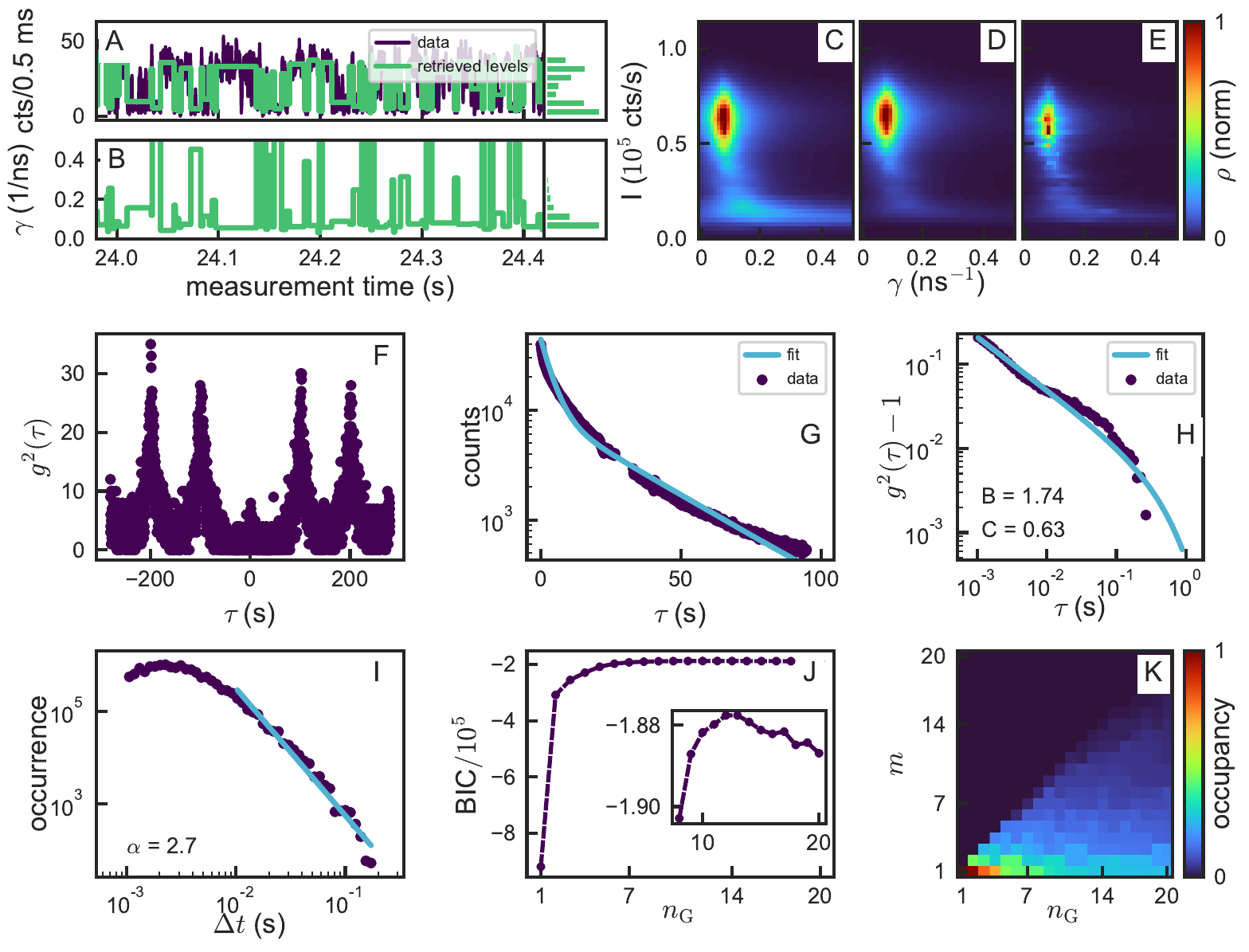}
 \end{framed}\caption{Summary of observations on dot 9 of 40}
\end{figure*}\clearpage 

\begin{figure*}\begin{framed}
  \includegraphics[angle=90,width=1.0\linewidth]{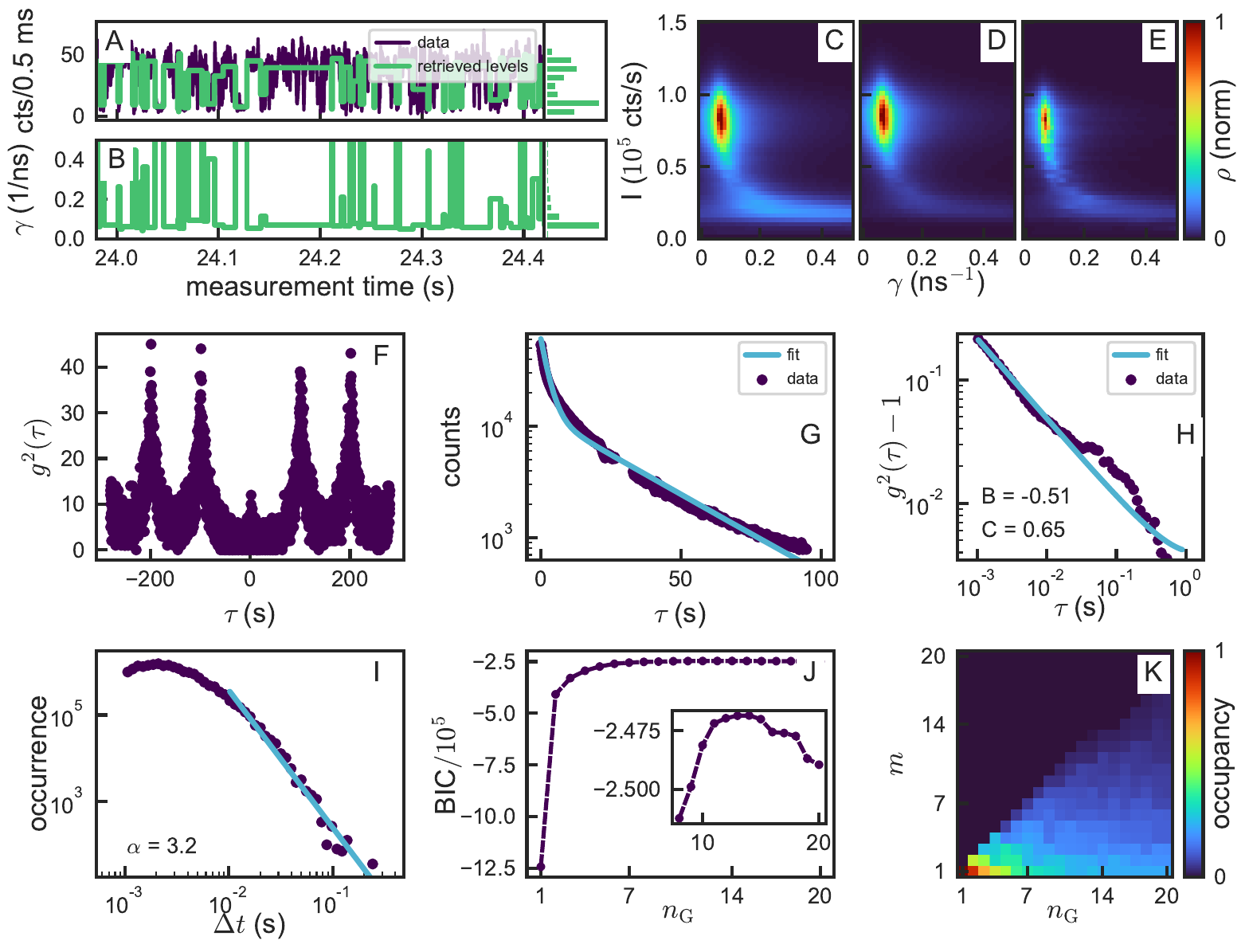}
 \end{framed}\caption{Summary of observations on dot 10 of 40}
\end{figure*}\clearpage 


\begin{figure*}\begin{framed}
  \includegraphics[angle=90,width=1.0\linewidth]{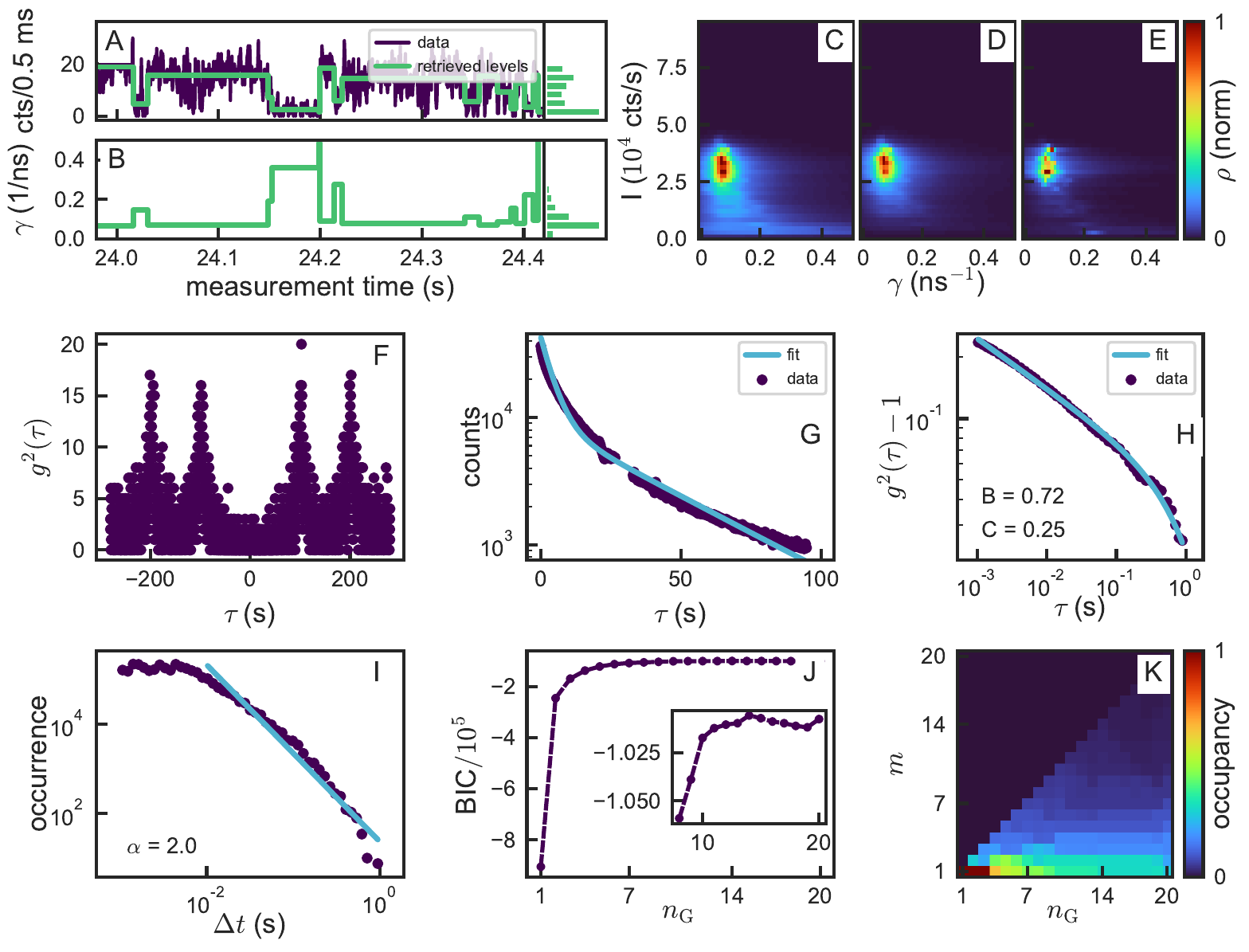}
 \end{framed}\caption{Summary of observations on dot 11 of 40}
\end{figure*}\clearpage 

\begin{figure*}\begin{framed}
  \includegraphics[angle=90,width=1.0\linewidth]{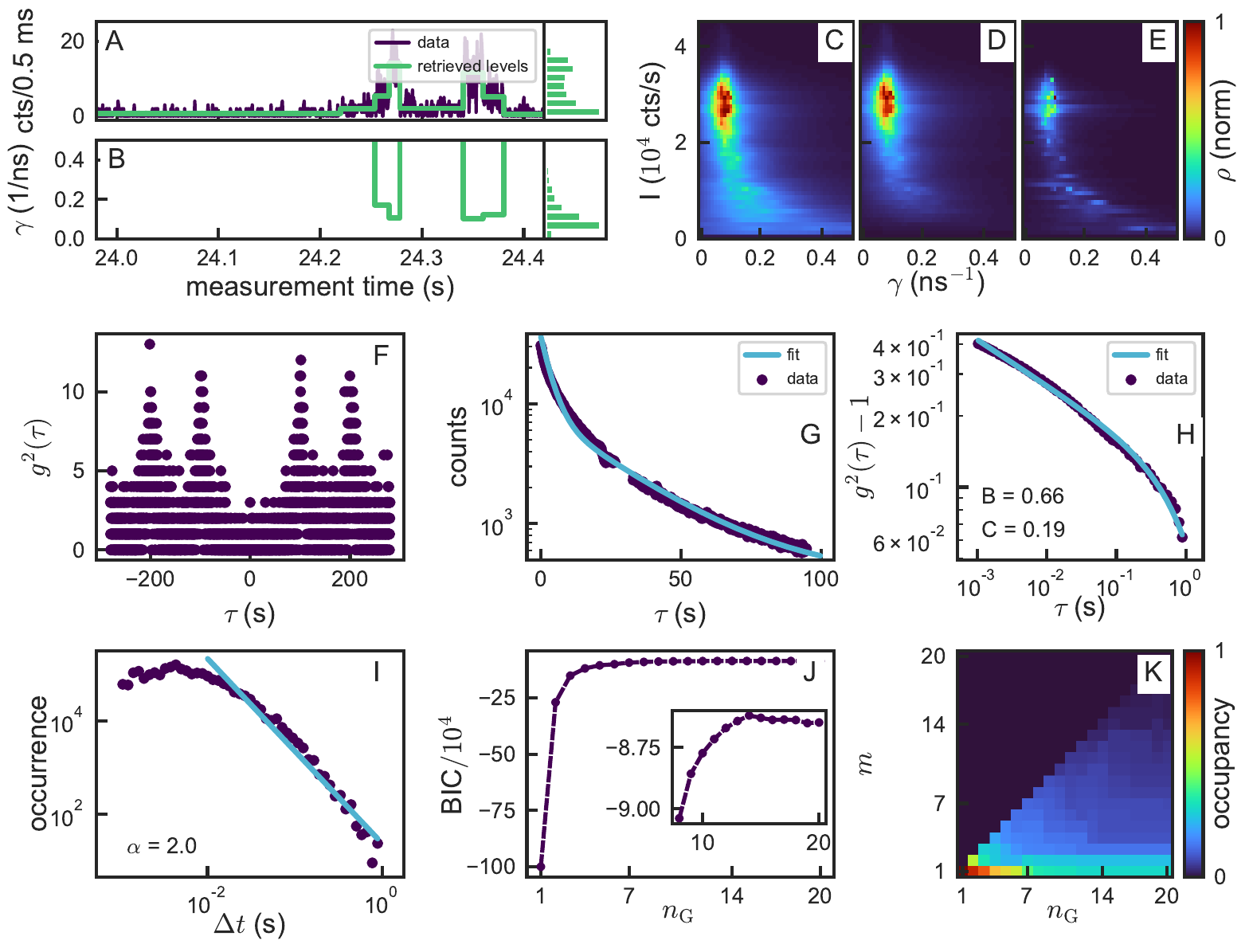}
 \end{framed}\caption{Summary of observations on dot 12 of 40}
\end{figure*}\clearpage 

\begin{figure*}\begin{framed}
  \includegraphics[angle=90,width=1.0\linewidth]{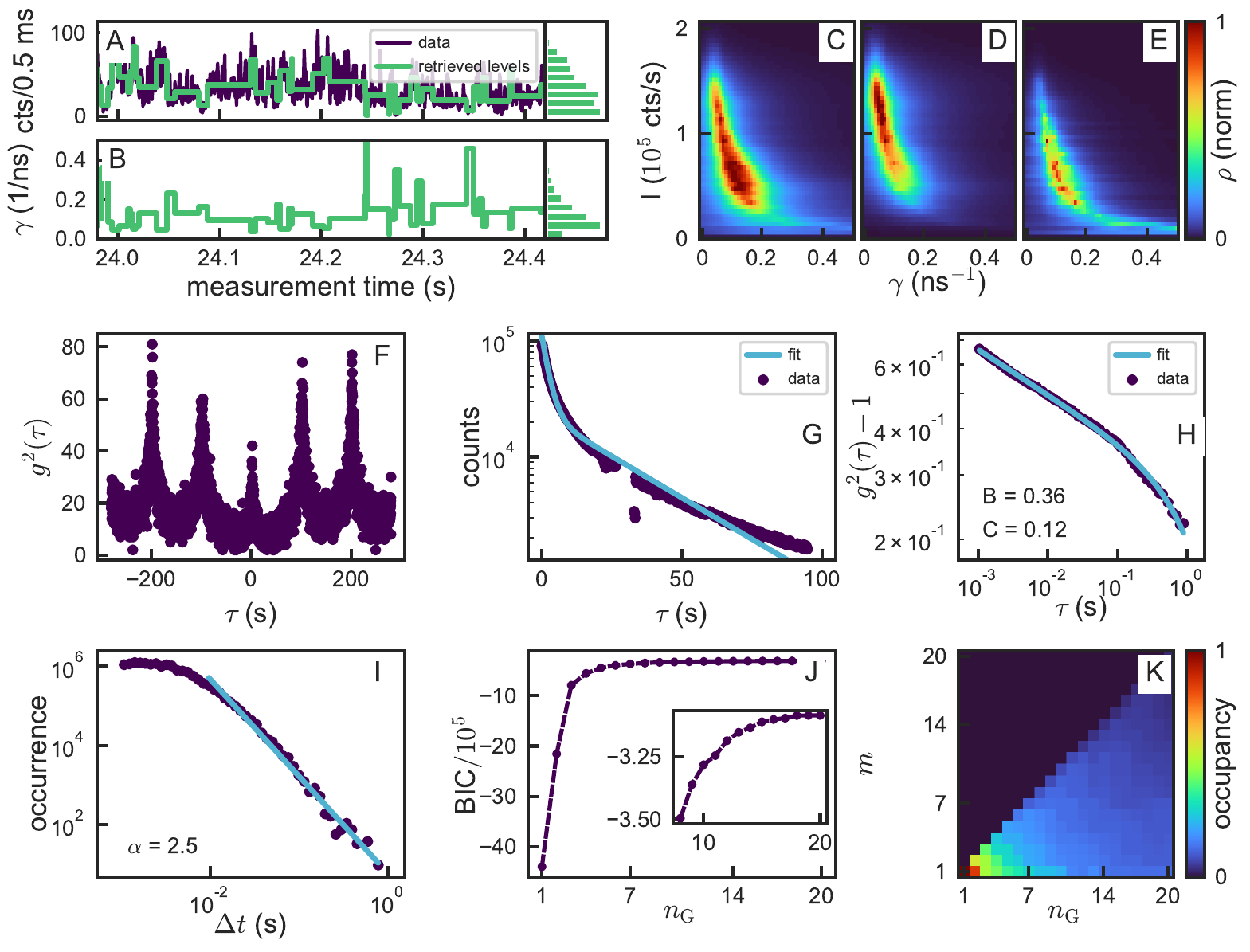}
 \end{framed}\caption{Summary of observations on dot 13 of 40}
\end{figure*}\clearpage 

\begin{figure*}\begin{framed}
  \includegraphics[angle=90,width=1.0\linewidth]{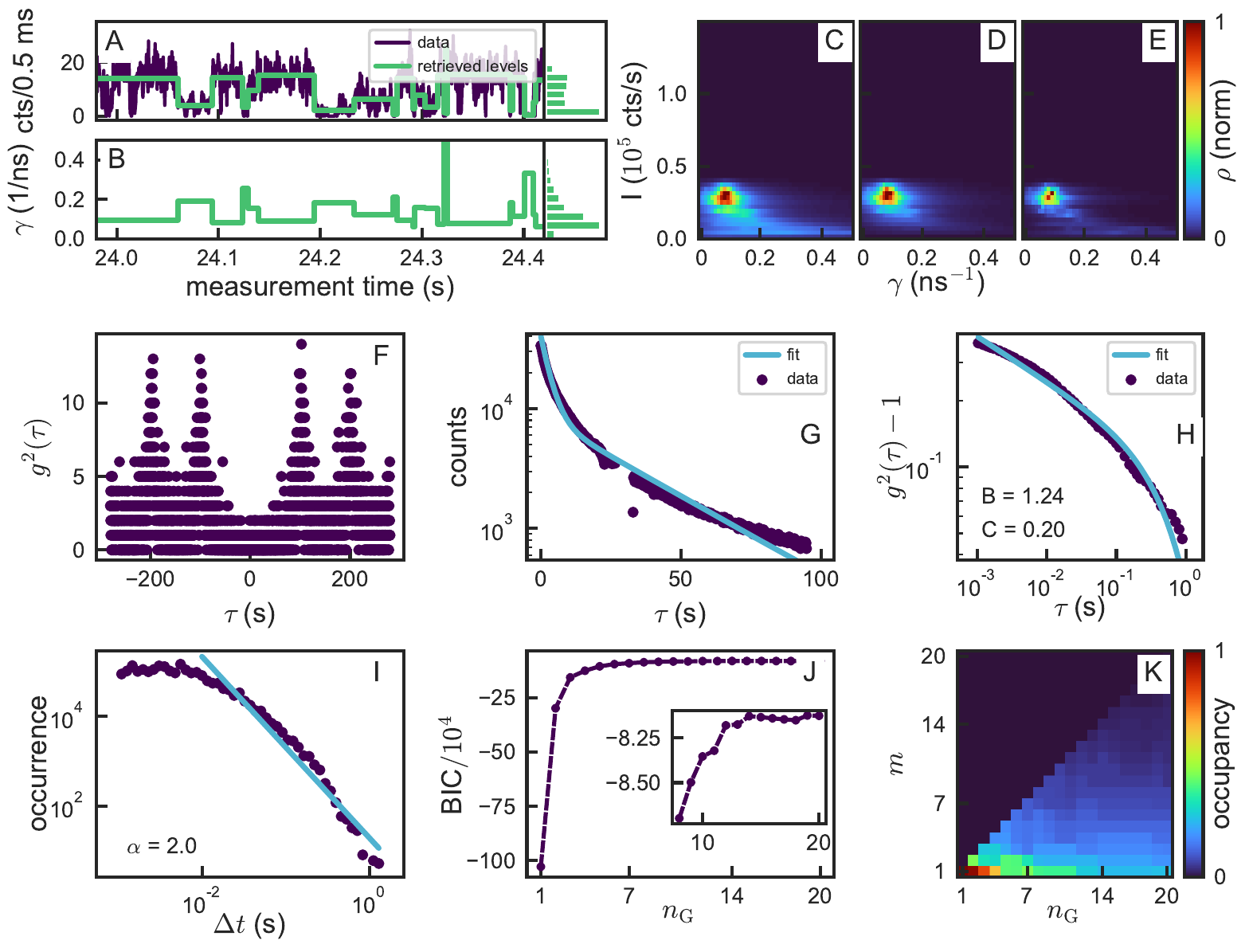}
 \end{framed}\caption{Summary of observations on dot 14 of 40}
\end{figure*}\clearpage

\begin{figure*}\begin{framed}
  \includegraphics[angle=90,width=1.0\linewidth]{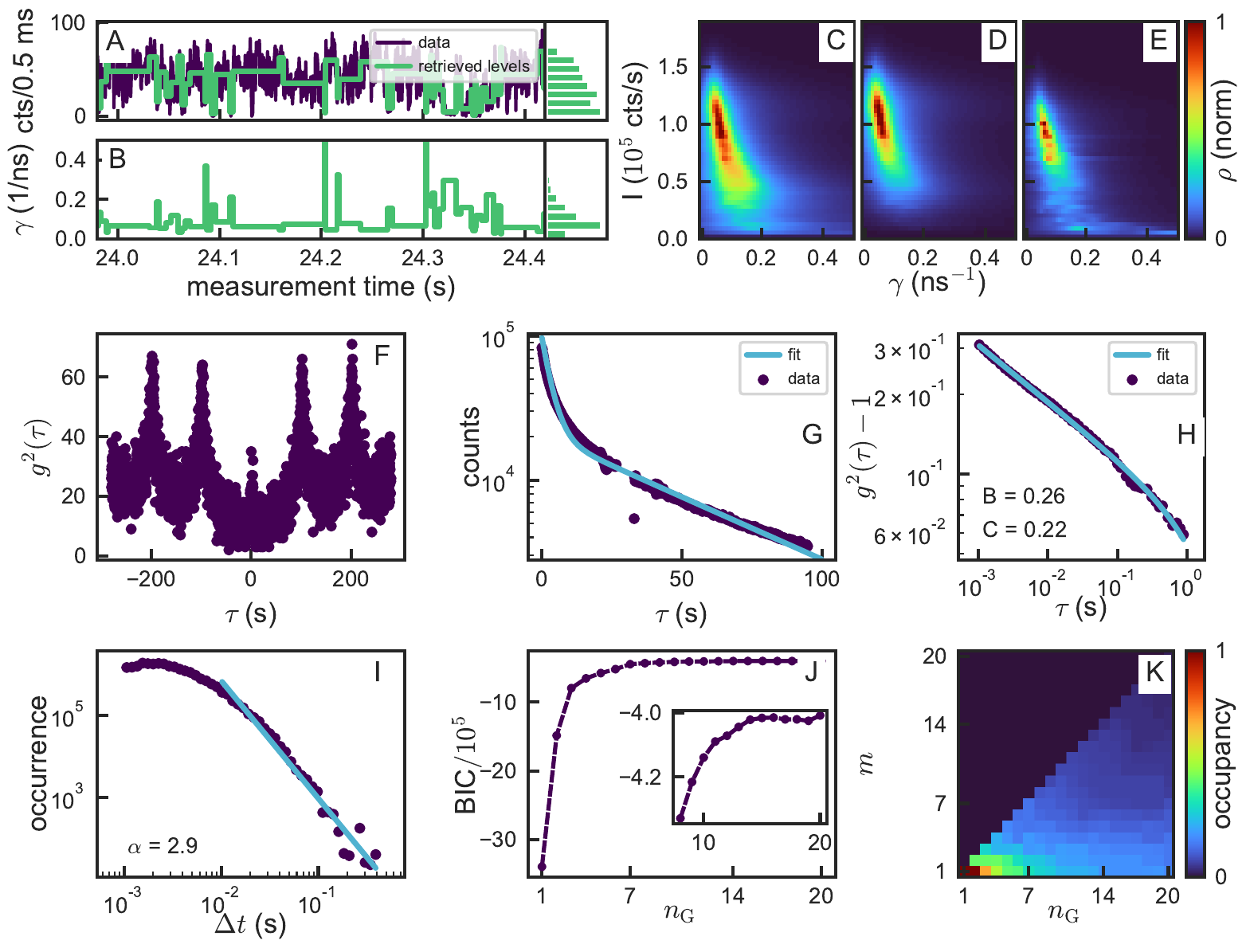}
 \end{framed}\caption{Summary of observations on dot 15 of 40}
\end{figure*}\clearpage 

\begin{figure*}\begin{framed}
  \includegraphics[angle=90,width=1.0\linewidth]{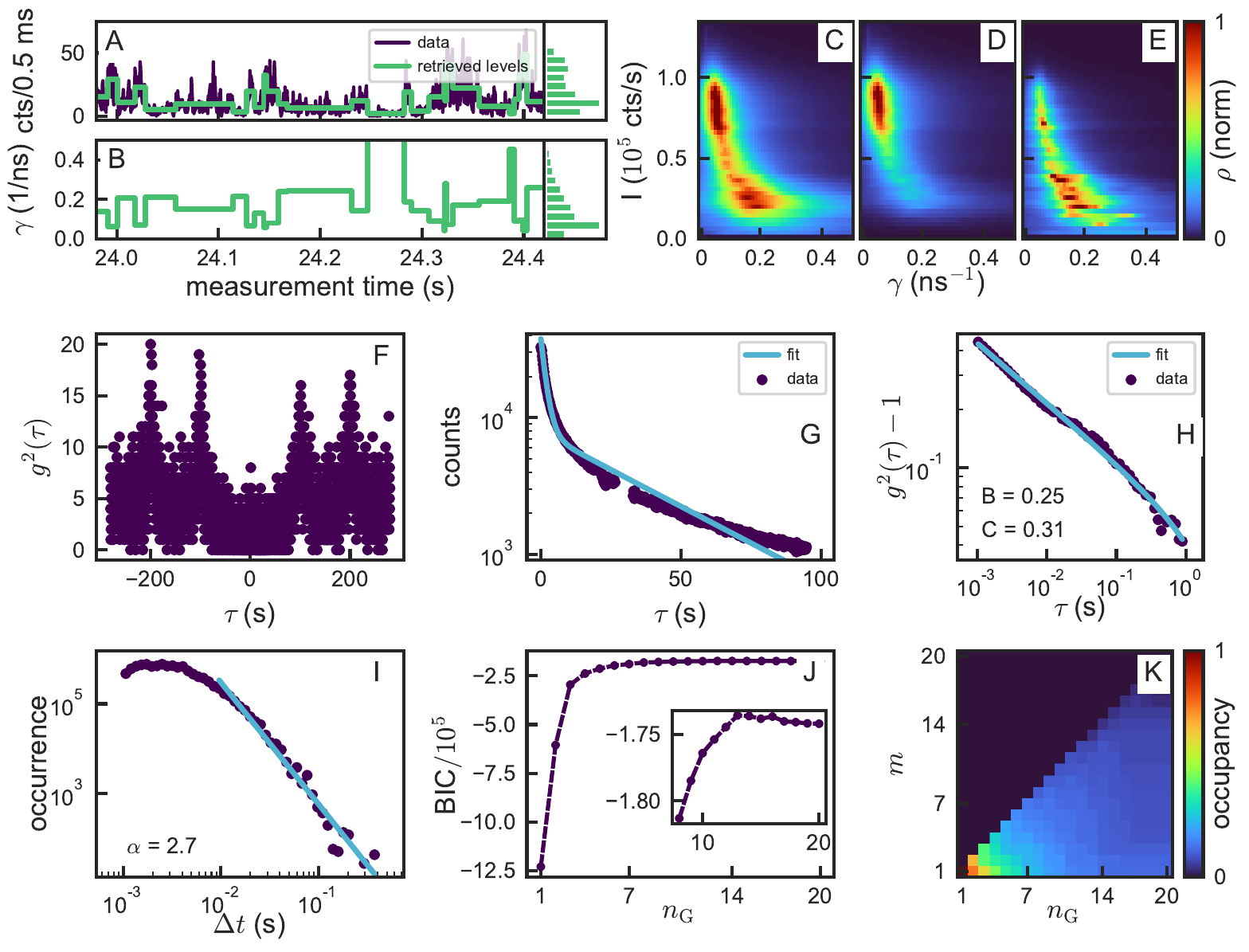}
 \end{framed}\caption{Summary of observations on dot 16 of 40}
\end{figure*}\clearpage 

\begin{figure*}\begin{framed}
  \includegraphics[angle=90,width=1.0\linewidth]{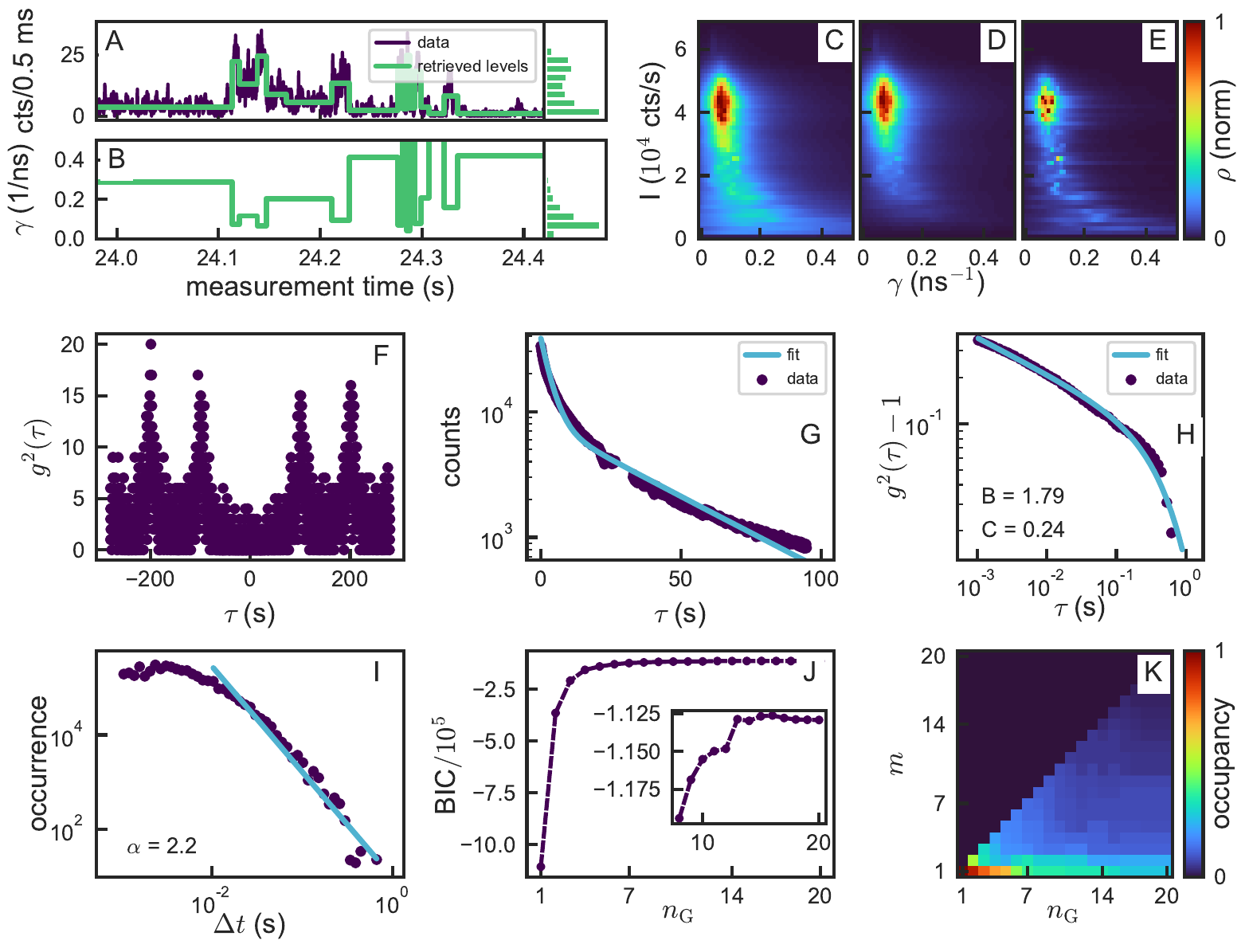}
 \end{framed}\caption{Summary of observations on dot 17 of 40}
\end{figure*}\clearpage 

\begin{figure*}\begin{framed}
  \includegraphics[angle=90,width=1.0\linewidth]{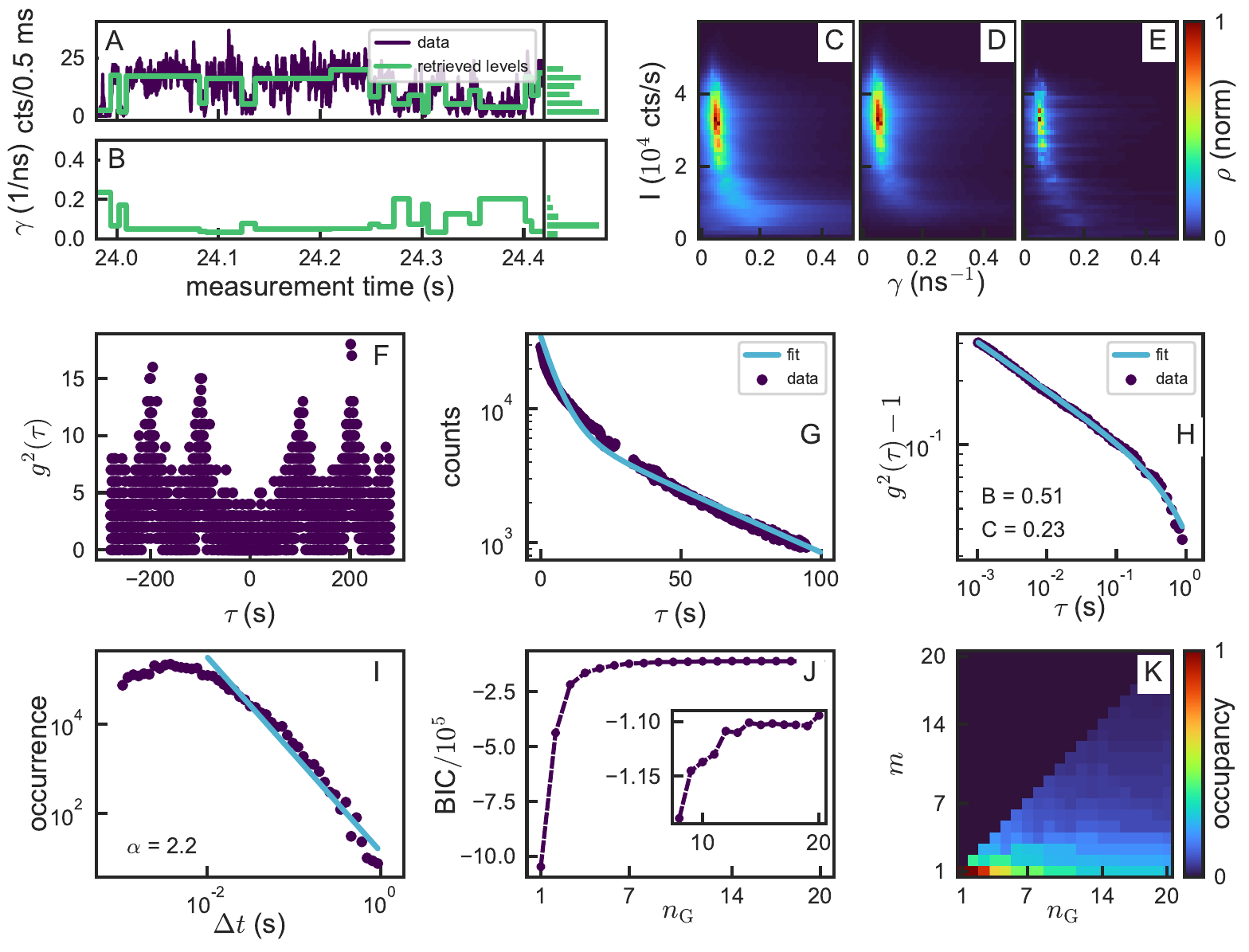}
 \end{framed}\caption{Summary of observations on dot 18 of 40}
\end{figure*}\clearpage

\begin{figure*}\begin{framed}
  \includegraphics[angle=90,width=1.0\linewidth]{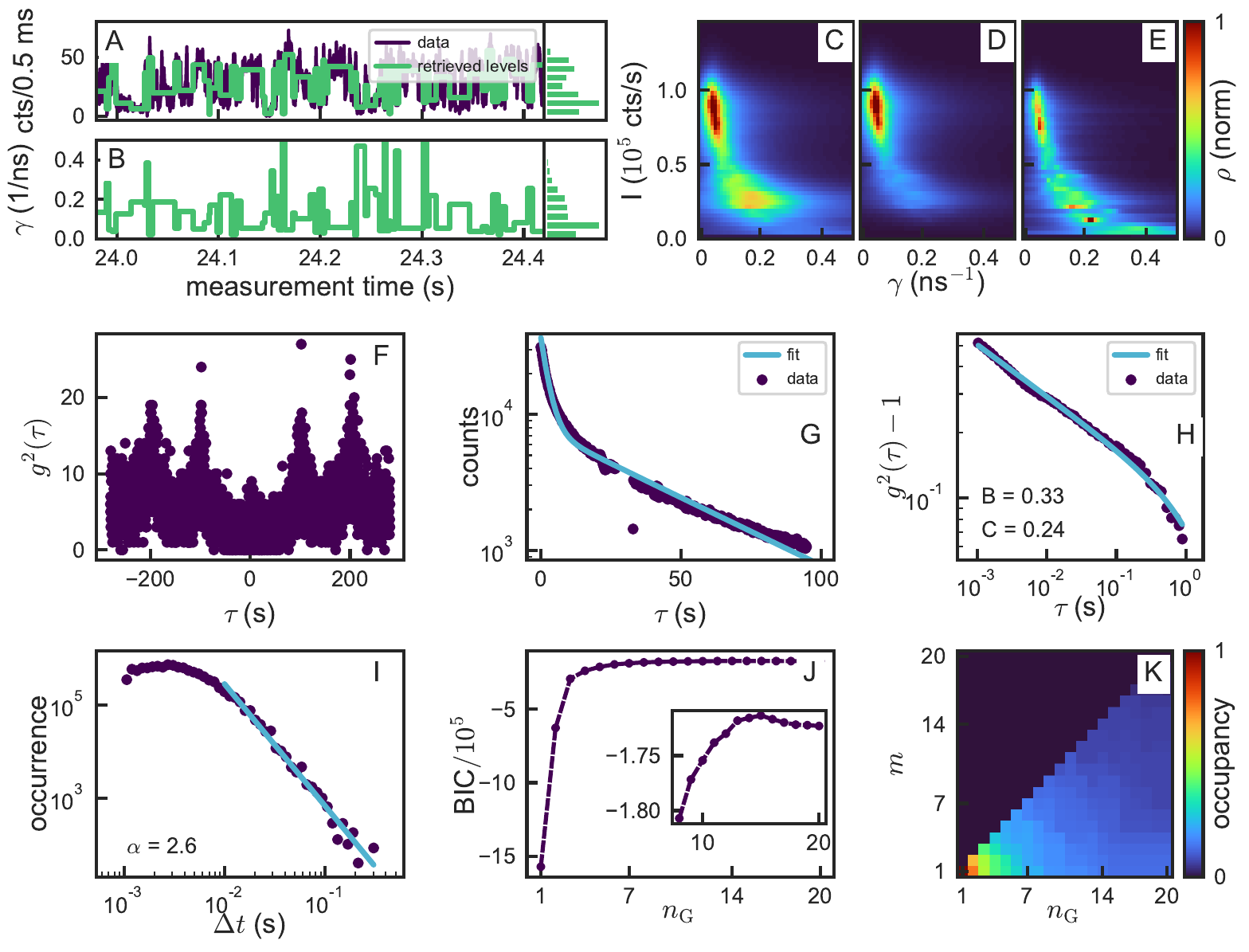}
 \end{framed}\caption{Summary of observations on dot 19 of 40}
\end{figure*}\clearpage 

\begin{figure*}\begin{framed}
  \includegraphics[angle=90,width=1.0\linewidth]{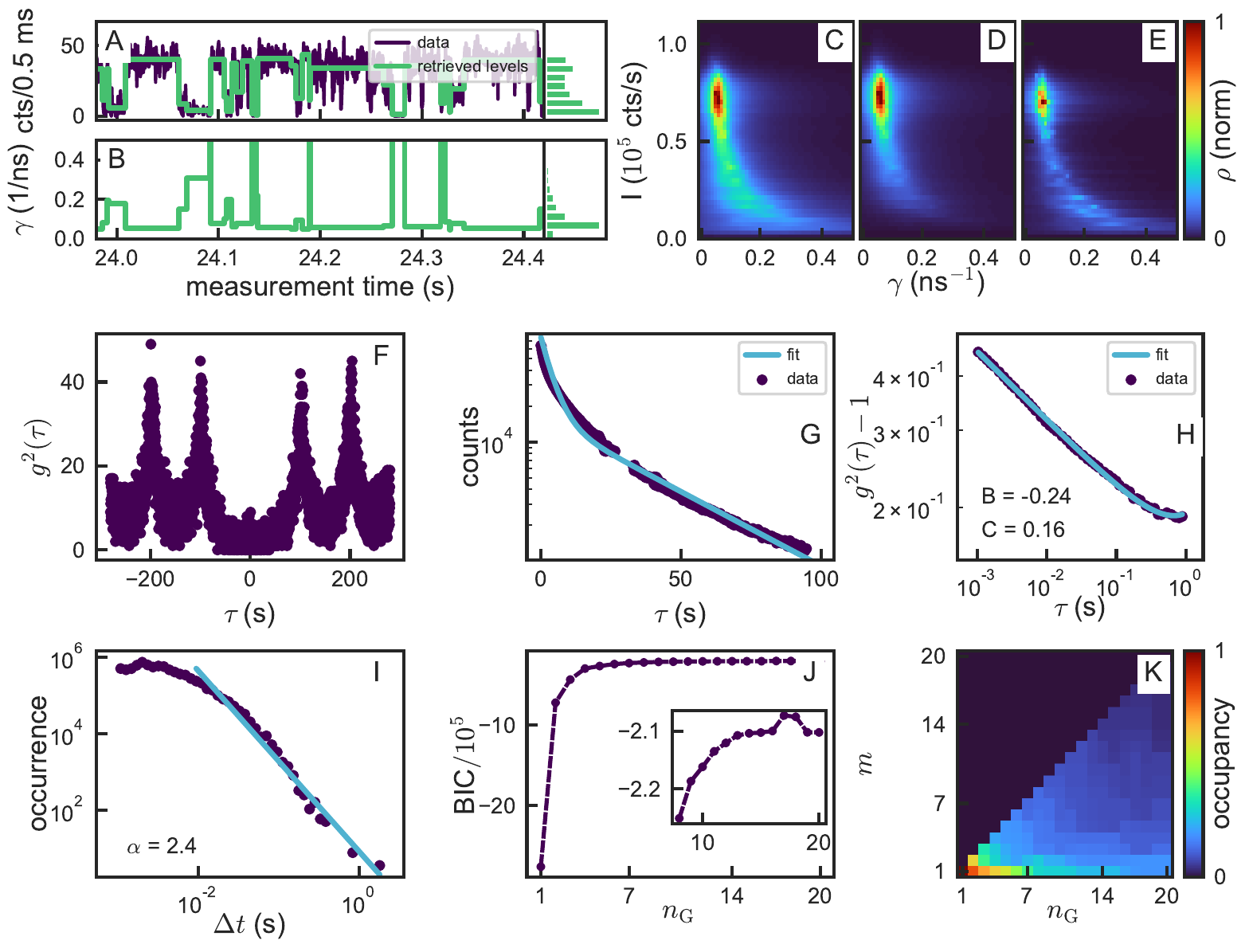}
 \end{framed}\caption{Summary of observations on dot 20 of 40}
\end{figure*}\clearpage


\begin{figure*}\begin{framed}
  \includegraphics[angle=90,width=1.0\linewidth]{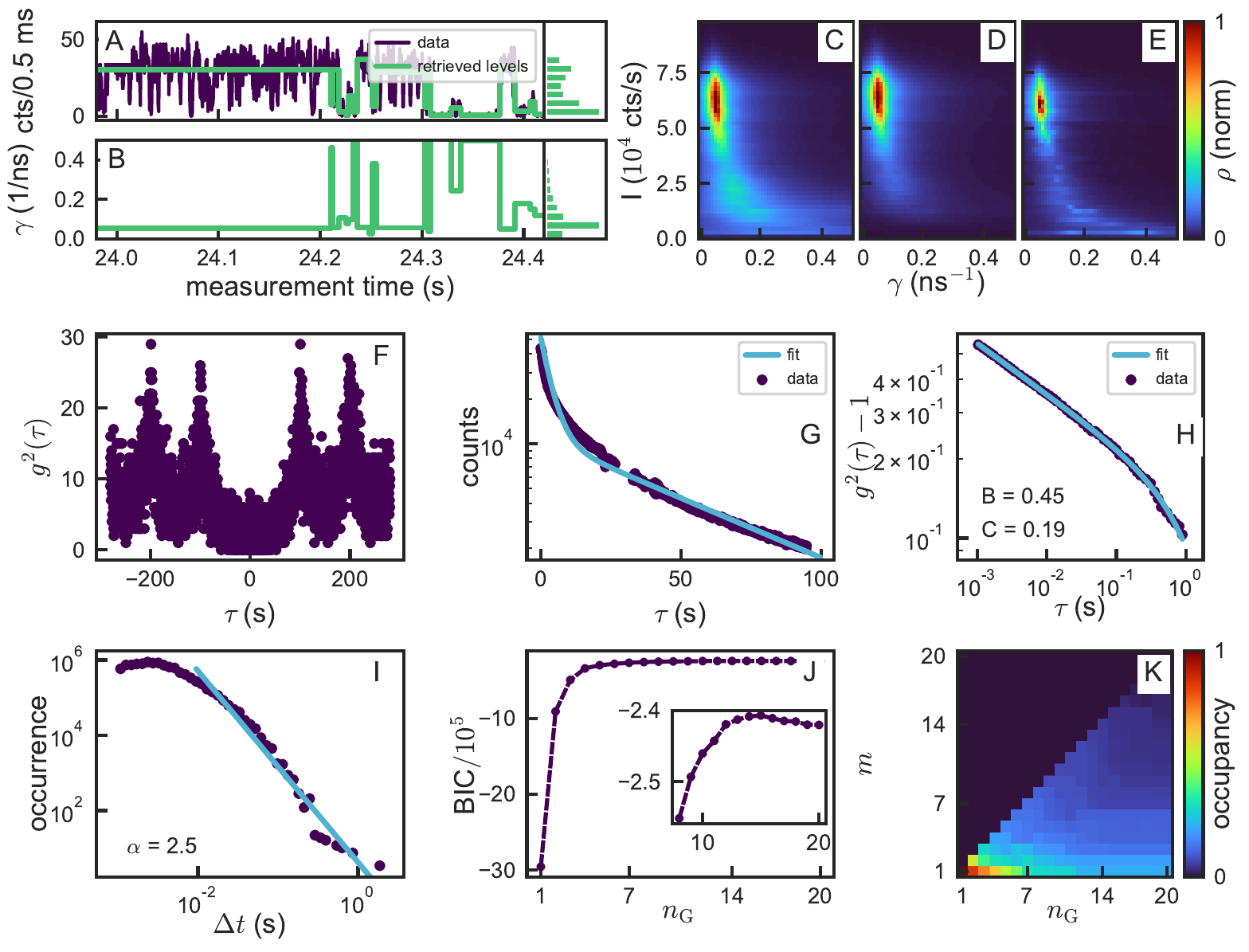}
 \end{framed}\caption{Summary of observations on dot 21 of 40}
\end{figure*}\clearpage 

\begin{figure*}\begin{framed}
  \includegraphics[angle=90,width=1.0\linewidth]{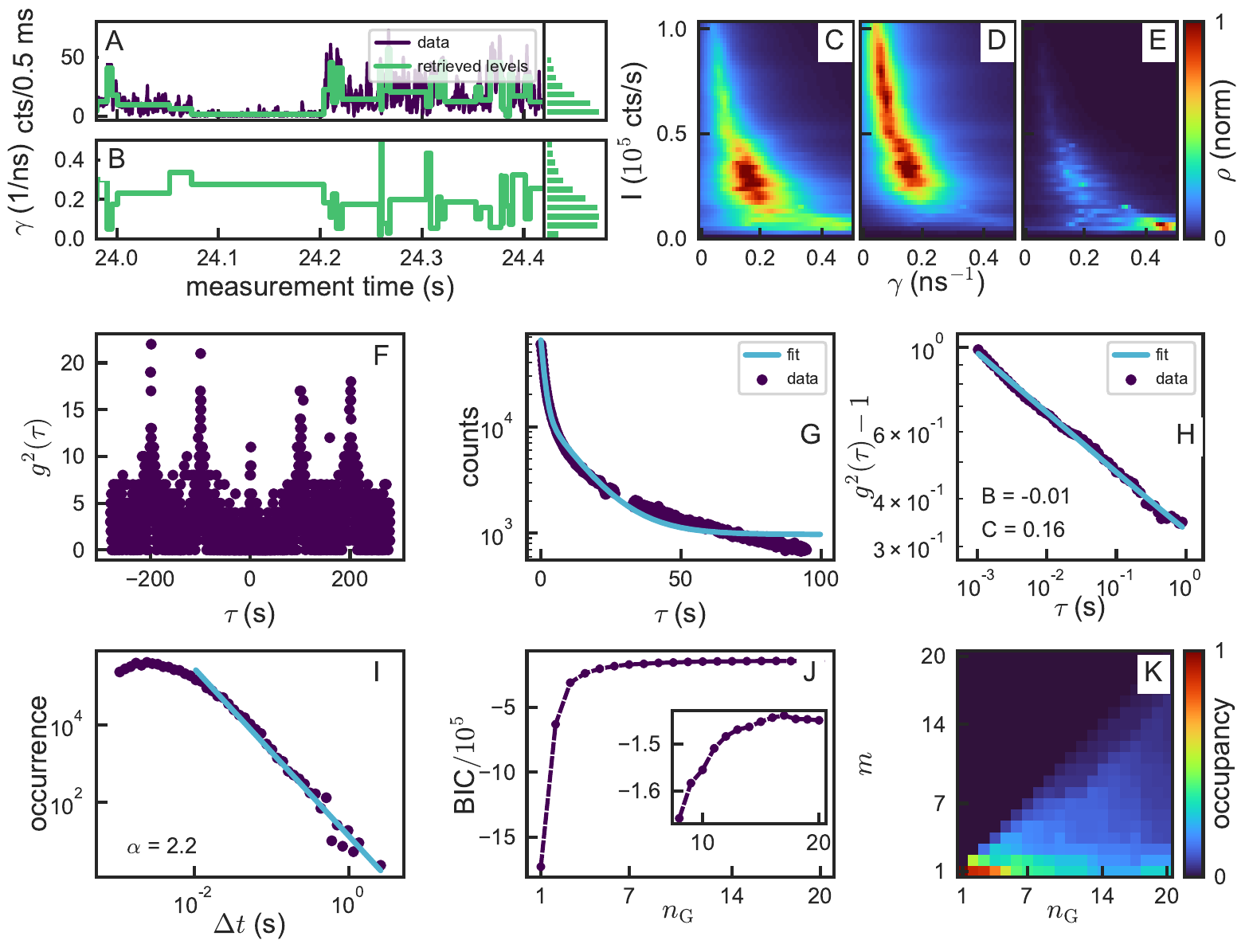}
 \end{framed}\caption{Summary of observations on dot 22 of 40}
\end{figure*}\clearpage 

\begin{figure*}\begin{framed}
  \includegraphics[angle=90,width=1.0\linewidth]{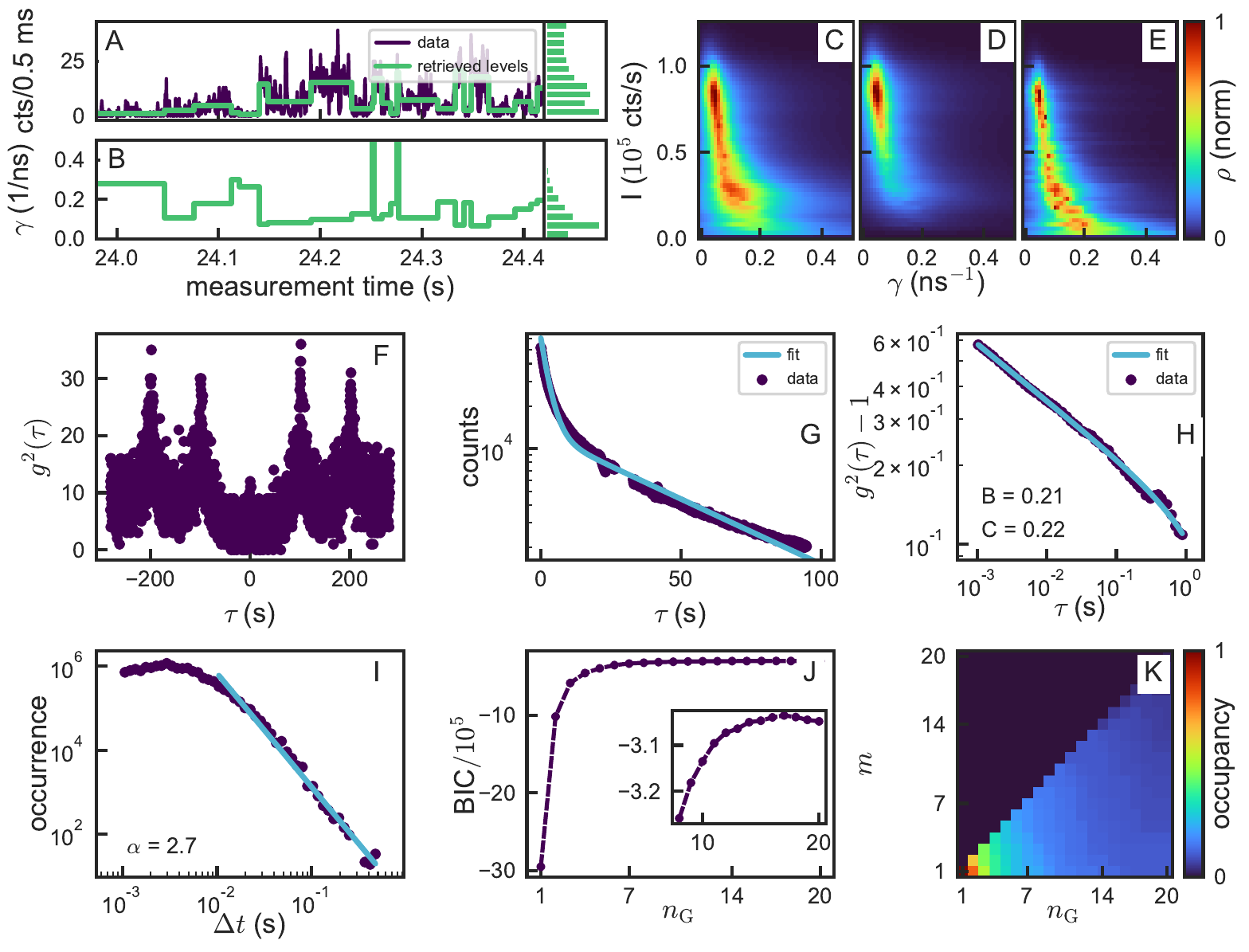}
 \end{framed}\caption{Summary of observations on dot 23 of 40}
\end{figure*}\clearpage 

\begin{figure*}\begin{framed}
  \includegraphics[angle=90,width=1.0\linewidth]{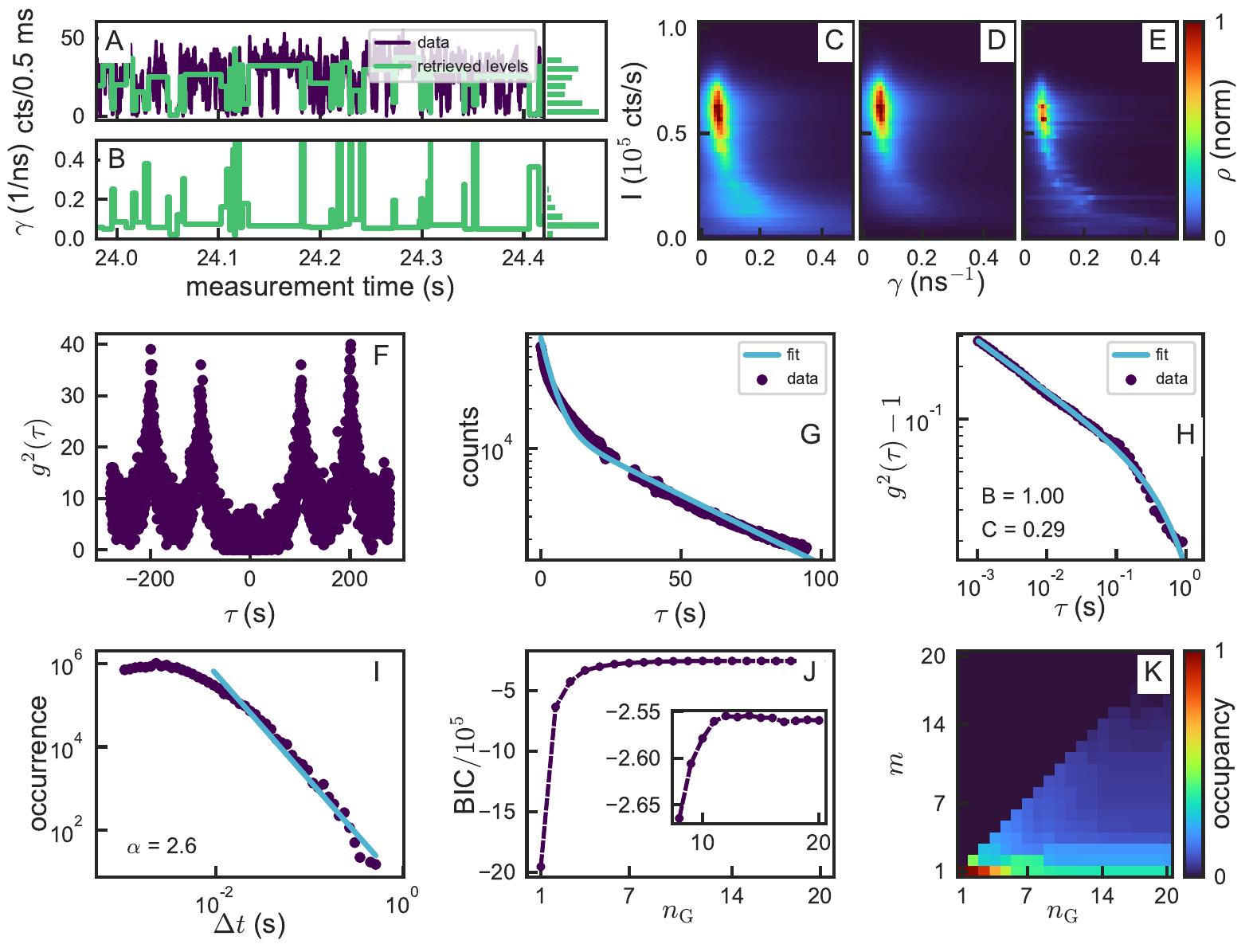}
 \end{framed}\caption{Summary of observations on dot 24 of 40}
\end{figure*}\clearpage

\begin{figure*}\begin{framed}
  \includegraphics[angle=90,width=1.0\linewidth]{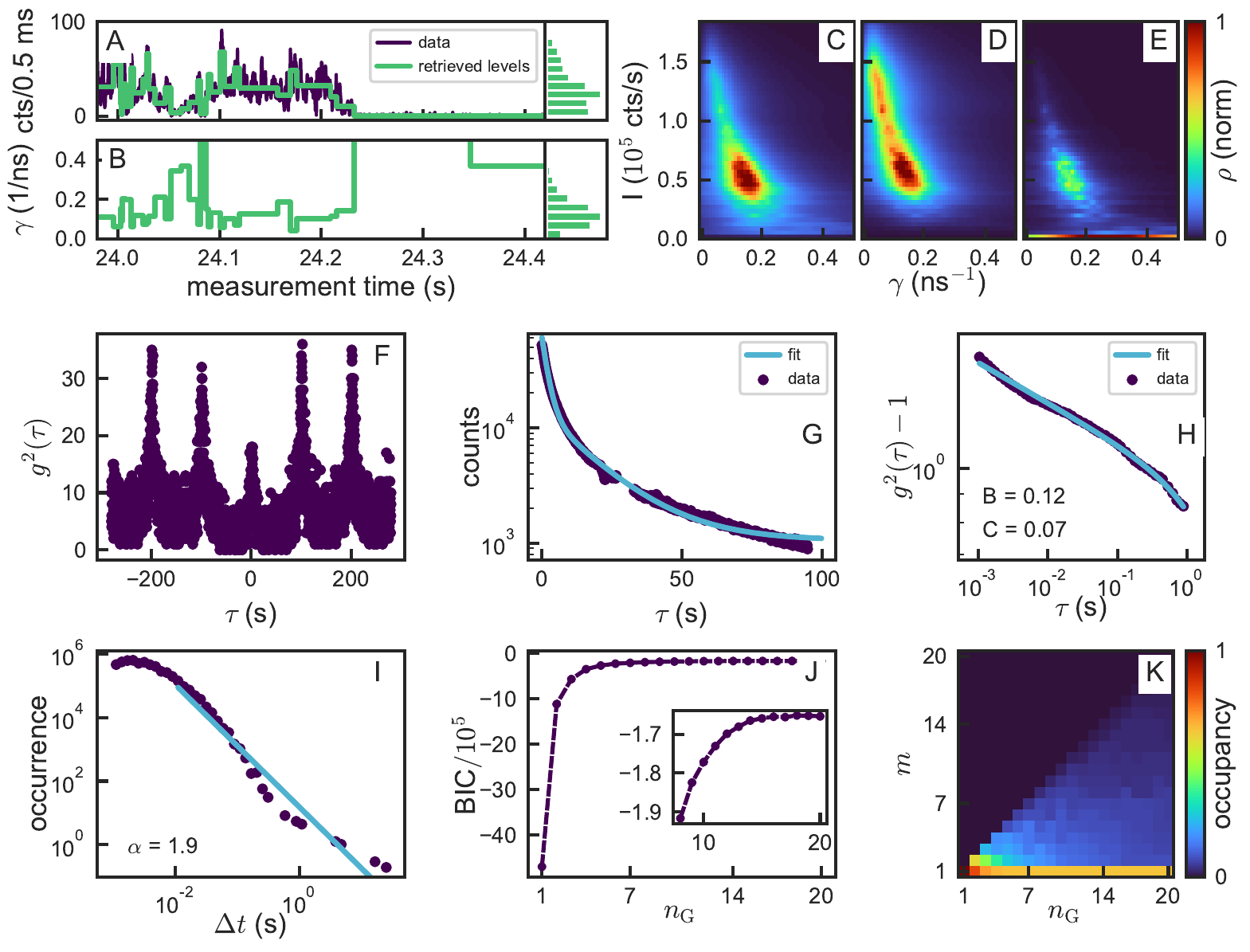}
 \end{framed}\caption{Summary of observations on dot 25 of 40}
\end{figure*}\clearpage 

\begin{figure*}\begin{framed}
  \includegraphics[angle=90,width=1.0\linewidth]{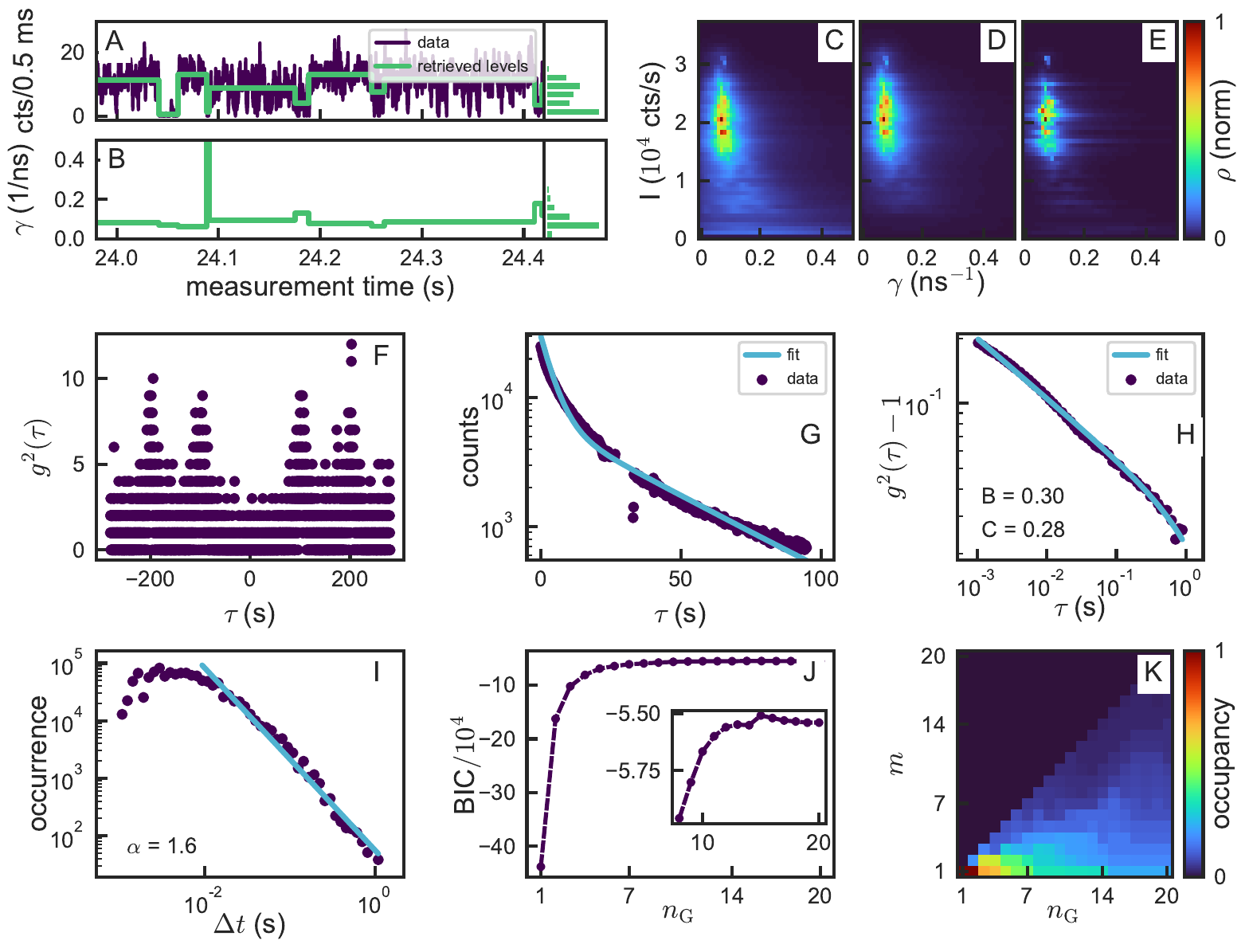}
 \end{framed}\caption{Summary of observations on dot 26 of 40}
\end{figure*}\clearpage 

\begin{figure*}\begin{framed}
  \includegraphics[angle=90,width=1.0\linewidth]{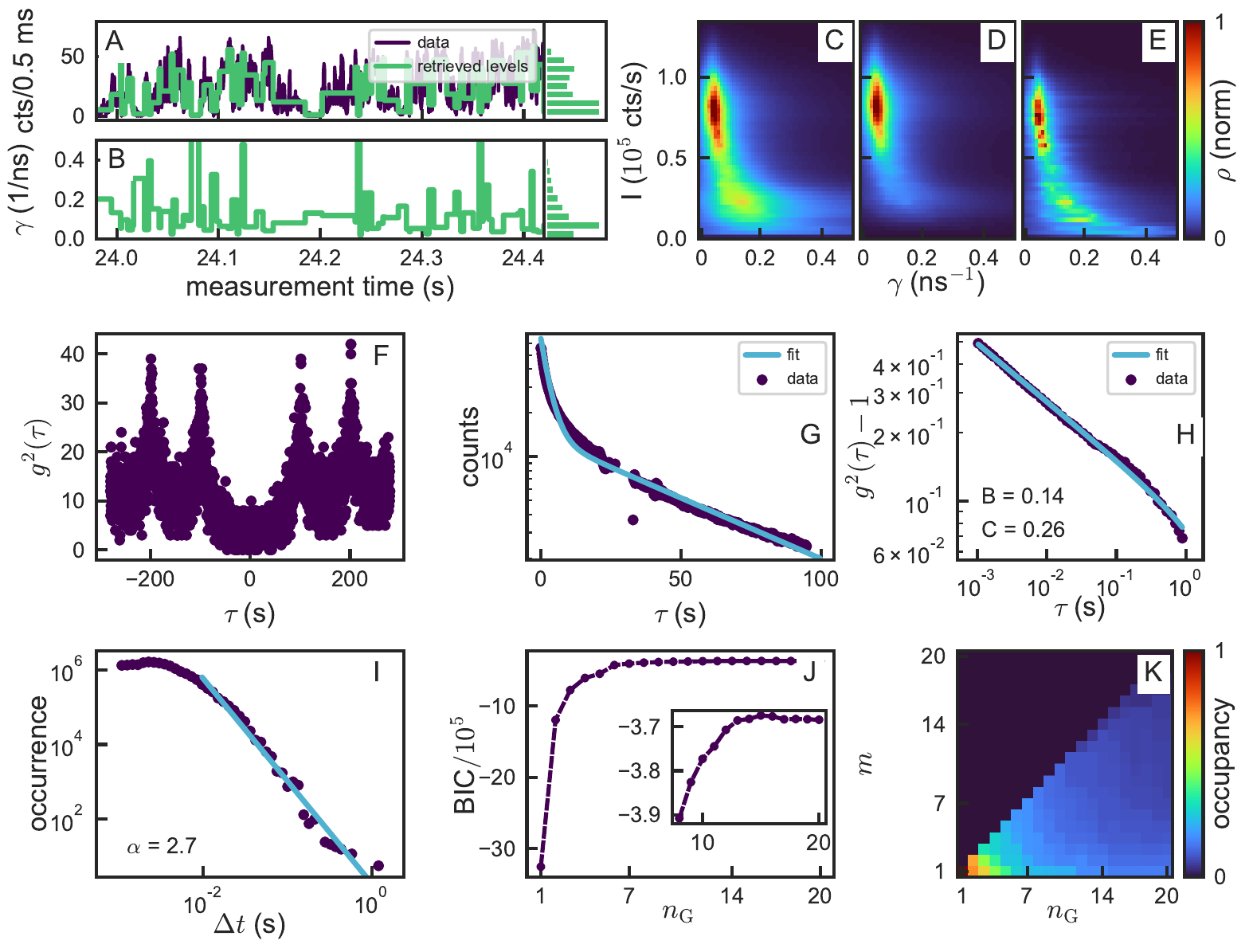}
 \end{framed}\caption{Summary of observations on dot 27 of 40}
\end{figure*}\clearpage 

\begin{figure*}\begin{framed}
  \includegraphics[angle=90,width=1.0\linewidth]{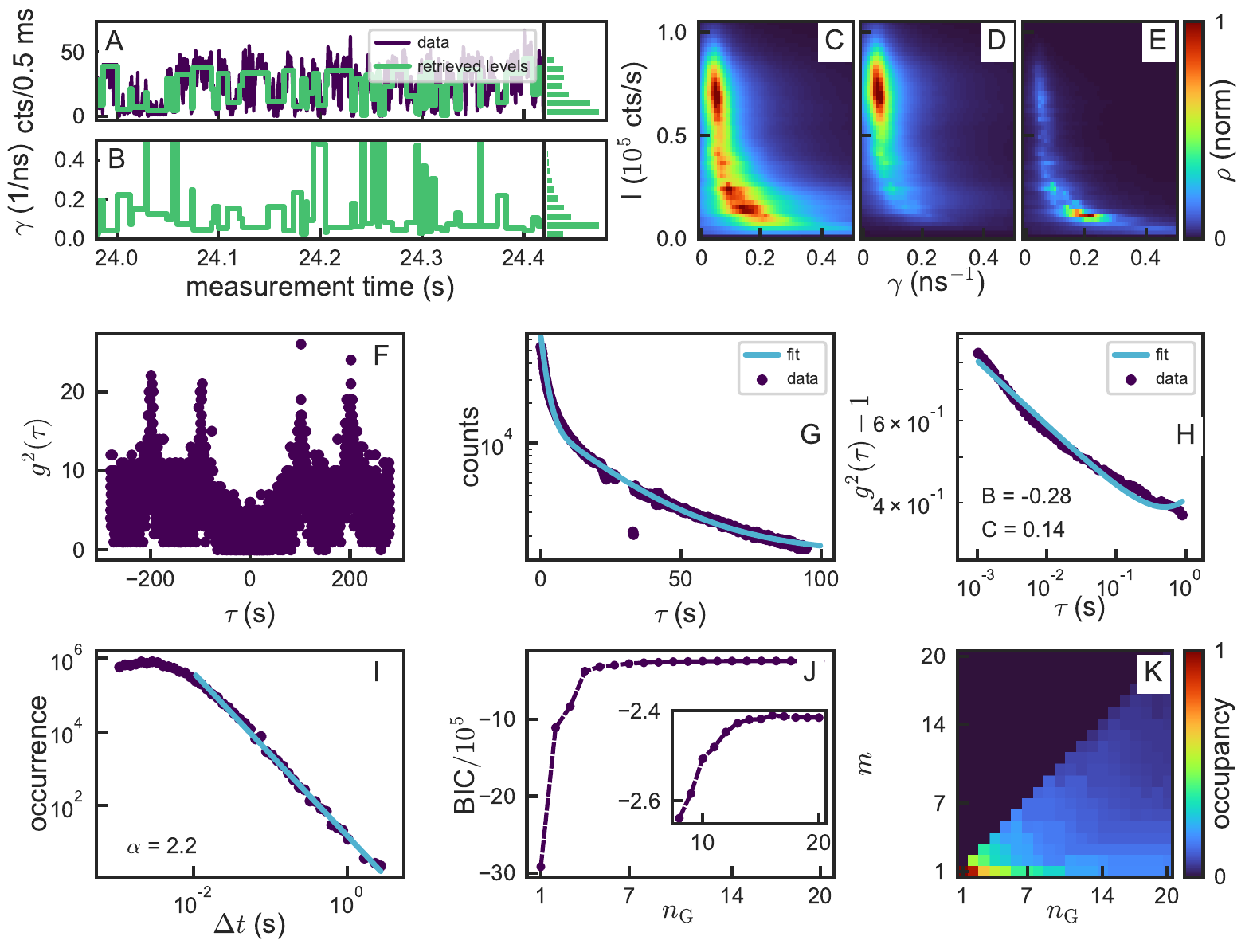}
 \end{framed}\caption{Summary of observations on dot 28 of 40}
\end{figure*}\clearpage

\begin{figure*}\begin{framed}
  \includegraphics[angle=90,width=1.0\linewidth]{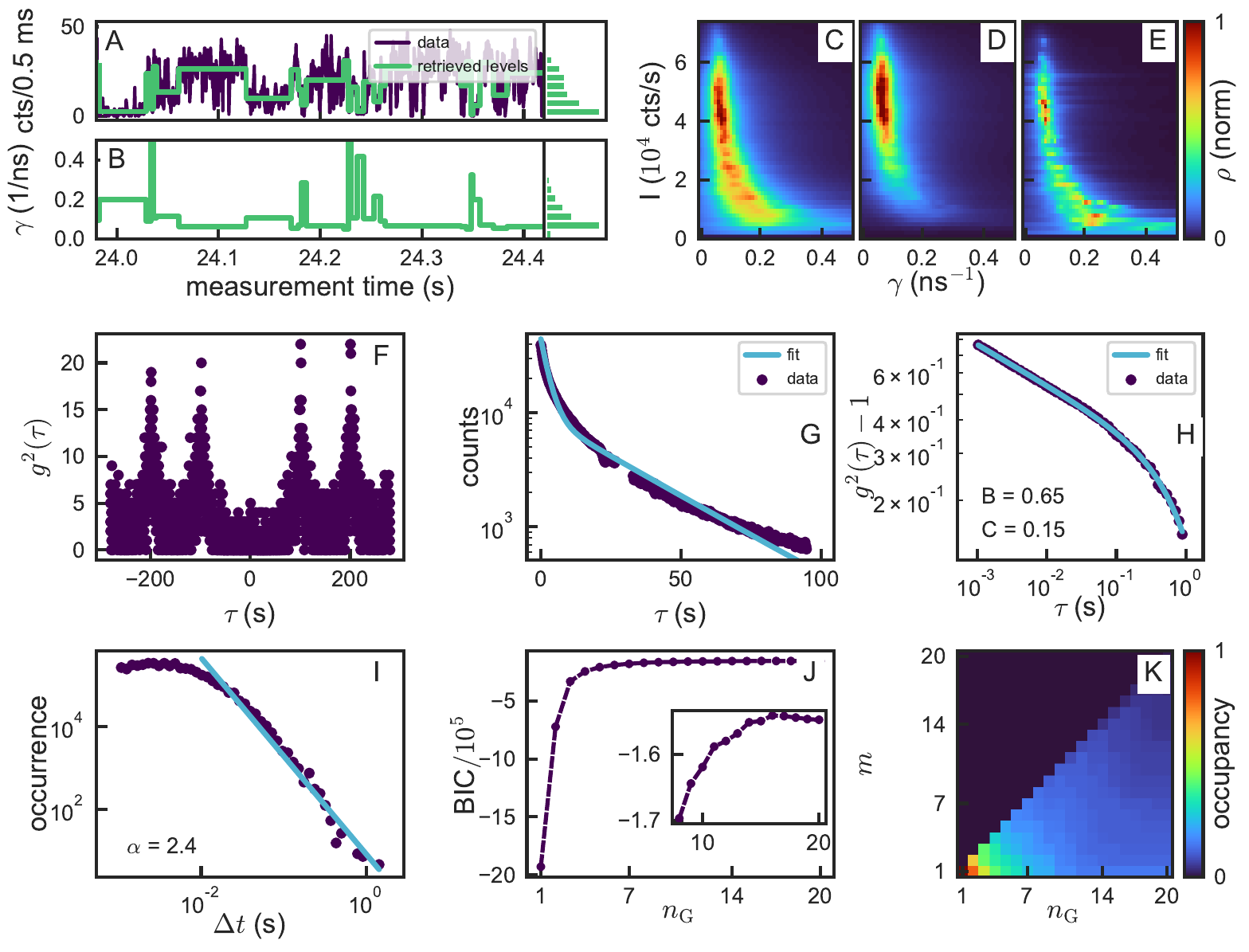}
 \end{framed}\caption{Summary of observations on dot 29 of 40}
\end{figure*}\clearpage

\begin{figure*}\begin{framed}
  \includegraphics[angle=90,width=1.0\linewidth]{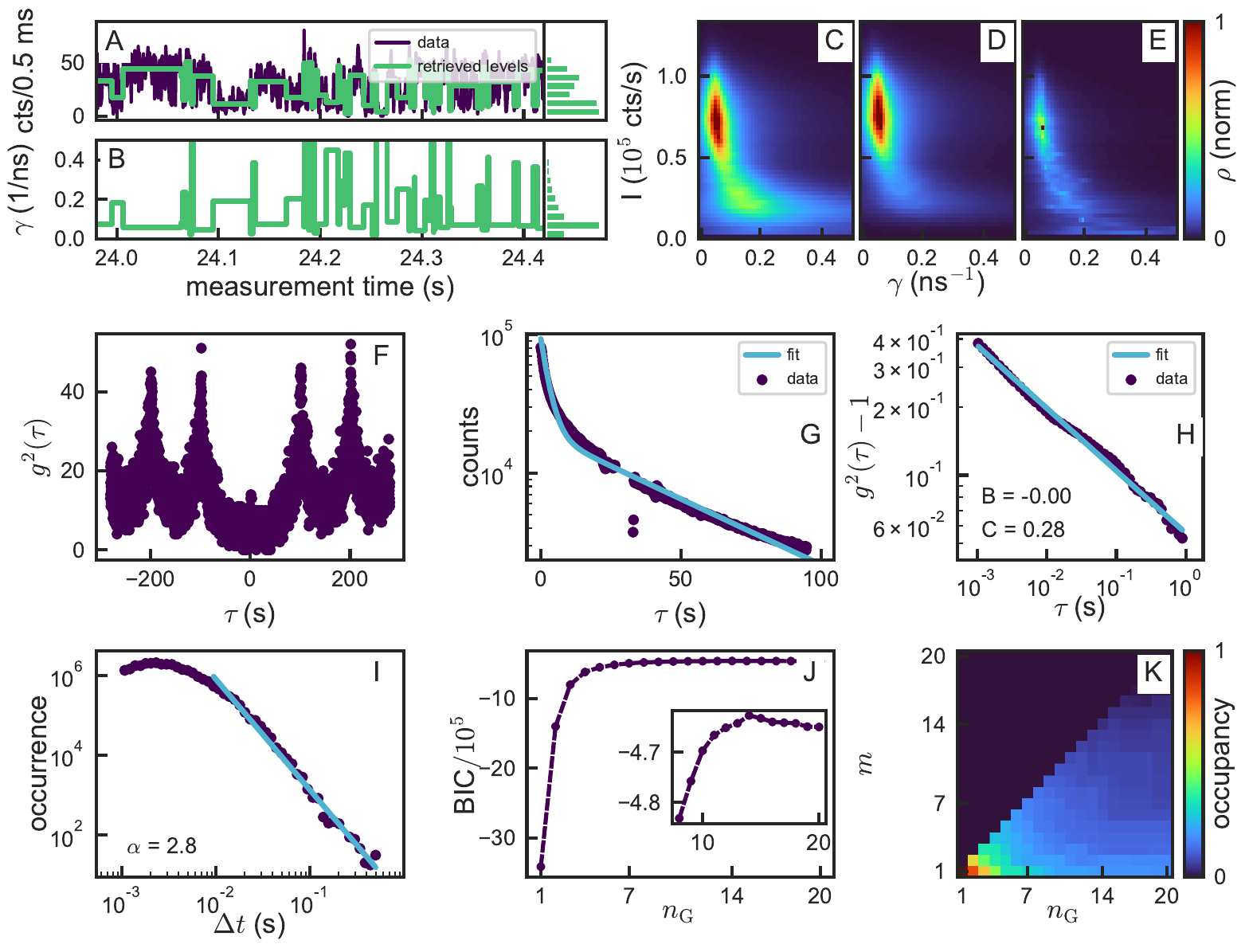}
 \end{framed}\caption{Summary of observations on dot 30 of 40}
\end{figure*}\clearpage 


\begin{figure*}\begin{framed}
  \includegraphics[angle=90,width=1.0\linewidth]{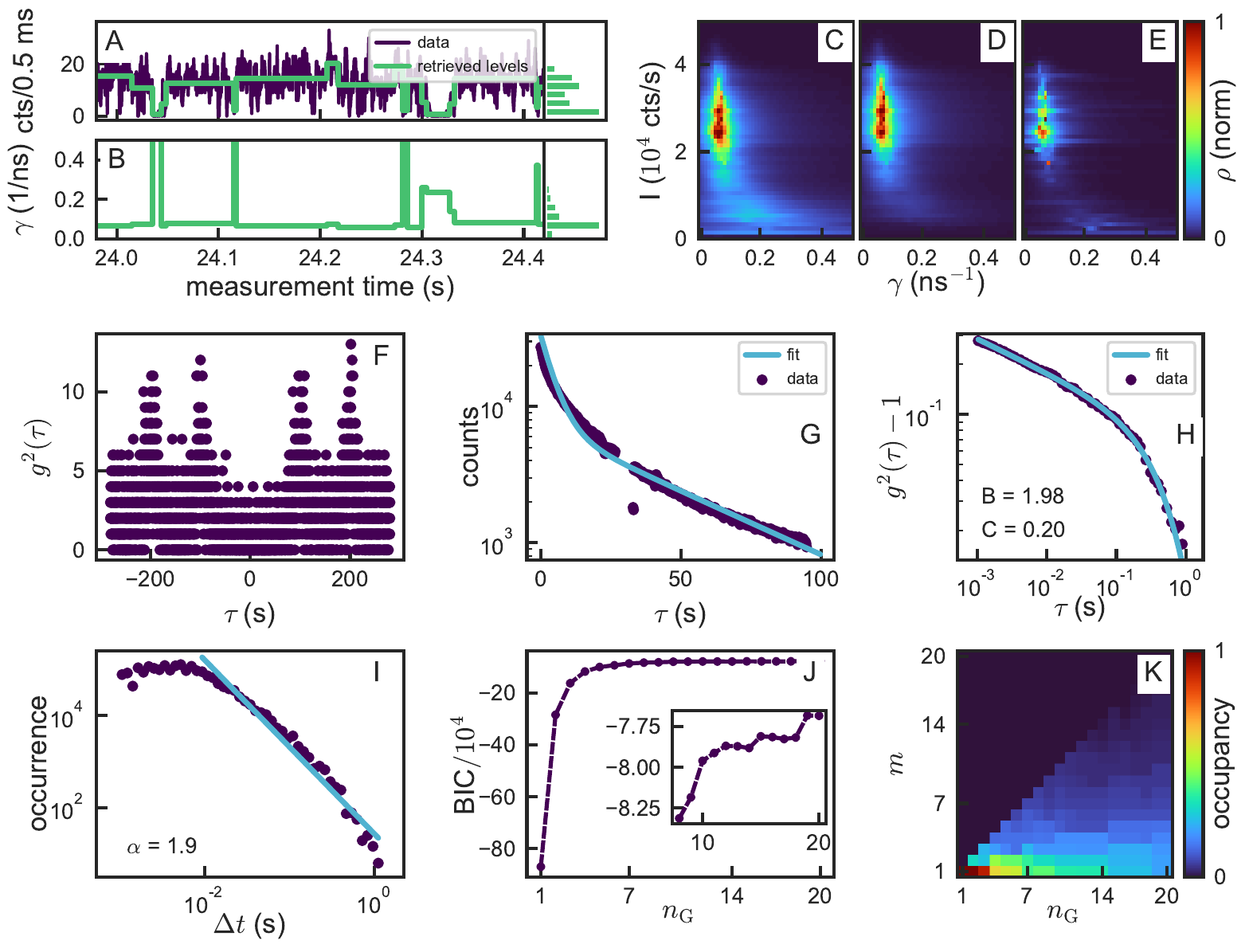}
 \end{framed}\caption{Summary of observations on dot 31 of 40}
\end{figure*}\clearpage 

\begin{figure*}\begin{framed}
  \includegraphics[angle=90,width=1.0\linewidth]{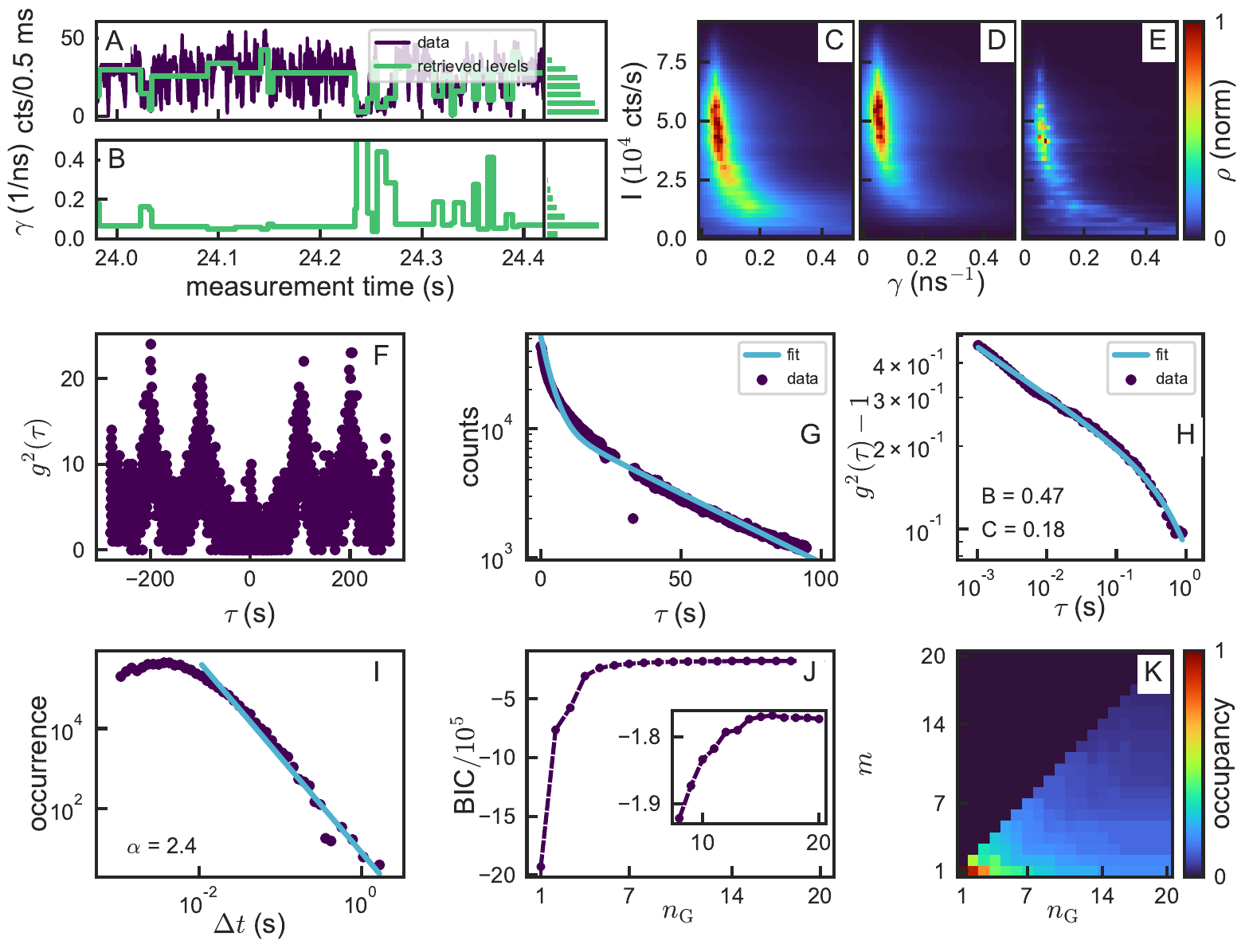}
 \end{framed}\caption{Summary of observations on dot 32 of 40}
\end{figure*}\clearpage 

\begin{figure*}\begin{framed}
  \includegraphics[angle=90,width=1.0\linewidth]{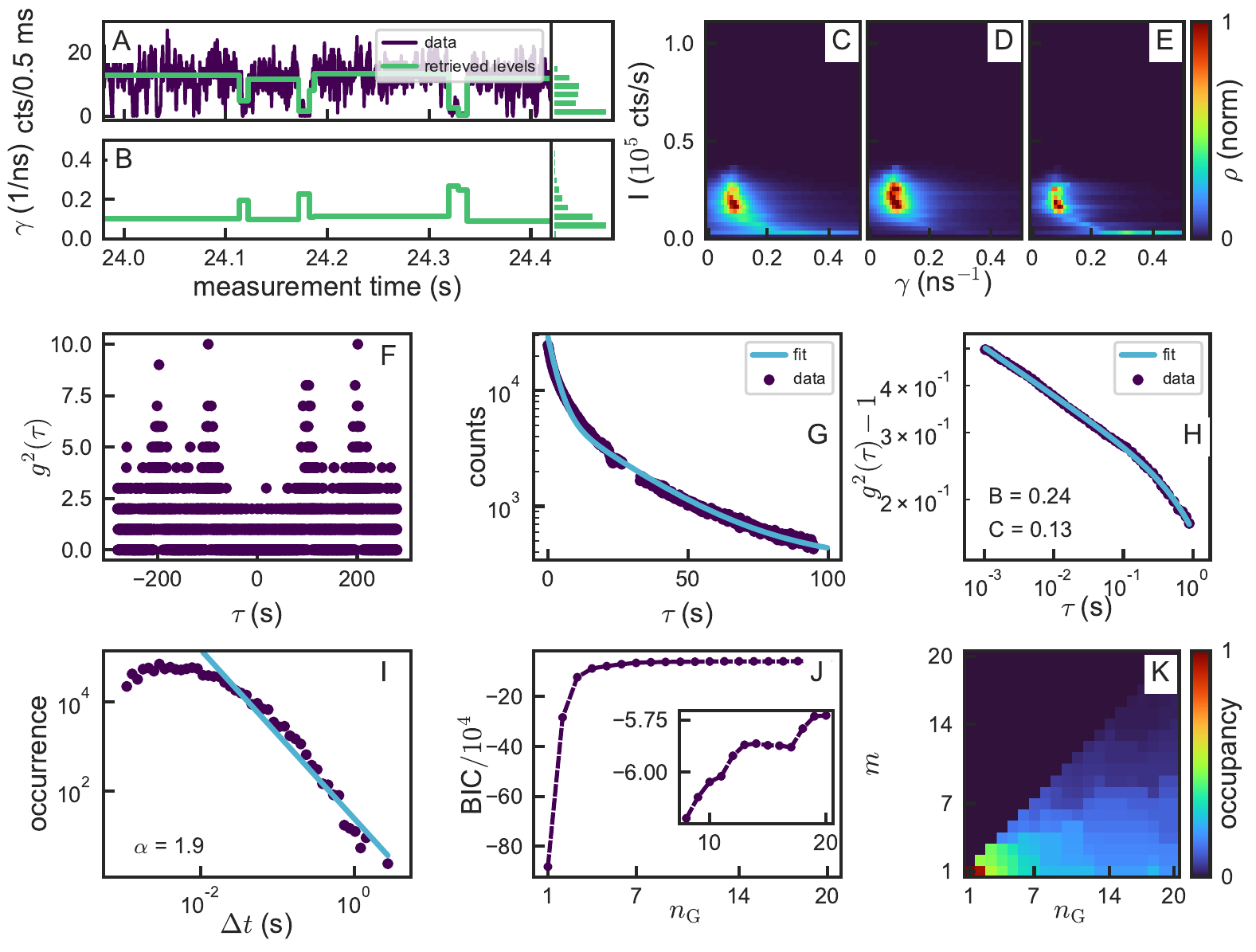}
 \end{framed}\caption{Summary of observations on dot 33 of 40}
\end{figure*}\clearpage 

\begin{figure*}\begin{framed}
  \includegraphics[angle=90,width=1.0\linewidth]{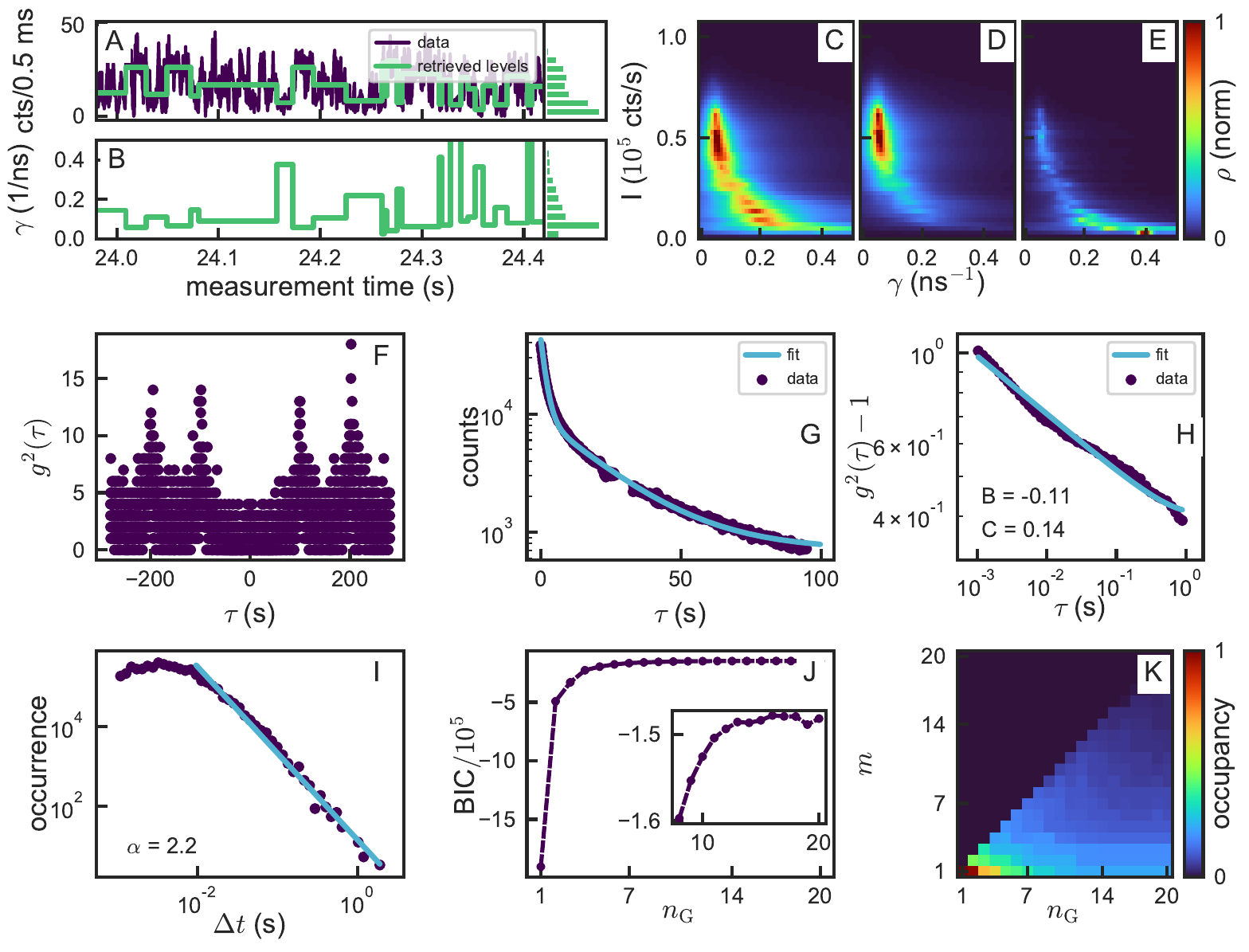}
 \end{framed}\caption{Summary of observations on dot 34 of 40}
\end{figure*}\clearpage

\begin{figure*}\begin{framed}
  \includegraphics[angle=90,width=1.0\linewidth]{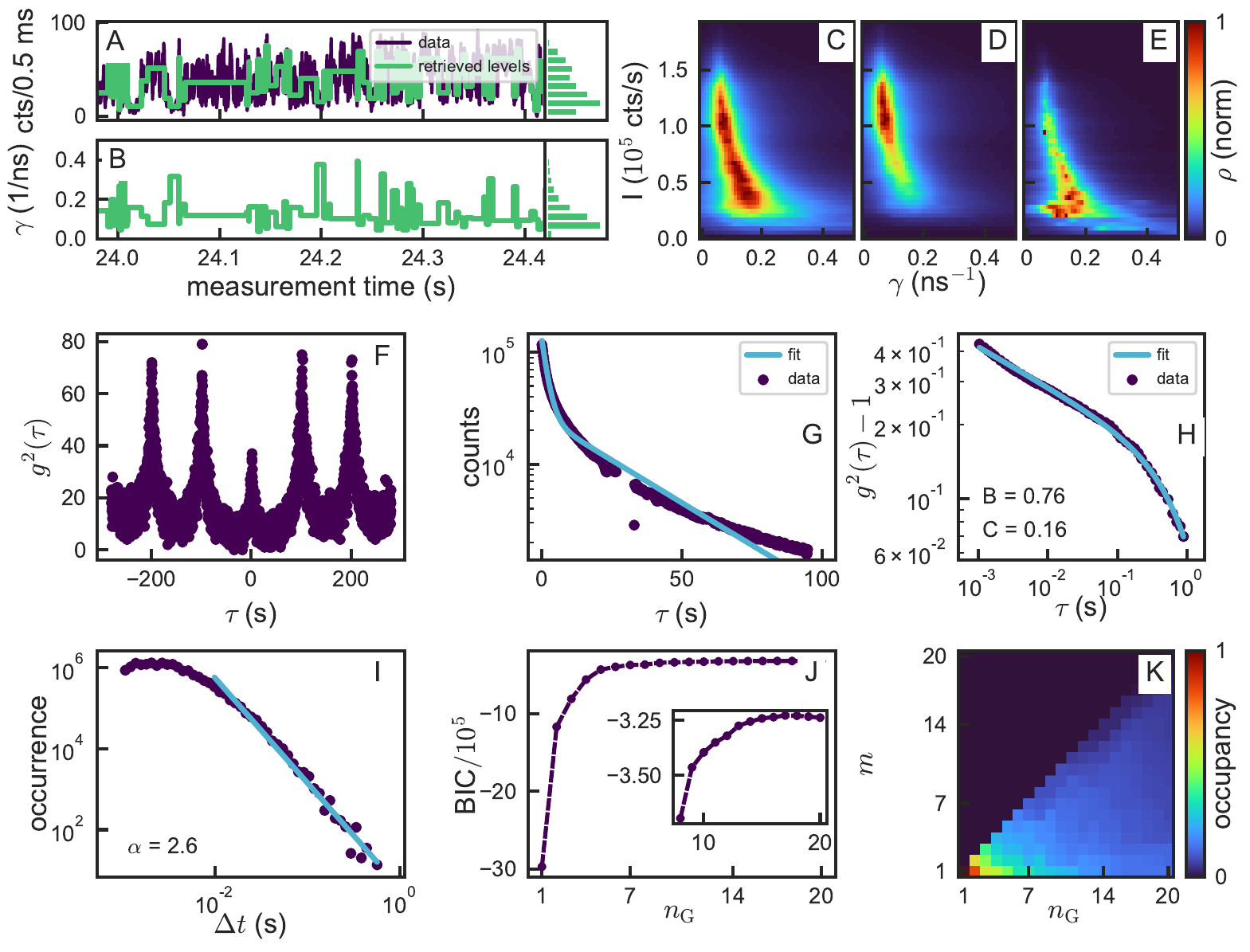}
 \end{framed}\caption{Summary of observations on dot 35 of 40}
\end{figure*}\clearpage 

\begin{figure*}\begin{framed}
  \includegraphics[angle=90,width=1.0\linewidth]{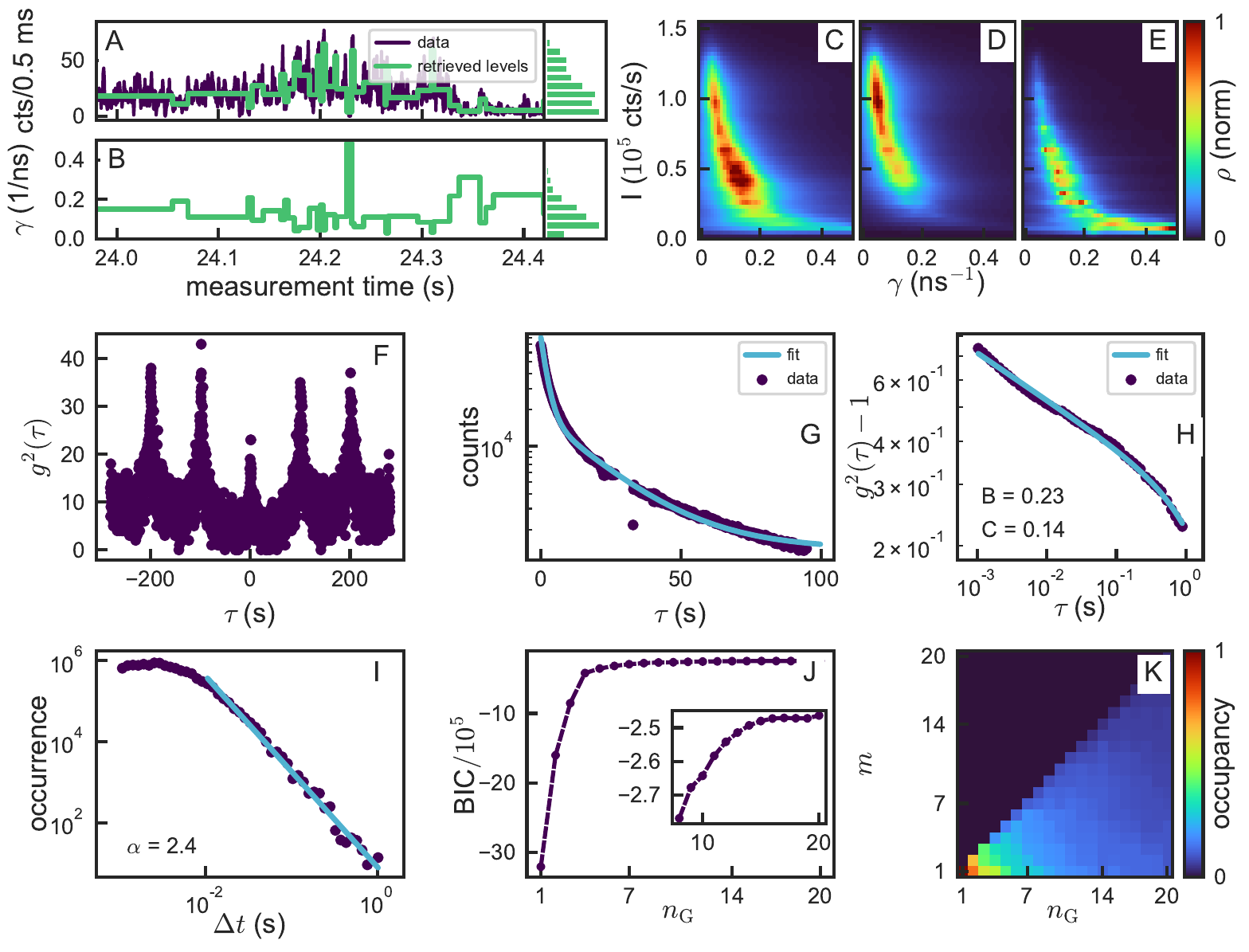}
 \end{framed}\caption{Summary of observations on dot 36 of 40}
\end{figure*}\clearpage 

\begin{figure*}\begin{framed}
  \includegraphics[angle=90,width=1.0\linewidth]{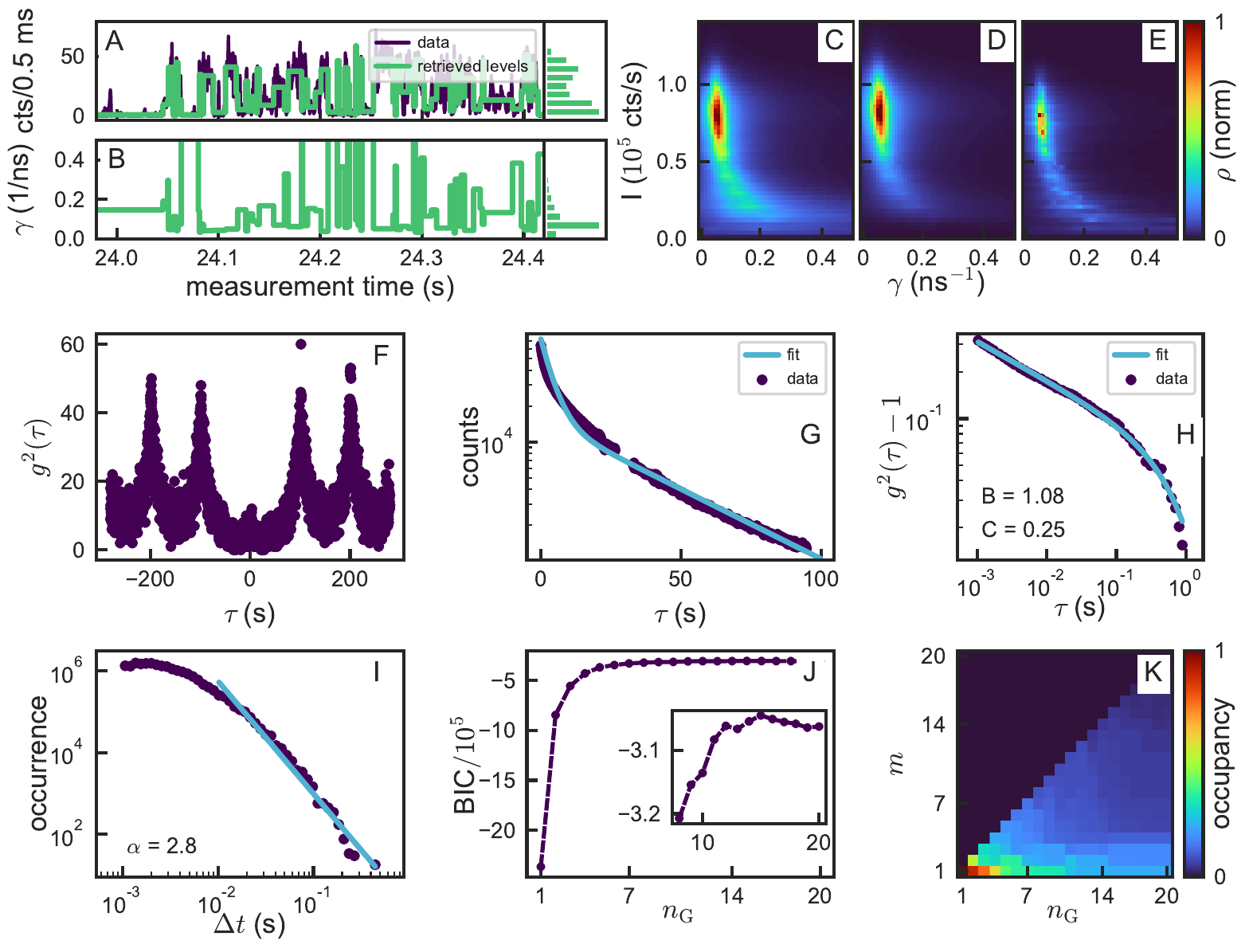}
 \end{framed}\caption{Summary of observations on dot 37 of 40}
\end{figure*}\clearpage 

\begin{figure*}\begin{framed}
  \includegraphics[angle=90,width=1.0\linewidth]{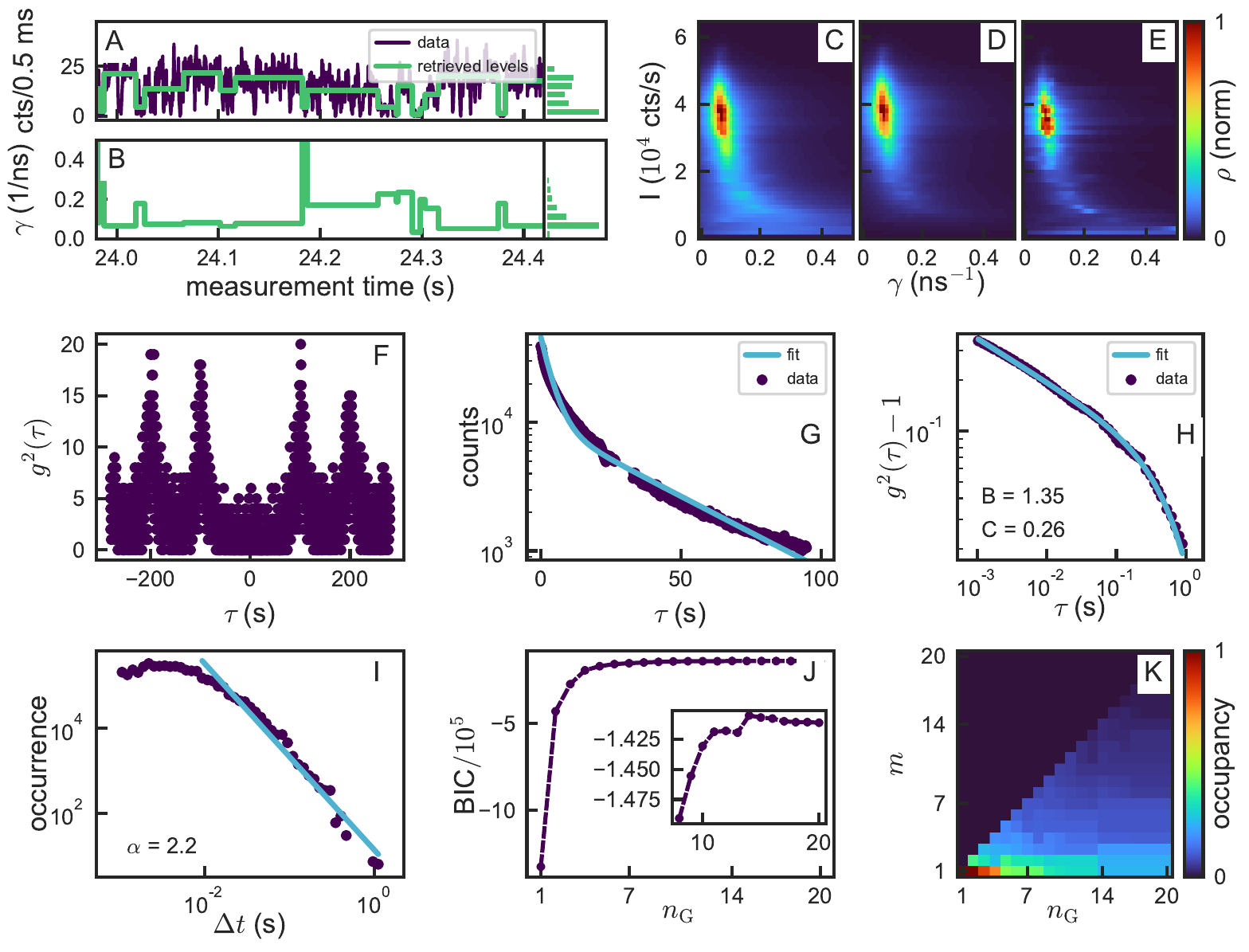}
 \end{framed}\caption{Summary of observations on dot 38 of 40}
\end{figure*}\clearpage 

\begin{figure*}\begin{framed}
  \includegraphics[angle=90,width=1.0\linewidth]{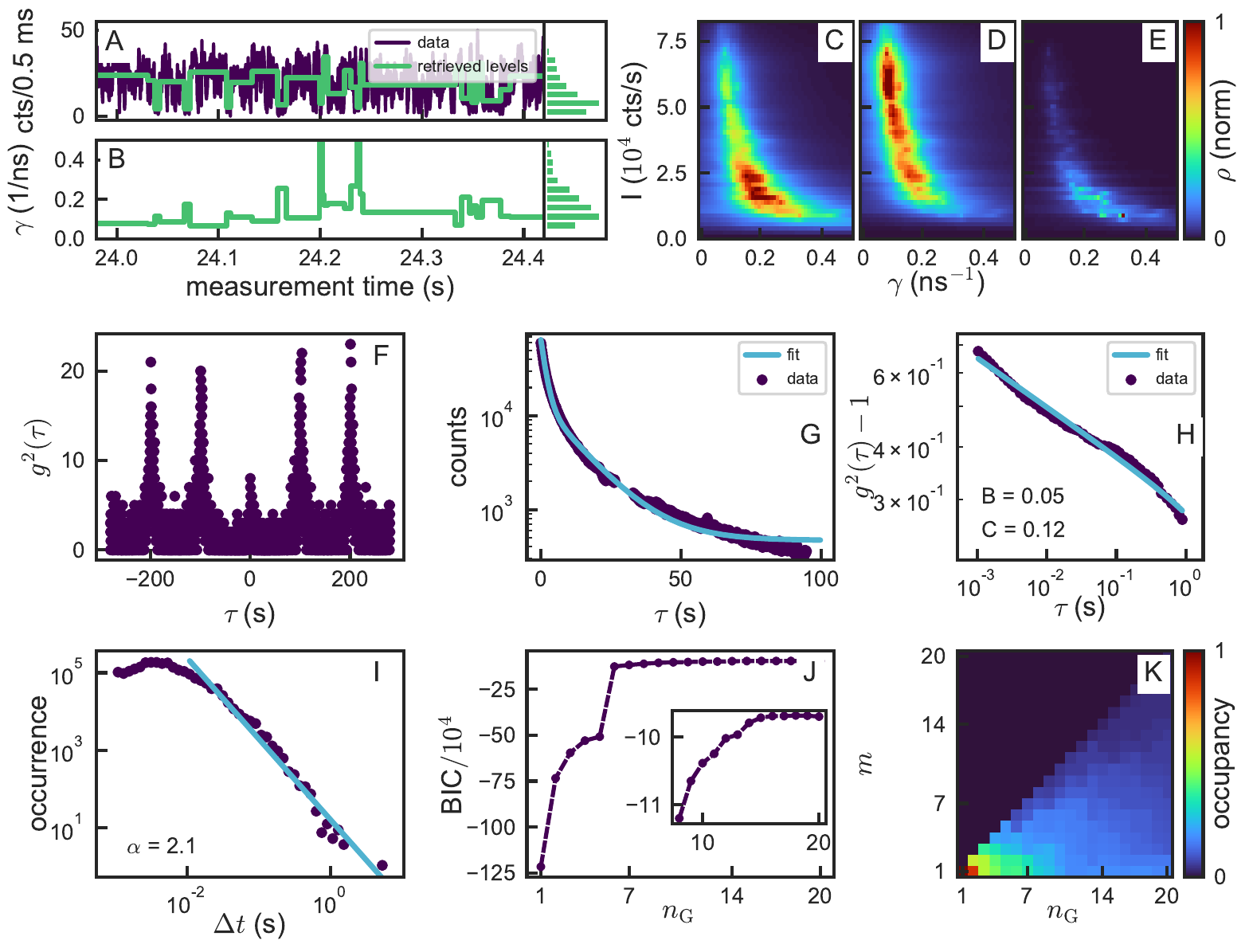}
 \end{framed}\caption{Summary of observations on dot 39 of 40}
\end{figure*}\clearpage 

\begin{figure*}\begin{framed}
  \includegraphics[angle=90,width=1.0\linewidth]{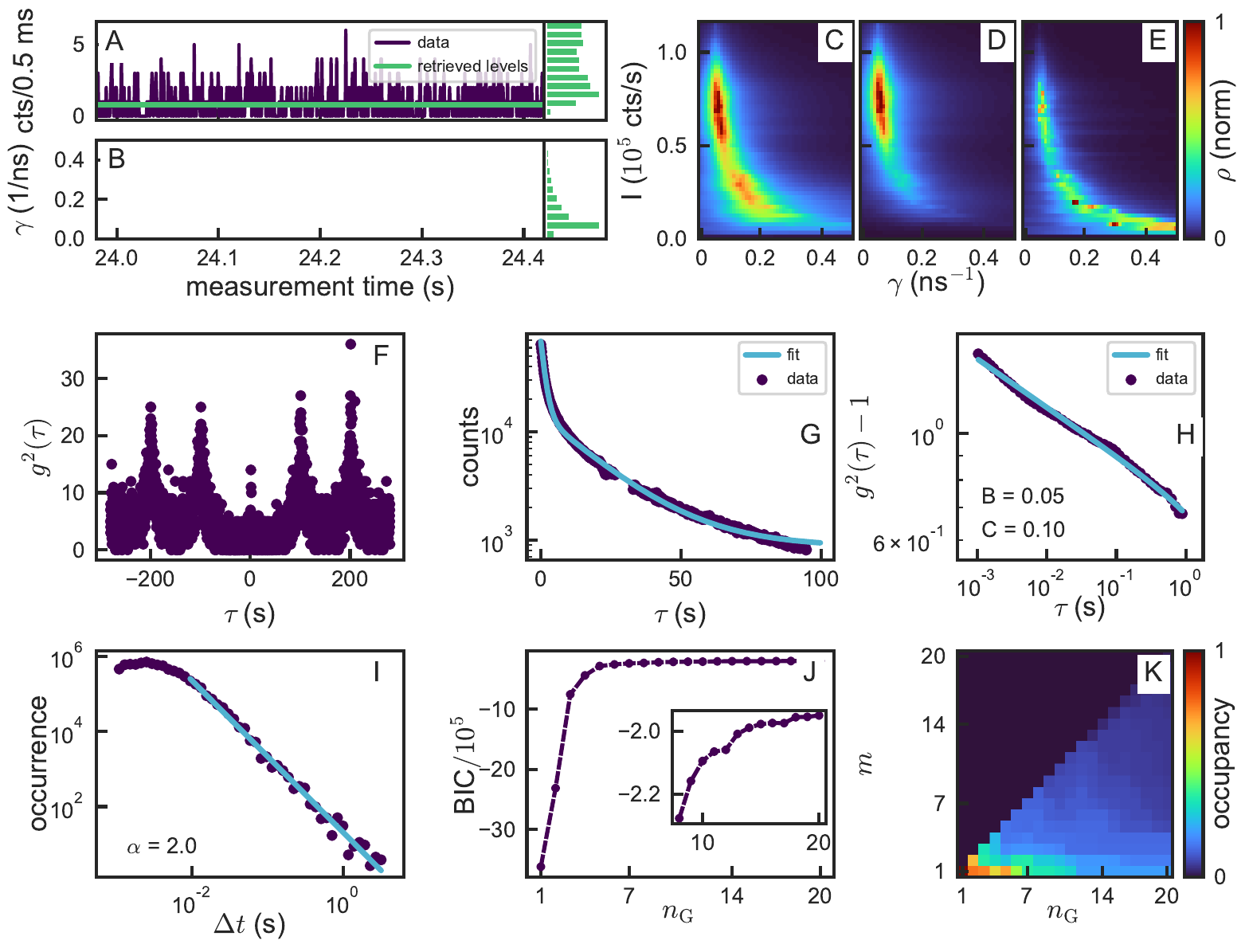}
 \end{framed}\caption{Summary of observations on dot 40 of 40}
\end{figure*}\clearpage

 \end{document}